\documentclass[aps,preprint,nofootinbib,preprintnumbers,eqsecnum]{revtex4-1}

\usepackage{color}

\newcommand{\nb}[1]{\color{blue}}

\newcommand{\HL}[1]{{\bf \textcolor{magenta}{#1}}}
\newcommand{\hl}[1]{\color{magenta}}

\usepackage[
	  pagebackref=false,
	  colorlinks=true,
      linkcolor=blue,
      urlcolor=blue,
      filecolor=black,
      citecolor=red,
      pdfstartview=FitV,
      pdftitle={},
        pdfauthor={Thomas Faulkner, Hong Liu, Mukund Rangamani},
        pdfsubject={},
        pdfkeywords={},
        pdfpagemode=None,
        bookmarksopen=true
      ]{hyperref}

\usepackage[normalem]{ulem}
\usepackage{amsmath}
\usepackage{enumerate}
\usepackage{amsfonts}
\usepackage{epsfig}
\usepackage{mathbbol}

\def\Tr{\mathop{\rm Tr}}
\def\tr{\mathop{\rm tr}}

\newcommand\half{{\ensuremath{\frac{1}{2}}}}
\newcommand\p{\ensuremath{\partial}}

\newcommand\vev[1]{{\ensuremath{\left\langle{#1}\right\rangle}}}

\newcommand{\be}{\begin{equation}}
\newcommand{\ee}{\end{equation}}
\newcommand{\bea}{\begin{eqnarray}}
\newcommand{\eea}{\end{eqnarray}}
\newcommand{\bega}{\begin{gather}}
\newcommand{\eega}{\end{gather}}

\newcommand{\bi}{\begin{itemize}}
\newcommand{\ei}{\end{itemize}}
\newcommand{\ben}{\begin{enumerate}}
\newcommand{\een}{\end{enumerate}}
\newcommand{\bca}{\begin{cases}}
\newcommand{\eca}{\end{cases}}
\newcommand{\bln}{\begin{align}}
\newcommand{\eln}{\end{align}}
\newcommand{\bst}{\begin{split}}
\newcommand{\est}{\end{split}}
\def\ie{\begin{equation}\begin{aligned}}
\def\fe{\end{aligned}\end{equation}}
\newcommand{\bma}{\le(\begin{matrix}}
\newcommand{\ema}{\end{matrix}\ri)}

\newcommand\al{{\alpha}}
\newcommand\ep{\epsilon}
\newcommand\sig{\sigma}
\newcommand\Sig{\Sigma}
\newcommand\lam{\lambda}
\newcommand\Lam{\Lambda}
\newcommand\om{\omega}
\newcommand\Om{\Omega}

\newcommand\ga{{\ensuremath{{\gamma}}}}
\newcommand\Ga{{\ensuremath{{\Gamma}}}}
\newcommand\de{{\ensuremath{{\delta}}}}
\newcommand\De{{\ensuremath{{\Delta}}}}
\newcommand\vp{\varphi}

\newcommand\ze{\zeta}

\newcommand\da{{\dagger}}
\newcommand\nab{{\nabla}}
\newcommand\Th{{\Theta}}
\def\th{{\theta}}

\newcommand\ov{\over}
\newcommand\ha{{\half}}

\def\le{\left}
\def\ri{\right}

\newcommand\sB{{\ensuremath{{\mathcal B}}}}
\newcommand\sC{{\ensuremath{{\mathcal C}}}}

\newcommand\sG{{\ensuremath{{\mathcal G}}}}
\newcommand\sH{{\ensuremath{{\mathcal H}}}}
\newcommand\sK{{\ensuremath{{\mathcal K}}}}
\newcommand\sL{{\ensuremath{{\mathcal L}}}}

\newcommand\sN{{\ensuremath{{\mathcal N}}}}
\newcommand\sO{{\ensuremath{{\mathcal O}}}}
\newcommand\sP{{\ensuremath{{\mathcal P}}}}

\newcommand\sT{{\mathcal T}}

\newcommand\ba{{\bar a}}

\newcommand\bd{{\bar d}}

\newcommand\vx{{\vec x}}
\newcommand\vk{{\vec k}}

\newcommand\ta{{\tilde a}}
\newcommand\tb{{\tilde b}}
\newcommand\tc{{\tilde c}}

\newcommand\te{{\tilde e}}
\newcommand\tf{{\tilde f}}
\newcommand\tg{{\tilde g}}

\newcommand\tk{{\tilde k}}

\newcommand{\hmu}{{\hat \mu}}

\newcommand{\lra}{{\leftrightarrow}}

\newcommand{\fb}{{\mathfrak{b}}}
\newcommand{\fc}{{\mathfrak{c}}}

\newcommand{\ft}{{\mathfrak t}}

\newcommand{\rt}{{\rm t}}

\newcommand{\heff}{{\hbar_{\rm eff}}}

\begin{document}

\title{Effective field theory of dissipative fluids}

\preprint{MIT-CTP/4734}

\author{Michael Crossley, Paolo Glorioso and Hong Liu}
\affiliation{Center for Theoretical Physics, \\
Massachusetts
Institute of Technology,
Cambridge, MA 02139 }

\begin{abstract}

\noindent
We develop an effective field theory for dissipative fluids which governs the dynamics of long-lived gapless modes associated with conserved quantities. The resulting theory gives a path integral formulation of fluctuating hydrodynamics which systematically incorporates  nonlinear interactions of noises.
The dynamical variables are mappings between a ``fluid spacetime''
and the physical spacetime and an essential aspect of our formulation is to identify the appropriate symmetries in the
%co-moving
fluid spacetime. The theory applies to nonlinear disturbances around a general density matrix. For a thermal density matrix, we require an additional $Z_2$ symmetry, to which we refer as the local KMS condition. This leads to the standard constraints of hydrodynamics, as well as a nonlinear generalization of the Onsager relations. It also leads to an emergent supersymmetry in the classical statistical regime, and a higher derivative deformation of supersymmetry in the full quantum regime.

\end{abstract}

\today

\maketitle

\tableofcontents

\section{Introduction}

\subsection{Motivations}

Hydrodynamical phenomena are ubiquitous in nature, governing essentially all aspects of
life. Hydrodynamics has also found important applications in many areas of modern physics, from evolution of galaxies, to heavy ion collisions, to classical and quantum phase transitions.  More recently, deep connections have also emerged between hydrodynamics and the Einstein equations around black holes in
holographic duality (see e.g.~\cite{Son:2007vk,Rangamani:2009xk,Hubeny:2011hd}).

Despite its long and glorious history, hydrodynamics has so far been formulated only at the level of the equations of motion (except for the case of ideal fluids), which cannot capture effects of fluctuations.  In a fluid, however, fluctuations occur spontaneously and continuously, at both the quantum and statistical levels, the understanding of which is important for a wide variety of physical problems, including equilibrium time correlation functions (see e.g.~\cite{Boonyip,Pomeau}),
dynamical critical phenomena in classical and quantum phase transitions (see e.g.~\cite{Halperin,Kirkpatrick}),
non-equilibrium steady states (see e.g.~\cite{Senger}), and possibly turbulence (see e.g.~\cite{forster}).
In holographic duality, hydrodynamical fluctuations can help probe quantum gravitational fluctuations of a black hole. Currently, the framework for dealing with hydrodynamical fluctuations is to add fluctuating dissipative fluxes with local Gaussian distributions to the stress tensor and other conserved currents~\cite{landau1,LL} (see e.g.~\cite{Senger,Kovtun:2012rj} for recent reviews). Such a formulation does not capture nonlinear interactions among noises, nor nonlinear interactions between dynamical variables and noises, nor fluctuations of dynamical variables. The situation becomes more acute for fluctuations around non-equilibrium steady states or dynamical flows, where the presence of nontrivial backgrounds of dynamical variables
could induce new couplings and long-range correlations~\cite{Senger}.

%The lack of an action principle also makes
Another unsatisfactory aspect of the current formulation of hydrodynamics is that it is phenomenological in nature. While it works well in practice, the underlying theoretical structure is obscure. More explicitly, the equations of motion are constrained by various phenomenological conditions on the solutions.
One is that the second law of thermodynamics should be satisfied locally~\cite{LL}, namely, there should exist an entropy current whose divergence is non-negative when evaluated on any solutions. The entropy current constraint imposes inequalities on various transport parameters such as the non-negativity of viscosities and conductivities. It also gives rise to equalities relating transport coefficients. For example, for  a charged fluid at first derivative order, one of the transport coefficients is required to vanish, even though the corresponding term respects all symmetries. Another condition is the existence of a stationary equilibrium in the presence of stationary external sources, which again imposes various equalities among transport coefficients. A third condition is that the linear response matrix should be symmetric as a consequence of microscopic time reversal invariance, the so-called Onsager relations.
While these constraints appear to be enough to first order in the derivative expansion, it is not clear whether they are the complete set of constraints at higher orders.
Clearly a systematic formulation of the constraints from symmetry principles would be desirable.
%To first in derivative expansion, these conditions the hydrodynamical equations agree well with experiments to, but at %higher derivative orders it is not clear whether. Secondly, these conditions are phenomenological in nature. They %impose constraints on the equations of motion from properties of the solutions. This is not only inconvenient in %practice, but theoretically awkward.
Recently, an interesting observation was made in~\cite{Banerjee:2012iz,Jensen:2012jh,Bhattacharyya:2013lha,Bhattacharyya:2014bha} that
the equality constraints from the entropy current appear to be equivalent to those from requiring  that in a stationary equilibrium, the stress tensor and conserved currents can be derived from an equilibrium partition function. The physical origin of the coincidence, however, appeared mysterious.

In this paper, we develop a path integral formulation for dissipative fluids as a low energy effective field theory
of a general quantum statistical system, from symmetry principles.
This formulation provides a systematic treatment of statistical and quantum hydrodynamical
fluctuations at the full nonlinear level. With noises suppressed, it recovers the standard equations of motion for hydrodynamics with all the phenomenological constraints incorporated. Furthermore, we find a new set of constraints on the hydrodynamical equations of motion, which may be considered as nonlinear generalizations of  Onsager relations.
Truncating to quadratic order in noises in the action, we recover the previous formulation of fluctuating hydrodynamics based on Gaussian noises.  As illustrations, we derive actions which generalize (a variation of) the
stochastic Kardar-Parisi-Zhang equation and the relativistic stochastic Navier-Stokes equations to include nonlinear
interactions of noises.

%Our formulation should also yield a complete set of constraints on hydrodynamical equations, and elucidates the microscopic  origin of the phenomenological constraints mentioned above.

Interestingly, we also find unitarity of time evolution requires introducing in the low energy effective action additional anti-commuting fields and a BRST-type symmetry, which also survive in the classical limit. {\it Thus even incorporating classical statistical fluctuations consistently  requires anti-commuting fields.}

Our formulation also reveals connections between thermal equilibrium and supersymmetry
at a level much more general than that in the context of the Langevin equation.\footnote{See e.g.~\cite{parisi,feigelman,Gozzi:1983rk,Mallick:2010su}. See also Chap. 16 and 17 of~\cite{zinnjustin} for a nice review on supersymmetry and the Langevin equation.} In particular, we find hints of the existence of a ``quantum deformed''  supersymmetry involving an infinite number of time derivatives.
Connections between supersymmetry and hydrodynamics have also been conjectured recently in~\cite{Haehl:2015foa}.

The search for an action principle for fluids has a long history, dating back at least to~\cite{Herglotz}
and subsequent work including~\cite{Taub,Salmon}  (see~\cite{soper,Jackiw:2004nm,Andersson:2006nr} for reviews), essentially all of which were for ideal fluids. Recent investigations
include~\cite{Dubovsky:2005xd,Dubovsky:2011sj,Dubovsky:2011sk,Endlich:2012vt,Torrieri:2011ne,Bhattacharya:2012zx,Grozdanov:2013dba,Nicolis:2013lma,Haehl:2013hoa,Kovtun:2014hpa,Haehl:2015pja,Harder:2015nxa,Haehl:2015foa,Endlich:2010hf,Nicolis:2011ey,Nicolis:2011cs,Delacretaz:2014jka,Haehl:2013kra,Geracie:2014iva,Galley:2014wla,Burch:2015mea}.  We will discuss connections to these earlier works along the way.

%Recent discussions of action principle for hydrodynamics include~\cite{}. Among them~\cite{kovtun} and~\cite{loga} are close in spirit to our discussion, but~\cite{kovtun} only deals with a quadratic action while~\cite{loga} focuses on non-dissipative sector and does not consider fluctuations. The dynamical variables used in~\cite{dubovsky} for ideal fluids are also close to ours.

We will restrict our discussion to a charged fluid with a single global symmetry in the absence of anomalies.
Generalizations to more than one conserved current or non-Abelian global symmetries are immediate.
Anomalies, the non-relativistic formulation, superfluids, as well as study of physical effects of the
theory proposed here will be given elsewhere.
When a system is near a phase transition or has a Fermi surface, there are additional gapless modes, which will also be left for future work.

In the rest of this section, we outline the basic structure of our theory.

\subsection{Dynamical degrees of freedom}

We are interested in formulating a low energy effective field theory for a quantum many-body system in a macroscopic state described by some density matrix $\rho_0$. As usual, to describe the time evolution of a density matrix and expectation values in it, %(as compared to transition amplitudes)
 we need to double the degrees of freedom and use
%The appropriate formalism for describing unitary time evolution of a density matrix is
the so-called
closed time path integral (CTP) or the Schwinger-Keldysh formalism %(see e.g.~\cite{Chou:1984es,Wang:1998wg,Hubook})
\be \label{fpth}
\Tr \le(\rho_0  \cdots \ri) = \int_{\rho_0} D \psi_1 D \psi_2 \, e^{i S [\psi_1] - i S [\psi_2]}  \,  \cdots,
\ee
where $\psi_{1,2}$  collectively denote dynamical fields for the two legs of the path, $S [\psi]$ is the microscopic action of the system, and $\cdots$ denotes possible operator insertions.
In this formalism, both dissipation and fluctuations can be incorporated in an action form, which is thus ideal for formulating an effective field theory  for dissipative fluids.  Aspects of the CTP formalism important for this paper will be reviewed in Sec.~\ref{sec:CTP}.

 Now, assume that the only long-lived gapless modes of the system in $\rho_0$ are hydrodynamical modes, i.e. those associated with conserved quantities such as the stress tensor and conserved currents for some global symmetries.
We can then imagine integrating out all other modes in~\eqref{fpth}, and obtain a low energy effective theory  for hydrodynamical modes only:
\be\label{22j}
\Tr \le(\rho_0  \cdots \ri)  = \int D \chi_1 D \chi_2 \, e^{i S_{\rm hydro} [\chi_1, \chi_2; \rho_0]}  \cdots ,
\ee
where $\chi_{1,2}$ collectively denote hydrodynamical fields for the two legs of the path, and $ S_{\rm hydro}$ is the low energy effective action (hydrodynamical action) for them. Note that in the CTP formalism, there are two sets of hydrodynamical modes $\chi_{1,2}$, which will be important for incorporating dissipative effects and noises in an action principle.
Note that $ S_{\rm hydro}$  no longer has the factorized form of~\eqref{fpth}, and  $\rho_0$ is encoded
in the coefficients of the action. The standard formulation of hydrodynamics arises as the saddle point equation of the path integral~\eqref{22j}.

While such an integrating-out procedure cannot be performed explicitly, following the usual philosophy of effective field theories,  we should be able to write down $S_{\rm hydro}$ in a derivative expansion based on general symmetry principles. The challenges are basic ones: (i) what the hydrodynamical modes $\chi_{1,2}$ are, as it is clear that the standard hydrodynamical variables such as the velocity field and local chemical potential are not suited for writing down an action; (ii) what the symmetries are.

To answer the first question, a powerful tool is to put the system in a curved spacetime and to turn on external sources for the conserved currents. Due to (covariant) conservation of the stress tensor and currents, the corresponding generating functional  should be invariant under diffeomorphisms of the curved spacetime, and gauge symmetries of the external sources. These symmetries then suggest a natural definition of hydrodynamical modes as Stueckelberg-like fields associated to diffeomorphisms and gauge transformations.

To illustrate the basic idea, let us consider the generating functional for a single conserved current $J_\mu$
in a state described by some density matrix $\rho_0$,
\be\label{1gent1}
e^{ W[A_{1\mu} , A_{2 \mu}]} =
%{1 \ov Z_0}
\Tr \le(\rho_0 \sP e^{i \int d^d x \, A_{1\mu} J_1^\mu -  i \int d^d x \, A_{2\mu} J_2^\mu} \ri),
%=  \int_{\rho_0} D \Phi_1 D \Phi_2 \, e^{i S [\Phi_1]  + i \int A_{1\mu} J_1^\mu - i S [\Phi_2] -i \int  \, A_{2\mu} J_2^\mu  }
\ee
where $\sP$ denote the path orderings. Given that $J_{1,2}^\mu$ are conserved, we have
\be \label{1shifts}
W [A_{1 \mu} , A_{2 \mu}] =W [A_{1 \mu} + \p_\mu \lam_1 , A_{2 \mu} + \p_\mu \lam_2] \quad
\ee
for arbitrary functions $\lam_1, \lam_2$, i.e. $W$ is invariant under independent gauge transformations of $A_{1\mu}$ and $A_{2\mu}$. Since we do expect  presence  of terms
in $W$ at zero derivative order,
this implies that $W  [A_{1\mu} , A_{2 \mu}]$ can {\it not} be
written as a  local functional of $A_{1 \mu} , A_{2 \mu}$.
We interpret the non-locality as coming from integrating out certain gapless modes, which are identified with the hydrodynamic modes associated with conserved currents $J_{1,2}$.
In order to obtain a local action we need to un-integrate them.
From~\eqref{1shifts}
one can readily guess the answer: we can write $W$ as
%a path integral over a local action
\be \label{1newc}
e^{W [A_{1\mu} , A_{2 \mu}]} = \int D \vp_1 D \vp_2 \, e^{i I [B_{1\mu},  B_{2 \mu}]} ,\
\ee
where
\be\label{1Bvar}
B_{1 \mu}  \equiv A_{1 \mu} + \p_\mu \vp_1, \qquad B_{2 \mu} \equiv A_{2 \mu} + \p_\mu \vp_2, \
\ee
and $I$ is a local action for $B_{1 \mu} , B_{2 \mu}$.
The integrations over Stueckelberg-like fields $\vp_{1,2}$ remove the longitudinal part of $A_{1,2 \mu}$, and by definition,  $W$ obtained from~\eqref{1newc} satisfies~\eqref{1shifts}.
  We thus identify $\vp_{1,2}$ as the hydrodynamical modes associated with $J^\mu_{1,2}$.

This discussion can be generalized immediately to also include the stress tensor $T^{\mu \nu}$, turning on the source of which  corresponds to putting the system in a curved spacetime. The generating functional now becomes
\be \label{1ooen1}
e^{W  [ g_{1\mu \nu} , A_{1\mu}; g_{2 \mu \nu}, A_{2 \mu}] }%\equiv e^{i I [ g_{1\mu \nu} , A_{1\mu}; g_{2 \mu \nu}, A_{2 \mu}]}
= \Tr \le[U_1 (+\infty, -\infty; g_{1\mu \nu}, A_{1\mu}) \rho_0 U_2^\da (+\infty, -\infty; g_{2\mu \nu}, A_{2 \mu})\ri],
\ee
where $U_1$ is the evolution operator for the system in a curved spacetime with
metric $g_{1 \mu \nu}$ and external field $A_{1 \mu}$, and similarly with $U_2$.
Due to (covariant) conservation of the stress tensor and the current, $W$ is invariant
under independent diffeomorphisms of $g_{1,2}$ and ``gauge transformations'' of $A_{1,2}$:
\be  \label{1shift1}
W  [g_{1} , A_{1}; g_{2}, A_{2}] = W  [\tilde g_1, \tilde A_{1}; \tilde g_{2}, \tilde A_{2}],
%\equiv e^{i I [ g_{1\mu \nu} , A_{1\mu}; g_{2 \mu \nu}, A_{2 \mu}]}
\ee
where
\be \label{1diffg}
\tilde g_{s \mu \nu} (x) = {\p y^\sig_s \ov \p x^\mu} g_{s \sig \rho} (y_s(x))   {\p y^\rho_s \ov \p x^\nu}, \qquad
\tilde A_{s \mu} (x) = {\p y^\sig_s \ov \p x^\mu} A_{\sig} (y_s(x)) + \p_\mu \lam_s (x), \quad s =1,2,
%\tilde g_{1 \mu \nu} (x) = {\p x'^\lam \ov \p x^\mu} g_{1 \lam \rho} (x'(x))   {\p x'^\rho \ov \p x^\nu}
\ee
and  $y^\sig_{1,2} (x), \lam_{1,2}$ are arbitrary functions.

Due to~\eqref{1shift1}, for the same reason as in the vector case,
 $W$ can not be a local functional of $g_{1,2}$ and $A_{1,2}$. Again interpreting  the non-locality as coming from integrating out hydrodynamical modes, we can write $W$ as a path integral of a local action over gapless modes
obtained from promoting the symmetry transformation parameters of~\eqref{1diffg} to dynamical fields, i.e.
\be \label{qft0}
e^{W [ g_{1} , A_1; g_{2}, A_2]} = \int D X_1 D X_2 D \tau D \vp_1 D \vp_2 \, e^{i  I [h_{1},  B_1; h_2, B_{2}; \tau]},
\ee
where ($s=1,2$ and no summation over $s$)
\bega \label{hdef1}
h_{sab} (\sig)  %=e^{-2 \tau (\sig)} g_{sab} (\sig), \qquad g_{sab} (\sig)
=  {\p X^\mu_s \ov \p \sig^a} g_{s\mu \nu} (X_s (\sig)) {\p X_s^\nu \ov \p \sig^b} , \qquad
B_{sa} (\sig) = {\p X^\mu_s \ov \p \sig^a}  A_{s\mu} (X_s (\sig)) + \p_a \vp_s (\sig)
\end{gather}
and $I$ is a local action of $h_{1,2}, \tau, B_{1,2}$. As in the earlier example, integrations over the Stueckelberg-like fields $X_{1,2}^\mu (\sig^a)$ and $\vp_{1,2}$  guarantee  that $W$ as obtained from~\eqref{qft0} will automatically satisfy~\eqref{1shift1}.
Note that, except in the implicit dependence of background fields, $X^\mu_s, \vp_s$ always come with derivatives and
thus describe gapless modes.
We have also introduced a new scalar field $\tau (\sig)$ which will be interpreted as describing local temperatures.
  %$\tau_{1,2}$ do not come with derivatives and in a sense can be considered as gapped,
%but we find it convenient to keep them in the low energy theory.
%For notational simplicity, we have used the same symbol $g$ to denote the metrics in $\sig$ and $X$ coordinates, differentiating them only by their subscripts and arguments, and similarly with $A_a (\sig)$ and $A_{\mu} (X)$.

The low energy effective field theory on the right hand side of~\eqref{qft0} is unusual as the arguments $X_1^\mu, X_2^\mu$ of background fields  $g_1 (X_1), A_1 (X_1)$ and $g_2 (X_2), A_2 (X_2) $ are dynamical variables.\footnote{Such kind of  theories are often referred to as parameterized field theories and have been used as toy models for quantizing theories with diffeomorphisms~\cite{kuchar}.}
In particular, the spacetime $\sig^a$ where $h_{ab} (\sig)$ and $B_a (\sig^a)$ are defined is not the physical spacetime, as the physical spacetime is where background fields $g_{\mu \nu}$ and $A_\mu$ live.
The spacetime represented by $\sig^a$ is an ``emergent'' one arising from promoting the arguments of background fields to dynamical variables.

Despite the original microscopic theory~\eqref{fpth} being formulated on a closed time path integral in the physical spacetime, the effective field theory~\eqref{qft0}  is defined on a single ``emergent'' spacetime, {\it not on a Schwinger-Keldysh contour}. The CTP nature of the microscopic formulation is reflected in the doubled degrees of freedom and in various features of the generating functional $W$ which we will impose below.

\begin{figure}[!h]
\begin{center}
%$
%\begin{array}{cc}
\includegraphics[scale=0.8]{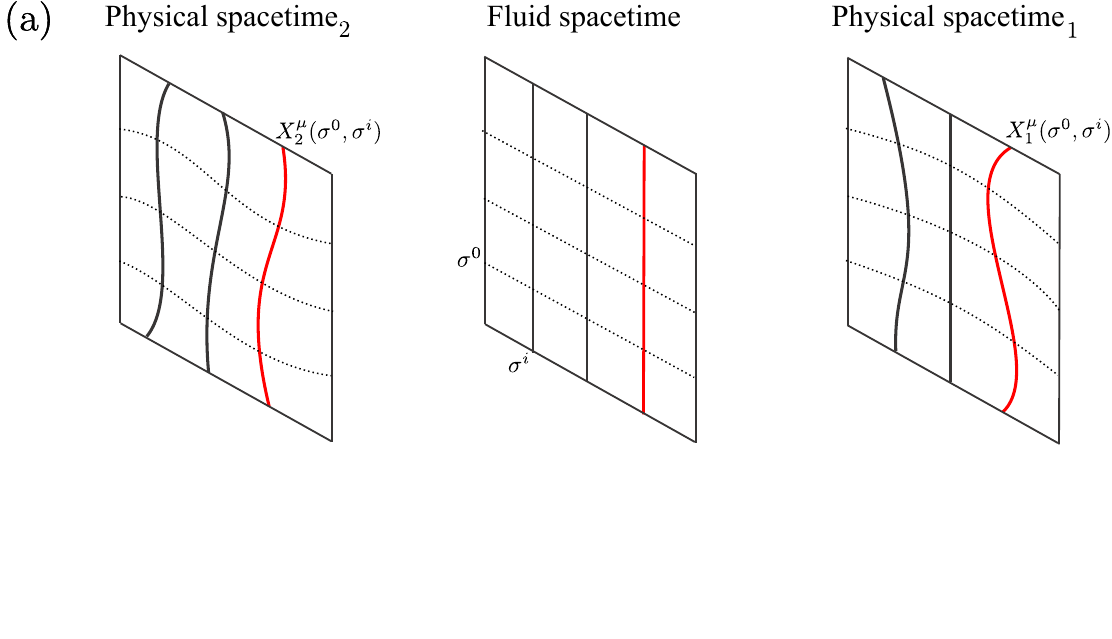} \quad
\includegraphics[scale=0.8]{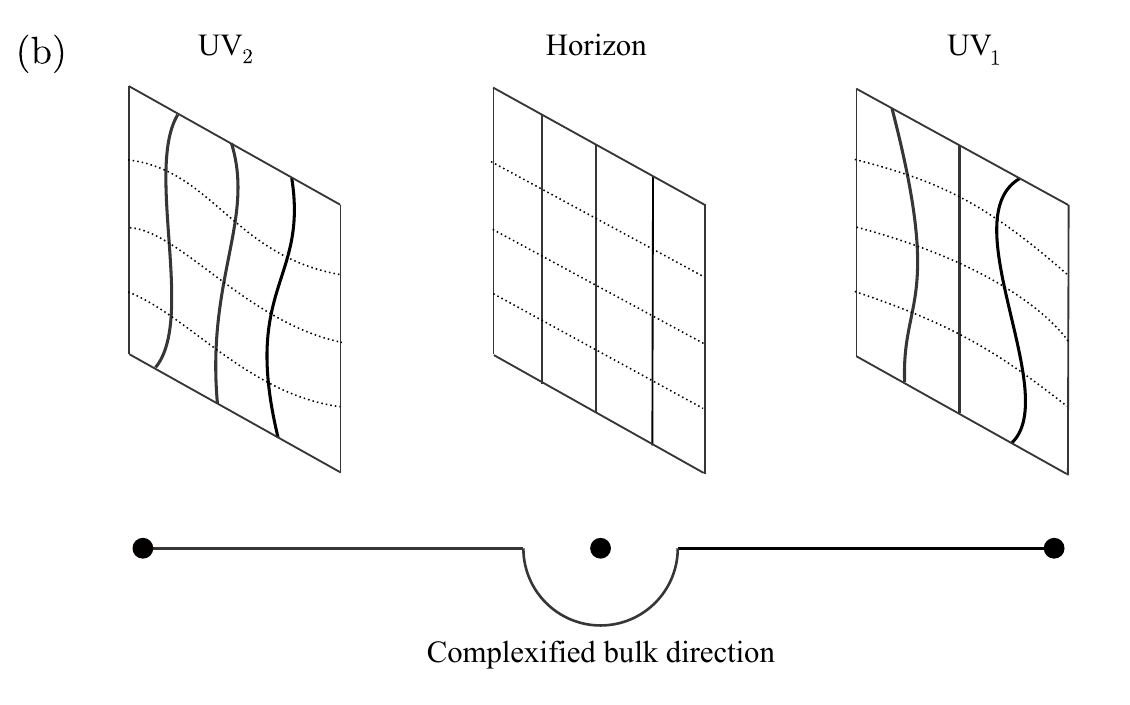}
%\end{array}
%$
\end{center}

\caption{Relations between the fluid spacetime and two copies of physical spacetimes.  The red straight
line in the fluid spacetime with constant $\sig^i$ is mapped by $X^\mu_{1,2} (\sig^0, \sig^i)$ to
physical spacetime trajectories (also in red) of the  corresponding fluid element. In the holographic context, the fluid spacetime corresponds to
the horizon hypersurface, and the two copies of physical spacetimes correspond to two asymptotic boundaries of AdS. $X^\mu_{1,2}$ describe relative embeddings of these hypersurfaces.
%and $\tau_{1,2}$ are the proper distances between the horizon and boundaries.
 }
 \label{fig:spaces}
\end{figure}

We will interpret the spacetime spanned by $\sig^a$ as that associated with fluid elements:
 the spatial part $\sig^i$ of $\sig^a$ labels fluid elements, while the time component $\sig^0$ serves as an ``internal clock'' carried by a fluid element.
In this interpretation, $X_{1,2}^\mu (\sig^a)$ then corresponds to the Lagrange description of fluid flows. With a fixed $\sig^i$, $X^\mu_{1,2} (\sig^0, \sig^i)$  describes how a fluid element labeled by $\sig^i$ moves in (two copies of) physical spacetime as the internal clock $\sig^0$ changes. This construction generalizes the standard Lagrange description, where $\sig^0$ coincides with the physical time.
In our current general relativistic context, it is more natural for a fluid element to be equipped with an internal time.
The relation between $\sig^a$ and $X_{1,2}^\mu (\sig)$ is summarized in Fig.~\ref{fig:spaces}.
Below, we will refer to $\sig^a$ as the fluid coordinates and the corresponding spacetime as the fluid spacetime.

While in hindsight, one could have directly started with a doubled version of the standard Lagrange description, the ``integration-in'' procedure described above shows that such a phenomenological description does arise naturally
as dynamical variables characterizing low energy gapless degrees of freedom of a general quantum many-body system.

Parts of these variables also have  been considered in the literature, although the starting points were different. For example, the fields $X^\mu(\sigma)$ already appeared
in~\cite{Taub,Salmon}. In the recent ideal fluid formulation of~\cite{Dubovsky:2005xd,Dubovsky:2011sj,Dubovsky:2011sk,Endlich:2012vt,Nicolis:2013lma,Endlich:2010hf,Nicolis:2011ey,Nicolis:2011cs,Delacretaz:2014jka}, a single set of $\sig^i (X^\mu)$ is used, which was subsequently generalized to the doubled version in the closed time path formalism in an attempt to include dissipation~\cite{Grozdanov:2013dba,Endlich:2012vt}.
The set $X^\mu (\sig), \vp (\sig)$ for a single side  arises  naturally in the holographic context as first pointed out in~\cite{Nickel:2010pr},  which along with~\cite{Dubovsky:2005xd,Dubovsky:2011sj} has been an important inspiration for our study. The doubled version of $X^\mu_{1,2} (\sig^a), \vp (\sig^a)$ in the closed time path formalism first appeared in~\cite{Haehl:2013hoa,Haehl:2015pja} (see also~\cite{Crossley:2015tka}).  In the holographic context, $X^\mu_{1,2} (\sig^a)$, correspond to the relative embeddings between the horizon hypersurface, which can be identified with the fluid spacetime, and the two asymptotic boundaries of AdS, which correspond to the physical spacetimes~\cite{Nickel:2010pr,Crossley:2015tka,deBoer:2015ija}. %{$\tau$ also appears in that context as the proper distances between the horizon and the boundaries~\cite{Crossley:2015tka}.
Similar variables were also employed in~\cite{Kovtun:2014hpa,Harder:2015nxa,Haehl:2015foa,Haehl:2013kra}.

%and
%the ``off-shell'' action  $ I$ in the exponent of~\eqref{newc} as the hydrodynamical action for them.
%Setting $A_{1 \mu} , A_{2 \mu}$ to zero in $I [ B_1,B_2]$  one obtains the hydrodynamical action
%for $\vp_{1,2}$ in the absence of external fields.

The interpretation of $\sig^a$ as the fluid spacetime immediately leads to an identification of the standard hydrodynamical variables %which are: velocity field $u_s^\mu (\sig)$, local temperature $T_s (\sig)$, and local chemical potential  $\mu_s (\sig)$ (with $s=1,2$ as there are two copies of them),
in terms of our variables $X^\mu_s, \tau, \vp_s$. With
$X^\mu_s (\sig^0, \sig^i)$ corresponding to the trajectory of a fluid element $\sig^i$ moving in physical spacetime, then
\be
- d \ell^2_s = g_{s \mu \nu} {\p X^\mu_s \ov \p \sig^0} {\p X^\nu_s \ov \p \sig^0} (d \sig^0)^2
\ee
is  the proper time square of the motion, and the fluid velocity is given by
\be \label{umu}
u^\mu_s (\sig) = {\de X^\mu_s \ov \de \ell_s} = {1 \ov b_s} {\p X^\mu_s \ov \p \sig^0}, \qquad b_s = \sqrt{- {\p X^\mu_s \ov \p \sig^0} g_{s\mu \nu} {\p X^\nu_s \ov \p \sig^0}}, \qquad g_{s \mu \nu} u^\mu_s u^\nu_s = -1 \ .
\ee
Similarly, interpreting  $B_{sa} (\sig)$ as the ``external sources'' for the currents of fluid elements
in fluid space, we can define the local chemical potential $\mu (\sig)$
\be \label{mub}
\mu_s (\sig) = {1 \ov b_s} B_{s0} (\sig) = u_s^\mu (\sig) A_{s\mu} (X_s (\sig)) + {1 \ov b_s} \p_0 \vp_s (\sig) \ .
\ee
The reason for the $1/b_s$ prefactor in~\eqref{mub} is the same as that in~\eqref{umu}: to convert from $dt$ to the local proper time $d \ell_s$. Finally we define the local proper temperature in fluid space as
\be \label{tay0}
T (\sig) = {1 \ov \beta (\sig)} = T_0 e^{-\tau (\sig)},
\ee
where $T_0 = {1 \ov \beta_0}$ is a reference temperature (e.g. the temperature at infinities). Note that there is only one $\tau$ field rather than two copies. In contrast to other fields, it is defined only in the fluid spacetime. It should be considered as an intrinsic property associated with each fluid element.

\subsection{Equations of motion} \label{sec:eom}

Given an action $I$ in~\eqref{qft0}, we define the ``off-shell hydrodynamical'' stress tensors and currents as
\bega \label{defst}
{\de I \ov \de g_{1\mu \nu} (x)}  \biggr|_{\tau, X} \equiv \ha \sqrt{-g_1} \hat T^{\mu \nu}_1 (x) ,\   \qquad
{\de I \ov \de A_{1\mu} (x)}  \equiv \sqrt{-g_1} \hat J^{\mu}_1 (x), \\
{\de I \ov \de g_{2\mu \nu} (x)}  \biggr|_{\tau, X} \equiv - \ha \sqrt{-g_2} \hat T^{\mu \nu}_2 (x) ,\   \qquad
{\de I \ov \de A_{2\mu} (x)}  \equiv - \sqrt{-g_2} \hat J^{\mu}_2 (x) \ .
 \label{defst1}
\end{gather}
In~\eqref{defst}--\eqref{defst1}, $x^\mu$ denotes the physical spacetime location at which $\hat T_s^{\mu \nu}, \hat J_s^\mu$ ($s=1,2$) are evaluated,
and should be distinguished from either $\sig$ or $X$, as $X$'s are dynamical variables and $\sig^a$
labels fluid elements.  $\hat T^{\mu \nu}_s$ and $\hat J^\mu_s$ are operators in the quantum effective field theory~\eqref{qft0}
of  $X^\mu_s, \tau$ and $\vp_s$. They are the low energy counterpart of the stress tensor $T^{\mu \nu}$ and current $J^\mu$ of the % original theory which are defined in terms of fundamental fields $\phi_{s}$ of the
microscopic theory~\eqref{fpth}. By definition, correlation functions of~\eqref{defst}--\eqref{defst1} in~\eqref{qft0}  should reproduce those of the microscopic theory in the long distance and time limit with choices of a finite number of parameters in~\eqref{qft0}.

By construction, $h_{sab}$ and $B_{sa}$, and so the action, are {\it invariant} under physical spacetime diffeomorphisms, which have the infinitesimal form
\be \label{1gdiff}
 \de X^\mu =  - \xi^\mu (X), \quad \de g_{ \mu \nu} (X) =  \nab_\mu \xi_\nu + \nab_\nu \xi_\mu, \quad
 \de A_\mu (X) = \p_\mu \xi^\nu A_\nu + \xi^\nu \p_\nu A_\mu,
\ee
where for notational simplicity we have suppressed the index $s=1,2$ for each quantity in the above equation, i.e. there are two identical copies of them. Similarly, $B_{sa}$ is invariant under a gauge transformation of $A_{s\mu}$ with a shift in $\vp_s$:
\be \label{1gauhe}
A_\mu \to A_\mu - \p_\mu \lam (X), \qquad \vp (\sig) \to \vp (\sig) + \lam (X (\sig)), \
\ee
with $s$ again suppressed. The invariance of the action under~\eqref{1gdiff}--\eqref{1gauhe}
immediately implies that the equations of motion for $\vp$'s are simply the conservation equations for currents in each segment of the contour, and the equations of motion for $X$'s are the conservation equations for the stress tensors (see also similar discussion in~\cite{Haehl:2015pja}),
\begin{gather}
\label{1eom11}
\vp_s \; {\rm eom}: \qquad  \nab_{s\mu} \hat J^\mu_s =0, \\
\label{1eom12}
X_s^\mu \; {\rm eom}: \qquad   \nab_{s\nu} \hat T_s^{\nu}{_\mu} - F_{s \mu \nu}\hat J^\nu_s = 0 \ .
\end{gather}
Note that in the above equations, $\nab_{s\mu}$ are covariant derivatives in physical spacetimes.

\subsection{Symmetry principles}

We now consider the symmetries which should be satisfied by the hydrodynamical action $I$ in~\eqref{qft0}.
Let us start with diffeomorphisms of $\sig^a$ and possible gauge symmetries of $B_{sa}$.
%under which $X^\mu_s (\sig), \tau_s (\sig), \vp_s (\sig)$ transform as scalar fields,  $h_{1ab}, h_{2ab}$  as tensors, and $B_{1a}, B_{2a}$ as vectors.
We require that $I$ should be invariant under:
 \ben
 \item  time-independent reparameterizations of spatial manifolds of $\sig^a$, i.e.
 \be \label{sdiff}
 \sig^i \to  \sig'^i (\sig^i), \qquad \sig^0 \to \sig^0  \ ;
 \ee

 \item time-diffeomorphisms of $\sig^0$, i.e.
 \be \label{tdiff}
 \sig^0 \to \sig'^0= f (\sig^0, \sig^i), \qquad \sig^i \to \sig^i   \ ;
 \ee
 %by choosing. F say, $\sig^0= \ha (X^0_1 + X^0_2)$.

 \item $\sig^0$-independent diagonal ``gauge'' transformations of $B_{sa}$, i.e.
 \be \label{daug}
 B_{1i} \to  B_{1i}' = B_{1i} - \p_i \lam (\sig^i), \qquad B_{2i} \to  B_{2i}' = B_{2i} - \p_i \lam (\sig^i), \
 \ee
or equivalently
\be
\vp_r \to \vp_r - \lam (\sig^i) , \quad \vp_a \to \vp_a  ,
\ee
with $\vp_r = \ha (\vp_1 + \vp_2), \, \vp_a = \vp_1 - \vp_2$.

\een

Equation~\eqref{sdiff} corresponds to a (time-independent) relabeling of fluid elements, while~\eqref{tdiff}
can be interpreted as reparameterizations of the internal time associated with fluid elements. Note that in~\eqref{tdiff} we allow time reparameterization to have arbitrary dependence on $\sig^i$, which physically can be interpreted as
each fluid element having its own choice of time. %\footnote{Note using~\eqref{tdiff} one of $X^0_{1,2}$ can be gauged away, say by setting $\sig^0 = \ha (X^0_1 + X_2^0)$.}
In contrast, we do not allow~\eqref{sdiff} to depend on $\sig^0$.
Requiring invariance under
\be \label{bei0}
\sig^i \to \sig'^i (\sig^i, \sig^0)
\ee
means allowing different labelings of fluid elements at different times. This would be too strong, as it would treat  some physical fluid motions as relabelings. The same conclusion can also be reached from the combination of~\eqref{bei0} with~\eqref{tdiff} amounting to full diffeomorphism invariance of $\sig^a$, under which one of the $X^\mu$'s can then be gauged away completely, which would be too strong.

The origin of~\eqref{daug} can be understood as follows. In a charged fluid,
each fluid element should have the freedom of making a phase rotation. As we are considering a global symmetry,
the phase cannot depend on time $\sig^0$, but since fluid elements are independent of one another,  they should have the freedom of making independent phase rotations, i.e. we should allow phase rotations of the form
$e^{i \lam (\sig^i)}$, with $\lam (\sig^i)$ an arbitrary function of $\sig^i$ only. As $B_{sa}$ are the ``gauge fields'' coupled to charged fluid elements in the fluid space, we thus
have the gauge symmetry~\eqref{daug} of $B_{sa}$. This consideration also makes it natural that in a superfluid, when the $U(1)$ symmetry is spontaneously broken,~\eqref{daug} should be dropped.

 We emphasize that~\eqref{sdiff}--\eqref{daug} are distinct from the physical spacetime diffeomorphisms~\eqref{1gdiff} and gauge transformations~\eqref{1gauhe}.
They are ``emergent" gauge symmetries which
arise from the freedom of relabeling fluid elements, choosing their clocks, and acting with independent phase rotations\footnote{Note that~\eqref{daug} can be considered as a generalization of the chemical shift symmetry introduced  in~\cite{Dubovsky:2011sj} for a single patch.}. These symmetries ``define'' what we mean by a fluid.
Indeed we will see later they are responsible for recovering the standard hydrodynamical constitutive relations including all dissipations.

The local symmetries~\eqref{sdiff}--\eqref{daug} are not yet enough to fix the
action $I$. By definition, the generating functional~\eqref{1ooen1} also has the following properties (see Sec.~\ref{sec:CTP} for their derivation)
\bea
\label{odd1}
& {\rm Reflectivity \; \; condition:}& \qquad W^* [g_1, A_1; g_2, A_2] = W[g_2, A_2;  g_1, A_1],  \\
 \label{top1}
& {\rm Normalization \;\; condition:} & \qquad W [g, A; g, A] =0  \
\eea
both of which have to do with unitarity of time evolution.

Let us first look at the reflectivity condition~\eqref{odd1} which is a $Z_2$ symmetry of the generating functional $W$. It can be achieved by requiring the off-shell action $I$ to satisfy:
\ben
\item[4.] a $Z_2$ reflection symmetry
\be\label{keyp3}
  I^* [h_1,  B_1; h_2, B_2;  \tau] = - I [h_2,  B_2; h_1,  B_1; \tau]  \ .
\ee
\een
 Equation~\eqref{keyp3} implies that the action $I$ must have complex coefficients, as all the fields are real. For the path integral~\eqref{qft0} to be well defined, we should also require that
\ben
\item[5.] the imaginary part of $I$ is non-negative.
\een
We will see later that this condition requires that noises have exponentially decaying distributions and leads to the non-negativity of various transport coefficients when combined
with the local KMS conditions to be discussed below.

Now consider the unitarity condition~\eqref{top1}, which implies that when setting
 \be \label{1yuen}
g_{1\mu \nu} = g_{2 \mu\nu} = g_{\mu \nu}, \qquad A_{1 \mu} = A_{2 \mu} = A_\mu \ ,
\ee
%$A_{1\mu} = A_{2 \mu} = A_\mu$
%and $g_{1\mu \nu} = g_{2 \mu\nu} = g_{\mu \nu}$
the path integral~\eqref{qft0}
%\be
%W [A, g] \equiv \int D \chi_1 D \chi_2 \, e^{i I [\chi_1, \chi_2; g, A]}
%\ee
becomes ``topological'', as $W$ is independent of $A_\mu$ and $g_{\mu \nu}$. In terms of correlation functions in the absence of sources, equation~\eqref{top1} implies that all correlation functions of $\hat T_a^{\mu \nu}$ and $\hat J_a^\mu$
vanish among themselves, where
\be
\hat T_a^{\mu \nu} \equiv \hat T_1^{\mu \nu} - \hat T_2^{\mu \nu} , \qquad \hat J_a^\mu \equiv \hat J_1^{\mu \nu} - \hat J_2^{\mu \nu}  \ .
\ee
To see this, let us adopt a simplified set of notation denoting the background fields (i.e. $g_{s\mu \nu}$ and $A_{s\mu}$) collectively as $\phi_s$ and dynamical variables  as $\chi_s$, with $\chi_{r,a}, \phi_{r,a}$ respectively symmetric and anti-symmetric combinations of various quantities, i.e.\footnote{There is only one $\tau$ which should be considered as a $r$-field.}
\be \label{1not}
\chi_r = \ha (\chi_1 + \chi_2), \quad \chi_a = \chi_1 - \chi_2, \quad \phi_r = \ha (\phi_1 + \phi_2), \quad \phi_a = \phi_1 - \phi_2 \  .
\ee
Similarly the currents associated with $\phi_s$ (i.e. $\hat T_s^{ \mu \nu}$ and $\hat J^\mu_s$) will be collectively denoted as $J_s$. We then have (schematically)
\be \label{j11}
J_1 = {\de I \ov \de \phi_1},
 \qquad
J_2 = - {\de I \ov \de \phi_2},
 \qquad J_a = {\de  I \ov \de \phi_r} , \qquad J_r =  {\de I \ov \de \phi_a}
 %, \qquad I = I [\chi_a, \chi_r; \phi_a, \phi_r]
  \ .
\ee
In terms of this notation, the path integral~\eqref{qft0} can be written as
\be \label{qft1}
e^{W [\phi_r, \phi_a]} = \int D \chi_r D \chi_a \, e^{i I [\chi_r, \chi_a; \phi_r, \phi_a]} \ ,
\ee
and~\eqref{top1} implies that when $\phi_a =0$,
\be
e^{W [\phi]} = \int D \chi_a D \chi_r \, e^{i I [\chi_a, \chi_r; \phi]} , \qquad
I [\chi_a, \chi_r; \phi] \equiv I [\chi_a, \chi_r; \phi_r = \phi, \phi_a =0],
\ee
should not depend on  $\phi = (g_{\mu \nu}, A_\mu)$ at all. Thus, from~\eqref{j11}, all correlation functions of $J_a$ must be zero.

We now show that at tree level of~\eqref{qft0} (or~\eqref{qft1}), this can be achieved by requiring
that:
\ben
\item[6.]  the action is zero when we set all the sources and dynamical fields of the two legs to be equal, i.e.
\be \label{1keyp2}
I [\chi_r, \chi_a =0; \phi_r, \phi_a =0] =0,
\ee
or, in our original notation,
\be \label{keyp2}
 I [h, B; h,  B;  \tau] = 0  \ .
\ee
\een
At tree-level, we have
\be
\label{class}
W_{\rm tree}  [\phi_r, \phi_a]  \equiv i  I_{\rm on-shell} [\phi_r , \phi_a]
= i I [\chi_a^{\rm cl}, \chi_r^{\rm cl}; \phi_r, \phi_a],
\ee
where $\chi_{a,r}^{\rm cl} [\phi_r, \phi_a]$ denote solutions to the equations of motion.
Given~\eqref{1keyp2}, when $\phi_a =0$, any term in $I$
must contain at least one power of $\chi_a$. Thus, $\chi_a^{\rm cl} = 0$ must always be a solution to the resulting equations of motion.
With the standard boundary conditions that $\chi_a$ must vanish at spatial and temporal infinities, this is the unique solution. It then follows that with $\phi_a =0$, the classical on-shell action always vanishes identically, i.e.
$W_{\rm tree}  [\phi_r, \phi_a=0] =0$.

It can readily be seen, however, that beyond the tree level~\eqref{keyp2} is not enough to ensure~\eqref{top1}. We will give a detailed discussion in the next subsection %Sec.~\ref{sec:brst}
and here just state the result.
To ensure~\eqref{top1} at the level of full path integrals, in addition to~\eqref{keyp2} we need to
\ben
\item[7.]  introduce a fermionic (``ghost'') partner $c_{r,a}$ for each of the dynamical fields $\chi_{r,a}$, and add a ``ghost'' action
$I_{\rm gh}$ to the original action:
\be \label{1tac}
I_{B} = I [\chi_a, \chi_r; \phi_a, \phi_r] + I_{\rm gh} [c_a, c_r, \chi_a, \chi_r; \phi_a , \phi_r],
\ee
so that when $\phi_a = 0$, the full action $I_B$ is invariant under the following BRST-type transformation (to which below we will simply refer as BRST transformation):
\be \label{1brst}
\de \chi_r^i = \ep c_r^i , \quad  \de c_r^i =0, \quad \de c_a^i = \ep  \chi^i_a , \quad \de \chi^i_a =0  \ .
\ee
Here, $\ep$ is a fermionic constant and $i$ labels different fields. Now the full path integral becomes
\be \label{2qft}
e^{W [\phi_r, \phi_a]} = \int D \chi_r D \chi_a  D c_a D c_r \, e^{i I_B  [c_a, c_r, \chi_r, \chi_a; \phi_r, \phi_a]} \ .
\ee
Note that the currents $J_{r,a}$ will now also
depend on the ghost fields.

\een
As will be discussed in the next subsection, given a bosonic action $I$ the condition of BRST invariance does not  fix the ghost action $I_{\rm gh}$ and the symmetric current $J_r$ uniquely, i.e.
there is freedom to parameterize them.

For a general density matrix $\rho_0$, we believe items $1-7$ listed above are the minimal set of symmetries needed to be imposed to describe a fluid. For specific $\rho_0$, there can be more symmetries.
We will describe the example of  thermal ensemble in Sec.~\ref{sec:thermal}.

Recent works~\cite{Kovtun:2014hpa,Haehl:2015pja,Harder:2015nxa} also share some
elements with our discussion here. In particular, Ref.~\cite{Harder:2015nxa} started from the CTP formulation of the generating functional to deduce a hydrodynamical action at quadratic level. Ref.~\cite{Haehl:2015pja} proposed a classification of  transports from entropy current using similar variables and also considered doubling degrees of freedom as in the CTP formulation.
While this paper was being finalized, reference~\cite{Haehl:2015foa} (see also~\cite{loga}) appeared which also pointed out that the path integral for hydrodynamical effective field theory should possess a topological sector and BRST invariance to ensure~\eqref{top1}. See also~\cite{Kovtun:2012rj,Harder:2015nxa,Kovtun:2014hpa}.
%Note that, in contrast to the discussion here, reference~\cite{Haehl:2015foa} proposes two BRST transformations.

\subsection{Ghost fields and BRST symmetry} \label{sec:brst}

We now elaborate on how to ensure the unitarity condition~\eqref{top1} beyond the tree level.
%In terms of simplified notations of~\eqref{qft1}, the requirement amounts to that when $\phi_a =0$,
%\be
%e^{W [\phi]} = \int D \chi_a D \chi_r \, e^{i I [\chi_a, \chi_r; \phi]} , \qquad
%I [\chi_a, \chi_r; \phi] \equiv I [\chi_a, \chi_r; \phi_r = \phi, \phi_a =0]
%\ee
%should not depend on  $\phi = (g_{\mu \nu}, A_\mu)$ at all.
To gain some intuition, let us first look at how to do this at one loop. With $\phi_a =0$, from~\eqref{keyp2},
$I$ can be expanded in powers of $\chi_a$ as
\be \label{ornc}
I = E_i (\chi_r, \phi) \chi^i_a + O(\chi_a^2), %\ha K_{2ij} (\chi_r, \phi)  \chi^i_a \chi^j_a + \cdots
\ee
where indices $i,j$ now collectively denote both field species and momenta.
At one loop order, only the terms linear in $\chi_a$ contribute, and we find\footnote{Note that $E_i =0$ are  in fact the standard hydrodynamical equations in the presence of background fields $\phi$, as will be clear
from the discussion of Sec.~\ref{sec:fluc}.}
\be
e^W  = \int D \chi_r D\chi_a \, e^{i \chi^i_a E_{i} + \cdots} = \int D \chi_r \, \le(\prod_i \de (E_{i} (\chi_r, \phi)) \ri) \ .
%e^{i S_1}
\ee
Clearly the above expression depends nontrivially on $\phi$ from the determinant in evaluating the delta functions. To cancel the determinant, we can add to the action an additional term $I_1$ of the following form
\be
e^{i I_1}  = \det  E_{ij}, \qquad E_{ij} \equiv {\p E_j \ov \p \chi_r^i}, %\equiv \p_i E_j (\chi_r, \phi)
\ee
so that the path integral from the full action
\be
I_{B} = I + I_1
\ee
is independent of $\phi$ at one-loop level. Now using a standard trick we can introduce ``ghost'' partners $c_r^i, c_a^i$ for $\chi_a^i, \chi_r^i$ to write
\be
e^{i I_1} = \int D c_r D c_a \, e^{i c_a^i E_{ij} c_r^j}  \ .
\ee
$c_{r,a}^i$ have the same quantum numbers as $\chi_{a,r}^i$, except that they are anti-commuting variables. The full path integral at one-loop order can then be written as
\be
e^W = \int D \chi_r D \chi_a D c_r D c_a \,  e^{i I_{B} },
\ee
with
\be \label{1}
I_{B}  =  \chi_a^i E_i +  c^i_r E_{ij} c^j_a + \cdots  \ .
\ee
Notice that $I_{B}$ has a BRST-type of symmetry
\be \label{brst0}
\de \chi^i_r = \ep c^i_r , \quad  \de c^i_r =0, \quad \de c^i_a = \ep  \chi^i_a , \quad \de \chi^i_a =0,
\ee
with $\ep$ an anti-commuting constant. We can write~\eqref{brst0} in terms of the action of
a nilpotent differential operator
\be
Q =  c^i_r {\de \ov \de \chi^i_r} +  \chi^i_a {\de \ov \de c^i_a}, \qquad Q^2 =0, \
\ee
and the action~\eqref{1} is BRST exact, i.e.
\be
I_{B} = Q \le(c^i_a E_i \ri) + \cdots \ .
\ee

Now it can be readily seen that if we can make the full action to be BRST invariant, and variation with respect to $\phi$ to be BRST exact, then $W$ will be independent
of $\phi$ to all loop orders.  Suppose
%\be \label{1exi}
$I_{B} [\phi_a = 0] $ is invariant
%=  Q_B V [\chi_r, \chi_a, c_r, c_a; \phi]
%\ee
under~\eqref{brst0} and under a variation of $\phi$ we have
\be
J_a = {\de I_{B} \ov  \de \phi}   = Q V,
\ee
for some operator $ V$. We then have under variation of $\phi$:
\be\label{argd}
e^W {\de W }   =  i \int  D \chi_r D \chi_a D c_r D c_a \, \le(Q   V \ri)\, e^{i I_{B}}
 =  i \int D \chi_r D \chi_a D c_r D c_a \, Q   \le( V e^{i I_{B}}  \ri)
 =0,
\ee
where in the second equality we have used that $I_{B}$ is BRST invariant and in
the third equality we have used that $Q$ can be written as a total derivative under the path integration.

To make the full action $I [\chi_r, \chi_a; \phi]$ BRST invariant, note that
%Finally we note that the construction~\eqref{1} can be readily generalized to
%any bosonic action $I [\chi_r, \chi_a; \phi]$ as follows.
%one generalize the argument can readily construct
%an action $I_{\rm tot}$ invariant under BRST transformation~\eqref{brst0} as follows.
from~\eqref{1keyp2} it contains at least one factor of $\chi_a$, i.e. we can write it as
\be \label{se}
I  [\chi_r, \chi_a; \phi] =  \chi^i_a  F_i (\chi_r, \chi_a; \phi) \  .
\ee
We can then construct a BRST invariant action:
\be \label{see}
I_{B}  [c_a, c_r, \chi_r, \chi_a; \phi]= \chi^i_a  F_i + c^i_r {\p  F_j  \ov \p \chi_r^i} c^j_a = Q \Psi , \qquad \Psi = c^i_a F_i \ .
\ee
Note that the choice of $F_i$ is not unique, as~\eqref{se} is invariant under the following redefinition of $F_i$:
\be \label{ambi}
F_i \to F_i +  \chi^j_a f_{ji}  (\chi_r, \chi_a; \phi)  , \qquad f_{ij} = - f_{ji}  \ .
\ee
Under~\eqref{ambi}, $\Psi$ and $I_{B} $ change as
\be \label{1adf}
 \Psi \to \Psi + \chi^i_a f_{ij} c^j_a, \qquad I_{B}  \to I_{B}  + c^k_r {\p  f_{ij}  \ov \p \chi^k_r}  \chi^i_a c^j_a  \ .
\ee

Clearly there is much more freedom  in writing down a BRST invariant action than~\eqref{1adf}. For example, in the construction above we set $\phi_a =0$ at the beginning. But we could have kept the
$\phi_a$ dependence, which could lead to a different BRST invariant action. More explicitly,
from~\eqref{1keyp2} we can write the full action as
\be
I [\chi_r, \chi_a; \phi_r, \phi_a] = \phi_a J_r^{(0)} + \chi_a^i G_i (\chi_a, \chi_r; \phi_r, \phi_a),
\ee
where $J_r^{(0)}$ does not contain any factors of $\chi_a$. We can then construct another action:
\be \label{1newib}
\tilde I_B = \phi_a J_r^{(0)} + \chi_a^i G_i (\chi_a, \chi_r; \phi_r, \phi_a) + c^i_r {\p  G_j  \ov \p \chi_r^i} c^j_a,
\ee
which is again BRST invariant for $\phi_a =0$. Note that in the absence of any background
fields,~\eqref{1newib} is equivalent to~\eqref{see} up to the freedom~\eqref{1adf} already noted, and they have the
same current $J_a$. But $J_r$ will in general differ by ghost dependent terms.

To summarize, with the requirements that the action be invariant under BRST-type symmetry~\eqref{brst0} and that currents $J_a$ be BRST exact,  the unitarity condition~\eqref{top1} is satisfied at the level of full path integral.
We also saw that the BRST symmetry does not fix the ghost action uniquely from the bosonic action, and there is
freedom in choosing ghost dependent terms in the definition of $J_r$.

We should also emphasize
that here  the BRST symmetry is a global symmetry; we do not require either physical operators or physical states to be BRST invariant. For example, $J_r$ is not BRST invariant.

\subsection{Thermal ensemble and KMS conditions} \label{sec:thermal}

Now let us take $\rho_0$ to be the thermal density matrix at some temperature $T_0 = {1 \ov \beta_0}$  and chemical potential $\mu_0$ for $Q = \int d^{d-1} \vx\, J^0$, i.e.
\be \label{emn}
\rho_0 = {1 \ov Z_0} e^{-\beta_0 (H - \mu_0 Q)}, \qquad Z_0 = \Tr e^{-\beta_0 (H - \mu_0 Q)}, \ .
\ee
%which is in fact what one normally refers to when talking about hydrodynamics.
In this case, the generating functional $W$ of~\eqref{1ooen1} additionally satisfies the so-called KMS condition~\cite{kubo57, mart59,Kadanoff}. The KMS condition can be considered as a $Z_2$ operation which relates
the generating functional $W$ to the corresponding $W_T$ for a time-reversed process:
\bln
\label{1newfdt}
W [\phi_{1} (x), \phi_{2} (x)] = W_T  [\phi_{2} (t - i \beta_0, \vx) , \phi_{1} (x)],
\end{align}
where we have again used the simplified notation of~\eqref{qft1} and $x = (t, \vx)$ denote the coordinates in physical spacetime.
See Sec.~\ref{sec:CTP} for the precise definition of $W_T$ and derivation of~\eqref{1newfdt}. In deriving~\eqref{1newfdt}, we also used that the stress tensor and current operators are neutral under~$Q$.
%\footnote{This is no longer the case for chemical potentials for some components of the stress tensor (like rotations) or for non-Abelian global symmetries. Our discussion can be easily generalized to such cases although the formulas are more complicated.}

At quadratic order in $\phi$'s,~\eqref{1newfdt}  gives the familiar fluctuation-dissipation theorem (FDT)
between retarded and symmetric Green functions
\be \label{1fdt4}
{\rm Im} G_R (k) = \tanh  {\beta_0 \om \ov 2}  G_{S} (k)  \ .
\ee
 At higher orders, $W_T$ cannot be expressed in terms of $W$, and
the KMS condition~\eqref{1newfdt} by itself does not impose constraints on $W$. However, in essentially all physical contexts, the Hamiltonian $H$ is $\sC \sP \sT$ invariant,
for which $\rho_0 (\beta_0, \mu_0)$ is mapped to $\rho_0 (\beta_0, -\mu_0)$
and $W_T (\mu_0)$ is related to $W (-\mu_0)$ by $\sC \sP \sT$. While our discussion can be applied to the most general cases, for simplicity here we will restrict to Hamiltonians
invariant under $\sP\sT$.\footnote{Here we treat different spacetime dimensions uniformly. By $\sP$ we simply invert all spatial directions. So for odd spacetime dimensions what we call $\sP \sT$
is in fact  $\sT$.}
 With $\sP \sT$ symmetry,
$W_T$ is related to $W$ as (see Sec.~\ref{sec:CTP} for a derivation, here for notational simplicity we have set free parameter $\th =0$)
\be \label{1newfdt0}
 W_T [\phi_{2} (t - i \beta_0, \vx) , \phi_{1} (x)] = W [\phi_{1}(- x), \phi_{2} (-t - i \beta_0, -\vx)  ],
 %W^* [\phi_{2i}^{PT} (t - i \beta_0),  \phi_{1i}^{PT} (t)]
\ee
and ~\eqref{1newfdt} can therefore be written as
\be \label{1kms}
W [\phi_{1} (x), \phi_{2} (x) ] = W [ \phi_{1} (-x) , \phi_{2} (- t - i \beta_0, -\vx )  ],
\ee
and in terms of our original notation,
\be \label{1newfdt1}
W [g_1 (x), A_1 (x); g_2 (x), A_2 (x)] = W [g_1 (-x), A_1 (-x); g_2 (- t - i \beta_0, -\vx ), A_2 (- t - i \beta_0, -\vx )
]   \ .
\ee
In the form of~\eqref{1newfdt1}, the KMS condition is now a $Z_2$ symmetry of $W$.

%Expanding~\eqref{1newfdt1} in power series of external fields, the equation gives rise to
%relations among various response and fluctuation functions, which give nonlinear generalizations of the fluctuation-dissipation theorem~\cite{Chou:1984es,Wang:1998wg}, and as we will
%show in Sec.~\ref{sec:imcon} also impose constraints on response functions
%themselves.

Now let us consider what symmetry to impose on the total action~\eqref{1tac} so as to ensure the KMS condition~\eqref{1newfdt1}.
For this purpose, first note that the bosonic action $I [\chi_r, \chi_a; \phi_r, \phi_a]$ can be split as
\be \label{1dec}
I [\chi_r, \chi_a; \phi_r, \phi_a] = I_{s} [\phi_r, \phi_a] + I_{s d} [\chi_r, \chi_a; \phi_r, \phi_a]
+ I_d [\chi_r, \chi_a],
\ee
where $ I_{s} [\phi_r, \phi_a]$ is obtained by setting all the dynamical fields to zero,
 $I_d [\chi_r, \chi_a]$ is obtained by setting all the background fields to zero\footnote{For spacetime metrics, zero external fields correspond to setting $g_{\mu \nu} = \eta_{\mu \nu}$.}, and
 $I_{sd}$ is the collection of remaining cross terms of $\chi$'s  and $\phi$'s.

$I_d [\chi_r, \chi_a]$ is the dynamical action
for  hydrodynamical modes $\chi$ in the absence of sources, while $I_{sd}$ describes
the coupling of dynamical modes to sources from which our off-shell hydrodynamical stress tensors and currents~\eqref{defst}--\eqref{defst1} are extracted. Given that $\chi$'s are gapless, path integrals of $I_d + I_{sd}$ generate nonlocal contributions to $W$, i.e. contributions which become singular in
the zero momentum/frequency limit.

The source action $ I_{s} [\phi_r, \phi_a]$ gives local terms in the generating functional $W$.
After differentiation, they give contributions to correlation functions of the stress tensor and current which are analytic in momentum and frequency, i.e. contact terms in coordinate space. In contrast to contact terms in vacuum correlation functions which are often discarded, these contact terms
are due to medium effects from finite temperature/chemical potential and contain important physical information.
For example, viscosities and conductivity can be extracted from them.

A remarkable fact of the structure of~\eqref{qft0}--\eqref{hdef1} is that
once the couplings of the source action $I_s$ are specified,
those of the dynamical action $I_d$ and the cross term action $I_{sd}$ are {\it fully} determined.
In other words, once the local terms in $W$ are fixed, the nonlocal parts are also fully determined.

Our proposal to ensure~\eqref{1newfdt1} consists of two parts. The first part concerns the bosonic
action $I$:
\ben
\item[8(a).]  we require that  the contact term action $I_s$ satisfies the KMS conditions~\eqref{1kms}, i.e. $I_s$ should satisfy the following  $Z_2$ symmetry:
\be \label{confdt}
I_s [\phi_{1} (x), \phi_{2} (x)] =- I_s [  \phi_{1} (-x), \phi_{2} (- t - i \beta_0, -\vx ) ], \
\ee
or in terms of our original variables\footnote{Note that in order to obtain the contact term action $I_s [g_{1}, A_1; g_2, A_2 ]$ from $I  [h_1,  B_1; h_2, B_2; \tau]$, we also need to specify a background value for $\tau$, which will be discussed in detail in Sec.~\ref{sec:kmsc}.},
\bln
 I_s [g_{1}, A_1; g_2, A_2 ] =- I_s [g_{1} (-x), A_1 (-x); g_{2} (- t - i \beta_0, -\vx ),  A_{2} (- t - i \beta_0, -\vx ) ] \ .
\label{keyp4}
\end{align}

\een
The motivations behind this proposal are: (i) nonlocal and local part of correlation functions should satisfy KMS conditions separately; (ii) Since the couplings of $I_d + I_{sd}$ are determined from those of $I_s$,~\eqref{confdt} imposes strong constraints on the couplings of the dynamical action as well as
the expressions of hydrodynamical stress tensors and currents, which may lead to~\eqref{1newfdt1} for full correlation functions. At tree level, where the ghost action can be ignored, it can be shown in the vector theory~\eqref{1newc} that~\eqref{confdt} ensures~\eqref{1newfdt1}. The proof requires introducing more specifics than the broad level at which we have been discussing so far, and will be left to Appendix~\ref{app:fdtar}. While we strongly suspect that the proof in Appendix~\ref{app:fdtar} can be generalized to a full charged fluid, the presence of $\tau$ fields make the story more tricky and a full proof will not be given here.

From now on, we will refer to~\eqref{keyp4} as the local KMS conditions. We will show in Sec.~\ref{sec:hydro} that the local KMS conditions~\eqref{keyp4} not only reproduce all the standard constraints on the hydrodynamical equations of motion (including the entropy condition constraints and those from linear Onsager relations), but also impose a new set of constraints which may be considered as nonlinear generalizations of Onsager relations.

To conclude let us remark that for general non-equilibrium situations  $\beta_0$ in~\eqref{confdt}--\eqref{keyp4} should be considered as the inverse temperature at spatial infinity, i.e. all dynamical modes including $\tau$ are assumed to fall off sufficiently rapidly approaching spatial infinities.

The importance of understanding macroscopic manifestations of the KMS condition
has been emphasized in~\cite{Haehl:2015pja,Haehl:2015foa}. There a different approach based on a $U(1)_T$ symmetry was proposed.

\subsection{KMS conditions and supersymmetry} \label{sec:susy}

We now consider %continue %the discussion of Sec.~\ref{sec:thermal} on
how to ensure
the KMS conditions~\eqref{1newfdt1} beyond the tree level, for which the situation becomes less clear.
 Currently
we have a concrete proposal only for the classical statistical limit of~\eqref{2qft}.
% discussed in Sec.~\ref{sec:hbar}.

%Beyond tree level the situation becomes less clear.
%As mentioned earlier there are freedom in both
%the ghost action and $J_r$ which are not fixed by the bosonic action. These freedom are likely
%important in ensuring~\eqref{1newfdt1} at all loop level.%Before describing that in detail
%in Sec.~\ref{sec:susy} let us first digress to discuss various limits and expansion schemes of~\eqref{2qft}.
%As mentioned at the end of Sec.~\ref{sec:thermal}
%the freedom in the ghost action and symmetric current $J_r$ should be relevant.
%Our results are still preliminary, but rather intriguing and suggestive.

Our understanding is mostly developed from the example of the hydrodynamics of a
single vector current~\eqref{1newc}, which we summarize here using the notation of~\eqref{1not}--\eqref{qft1}.
Details are given in Sec.~\ref{sec:diffusion}.  We believe the discussion below should apply, with small changes, to full charged fluids~\eqref{qft0} in the small amplitude expansion.  But the expressions become quite long and tedious, which we will leave for future investigation. Note that in both~\eqref{1newc} and the small amplitude expansion of~\eqref{qft0}, the physical and fluid spacetimes coincide, so we will not make this distinction below.

Consider the small amplitude expansion of external sources and dynamical modes, i.e.
\be
I_B = I_2 + I_3 + \cdots,
\ee
where $I_m$ contains altogether $m$ factors of sources and dynamical fields (but can be kept to all derivative orders).
We find that at quadratic order $I_2$, the ghost action is uniquely determined from the requirement of BRST
invariance for $\phi_a =0$, and there is no freedom in $J_r$.
After imposing the local KMS conditions~\eqref{keyp4}, with all external sources turned off, in addition to~\eqref{brst0}, the full action has an emergent fermonic symmetry, which can be written in a form
\be \label{1brst3}
\bar \de \chi_r =  c_a \bar \ep , \qquad
\bar \de c_r =  (\chi_a  + \Lam  \chi_r ) \bar \ep, \qquad
 \bar \de \chi_a = - \Lam c_a  \bar \ep,
\ee
where
\be
\Lam = 2 \tanh {i \beta_0 \p_t \ov 2} \ .
\ee
The appearance of $\Lam$ has its origin in the FDT relation~\eqref{1fdt4}.

It can be readily checked that $\de$ of~\eqref{brst0} and $\bar \de$ satisfy the following supersymmetric algebra
\be \label{11alg}
\de^2 = 0, \qquad \bar \de^2 = 0, \qquad [\de , \bar \de ] = \bar \ep \ep \Lam  \ .
\ee
In addition, the currents $J_{r,a}$, being linear in the dynamical fields, satisfy the following relations under $\de$ and $\bar \de$:
\bega \label{11a}
\de J_r = \ep \xi_r, \quad \bar \de J_r = \xi_a \bar \ep  , \quad
\de \xi_a = \ep J_a , \quad \bar \de \xi_r = (J_a +\Lam J_r) \bar \ep, \quad  \bar \de J_a = - \Lam \xi_a \bar \ep, \
%\label{12a}
\end{gather}
where $\xi_{a,r}$ are some fermonic operators which may be interpreted as fermionic partners of $J_{a,r}$. In other words, the current operators, $(J_a, J_r, \xi_a, \xi_r)$, transform in the same representation under~\eqref{11alg} as the fundamental multiplet $(\chi_a, \chi_r, c_a, c_r)$.

At cubic order $I_3$, there are a few new elements. Firstly, BRST invariance no longer fixes the ghost action or the ghost part of $J_r$. Secondly, the algebra~\eqref{1brst3} cannot remain a symmetry at nonlinear orders as there is a fundamental obstruction in applying the algebra~\eqref{11alg}
to a nonlinear action. By definition, acting on a product of fields, both $\de$ and $\bar \de$ are derivations, i.e. they
satisfy the Leibniz rule, and so does their commutator. But on the right hand side of~\eqref{11alg}, $\Lam$ does not satisfy the Leibniz rule. The contradiction does not cause a problem at quadratic level as
\be
\int dt \, (\Lam_1 + \Lam_2) \sL_2 = 0,
\ee
where $\Lam_1$ ($\Lam_2$) denotes that $\Lam$ is acting on the first (second) field of $\sL_2$. But this is no longer true at nonlinear orders.

Both of the above issues can be addressed in the classical statistical limit  $\hbar \to 0$, which we will explain in more detail in next subsection. For now it is enough to note that in this limit, the path integrals~\eqref{qft0} survive due to statistical fluctuations.

In the $\hbar \to 0$ limit
(restoring $\hbar$),
\be
\Lam = 2 \tanh {i \beta \hbar \p_0 \ov 2} \to i \beta \hbar \p_0, \qquad \hbar \to 0,
\ee
and equations~\eqref{11alg} become the standard supersymmetric
algebra,
\be \label{1susy}
\de^2 = 0, \qquad \bar \de^2 = 0, \qquad [\de , \bar \de ] = \bar \ep \ep i \beta_0 \p_t  \
\ee
after a rescaling of $\bar \ep$, and thus~\eqref{1susy} could persist to all nonlinear orders.
Indeed, we find that at cubic order in the $\hbar \to 0$
limit, the local KMS conditions gives a bosonic action which is supersymmetrizable, and in addition invariance under~\eqref{1susy} uniquely fixes the ghost action. %In particular, gauge symmetry~\eqref{daug} is crucial for the coefficients to be nonsingular in the zero frequency limit.
Furthermore, we find that requiring that the currents $J_{r,a}$ satisfy the $\hbar \to 0$ limit of~\eqref{11a}\footnote{
We should also scale $(J_a, \xi_a) \to \hbar (J_a, \xi_a)$ and $J_r , \xi_r \to J_r, \xi_r$.}, i.e.
\bega \label{12a}
\de J_r = \ep \xi_r, \quad \bar \de J_r = \xi_a \bar \ep  , \quad
\de \xi_a = \ep J_a , \quad \bar \de \xi_r = (J_a + i \beta_0 \p_t J_r) \bar \ep, \quad  \bar \de J_a = -  i \beta_0 \p_t \xi_a \bar \ep \
 \end{gather}
 uniquely fixes $J_r$.
It is thus tempting  to conjecture that in the $\hbar \to 0$ limit, {\it combined with local KMS conditions}, supersymmetry will be able to uniquely determine the ghost action and $J_r$ to all
nonlinear orders, and ensure the KMS conditions to all loops.

One can immediately conclude from~\eqref{12a} that supersymmetry ensures {\it one of the}
KMS conditions to be satisfied at the level of full path integral. From the fourth equation of~\eqref{12a}, we find that
$\tilde J_A \equiv J_a + i \beta_0 \p_t J_r = \bar Q \xi_r$ where $\bar Q$ is the operator which generates transformation $\bar \de$. Given that the action is invariant under $\bar Q$, then from manipulations exactly parallel to~\eqref{argd} (with $Q$ replaced by $\bar Q$) we conclude that correlation functions involving only $\tilde J_A$ all vanish.
 As discussed around~\eqref{ggg1}--\eqref{gggf} in Appendix~\ref{app:fdt} this is precisely one of the KMS conditions. In fact for two-point functions, it is the full KMS condition. Thus for two-point functions, supersymmetry~\eqref{12a} ensures KMS conditions at full path integral level. Perhaps not surprisingly, as we will see explicitly in Sec.~\ref{sec:cubic}, it is exactly the local version of this particular KMS condition (i.e. this KMS condition applied to $I_s$) that leads to the invariance of the action under $\bar \de$ and the supermultiplet structure~\eqref{12a}. It is still an open question at the moment for $n$-point functions with $n \geq 3$ whether local KMS and SUSY are enough to ensure other KMS conditions and how.

To summarize, in the classical statistical limit we can now state the second part of the symmetries which need to imposed to ensure the KMS conditions~\eqref{1kms}:
\ben
\item[8(b).]  The full action should be invariant under~\eqref{1susy}, which fixes the ghost action, and
the supersymmetric transformations of $J_{r,a}$ should satisfy~\eqref{12a}, which fixes $J_r$.
\een
We believe these are the full set of symmetries which need to be imposed for a full classical statistical
path integral.

For finite $\hbar$, the story is more tantalizing and potentially more exciting, as some theoretical structure beyond the standard supersymmetry algebra should be in operation. The algebra~\eqref{11alg} is reminiscent of higher spin symmetries and also possibly suggests a quantum group version of supersymmetry.\footnote{We would like to thank Guido Festuccia and Tom Banks for these interesting ideas.}

We have also only been looking at the situation where the fluid spacetime coincides with the physical spacetime. For~\eqref{qft0} at full nonlinear level,  supersymmetry (or whatever replaces it for finite $\hbar$) should be formulated in the fluid spacetime. When combined with time diffeomorprhism~\eqref{tdiff}, it should lead to a supergravity theory. We will leave this for future investigation.

We note that the emergence of supersymmetry in the classical statistical limit
is in some sense anticipated from that for pure dissipative Langevin equation (see e.g.~\cite{Gozzi:1983rk,Mallick:2010su}, and also~\cite{zinnjustin} for a review). But even at the level of hydrodynamics for a single current~\eqref{1newc}, the interplay between local KMS conditions and supersymmetry already goes far beyond the scope of a Langevin equation whose corresponding action is quadratic and the distribution of noise is independent of dynamical variables. Here we have a full interacting theory between noises and dynamical variables.

%We also note that supersymmetry is a consequence of the local KMS conditions, but the converse is
%not true. As we will see in detail in Sec.~\ref{sec:diffusion}, for the bosonic part of the action to be supersymmetrizable, only a subset of local KMS conditions are needed.  Nevertheless, as we discussed above, supersymmetry plays a key role in fixing the ambiguities in the ghost part of the effective action and currents.

 At a philosophical level, the interplay between local KMS conditions and supersymmetry may be understood as follows. The thermal ensemble~\eqref{emn} is thermodynamically stable, i.e. any perturbations result in a higher free energy. Furthermore, KMS conditions have been known to be equivalent to the stability conditions. It appears reasonable that such thermodynamical stability conditions are reflected as supersymmetry in the closed time path formalism.

While this paper was being finalized, reference~\cite{Haehl:2015foa} (see also~\cite{loga}) appeared, which conjectures similar supersymmetric algebra for the hydrodynamical action based on the analogue with stochastic Langevin systems.

%The formulation proposed there has some fundamental differences  from ours, as in~\cite{Haehl:2015foa}, supersymmetry is used as a requirement to constrain the bosonic action, along with a $U(1)_T$ symmetry which we do not require. It also has twice the number of supersymmetries, and the supersymmetric algebra differs from~\eqref{11alg}.

%They will be supplemented by further conditions
%which will provide additional constraints on the coefficients of $I$.
%We will go back to them after some further development.
%To introduce these additional conditions we will first have some preparations.

%In the rest of this section we consider general structure of~\eqref{qft0} which does not depend
%on explicit form of $I$, including some further conditions which should be imposed on $I$.

%hydrodynamical action.
% $I$ for excitations around any initial state $\rho_0$.
%

%The motivation for~\eqref{daug} is less intuitive.\footnote{As we will discuss separately elsewhere, it has a simple interpretation
%in holographic constructions.}  It can be deduced by a physical requirement: the
%leading equations of motion of the theory~\eqref{newc}--\eqref{Bvar} should be given by diffusion, as we will discuss in detail
%in Sec.~\ref{sec:diffusion}.

\subsection{Various limits and expansion schemes} \label{sec:hbar}

%\subsection{The thermodynamical limit}
%\label{sec:class}

In this subsection we discuss various limits and expansion schemes of~\eqref{2qft} which we copy here
for convenience with $\hbar$ reinstated
\be \label{3qft}
e^{W [\phi_r, \phi_a]} = \int D \chi_r D \chi_a  D c_a D c_r \, e^{{i  \ov \hbar} I_B  [c_a, c_r, \chi_r, \chi_a; \phi_r, \phi_a]} \ .
\ee
In a usual quantum field theory $\hbar$ controls
the loop expansion.  Here, however, the effective loop expansion constant $\heff$ is in general not $\hbar$, as
the action $I$ describes dynamics of macroscopic non-equilibrium configurations, which have both statistical
and quantum fluctuations. In particular, statistical fluctuations should persist even in the $\hbar \to 0$ limit,
i.e.  $\heff$ has a finite $\hbar \to 0 $ limit and  the path integral in~\eqref{3qft} survives. To emphasize the statistical aspect of it, from now on we will refer to the $\hbar \to 0$ limit as the {\it classical statistical limit}.

More explicitly, we define
the $\hbar \to 0$ limit in~\eqref{3qft} as
\be \label{1conl}
(c_a , \chi_a, \phi_a) \to \hbar (c_a , \chi_a, \phi_a) , \qquad c_r , \chi_r, \phi_r \to c_r , \chi_r, \phi_r  , \qquad
\hbar \to 0,
\ee
and the coefficients of the action $I_B$ should be scaled in a way that the whole action has a well-defined limit.
As an example, suppose $I_B$ contains the following terms:
\be \label{1hva}
I_B = \cdots + {G \ov 6} \chi_a^3 + {i \ov 2} H \chi_a^2 \chi_r + {K \ov 2} \chi_a \chi_r^2,
%- {i \ov 2} H c_a \vp_a c_r
- i f c_a \chi_a c_r
%- K c_a \vp_r c_r
 + \cdots
\ee
then $G, H,K, f$ should scale in the $\hbar \to 0$ limit as
\be
G \to {1 \ov \hbar^2} G , \quad H \to {1 \ov \hbar} H, \quad  K \to K, \quad f \to {1 \ov \hbar} f \ .
\ee
As will be seen in Sec.~\ref{sec:CSL}, the above scalings are indeed those dictated by the small $\hbar$ limit of various correlation functions. Below we will also use~\eqref{3qft} to refer to
its classical statistical limit. We also emphasize that while the ``ghost'' fields $c_{r,a}$ are introduced to satisfy the unitary
condition~\eqref{top1} which is a quantum condition, they survive in the classical limit. Thus to describe (classical) thermal fluctuations consistently we still need anti-commuting fields!

When $\heff$ is small, the path integral~\eqref{3qft} can be evaluated using the saddle point approximation, with
\be \label{hcbw}
W  [\phi_r, \phi_a]  = {1 \ov \heff} W_{\rm tree} + W_1 + \heff W_2 + \cdots,
\ee
where the leading contribution is the tree-level term~\eqref{class} discussed earlier. Note that
the ghost action can be ignored at tree-level. The most convenient choice of the effective loop expansion
parameter $\heff$ will in general depend on the specific system under consideration. On general grounds,
we expect it to be proportional to the energy or entropy density of a macroscopic system. In particular,
\be
\heff \propto {1 \ov \sN}
\ee
where $\sN$ is the number of degrees of freedom. From now on we will refer to $W_{\rm tree}$ as the
{\it thermodynamical limit} of $W$.

As usual in effective field theories, $I_B$ can contain an infinite number of terms, and for explicit calculations
one needs to decide an expansion scheme to truncate it.  In our current context, due to the doubled degrees of freedom and sources, there is also a new element. In this paper, the following expansions or their combinations will often be considered:

\ben

\item[a.] Derivative expansion. As usual the UV cutoff scale for the derivative expansion is the mean free path $\ell_{\rm mfp}$, whose explicit form of course depends on specific systems. For example, for a
strongly interacting theory at a finite temperature $T = {1 \ov \beta}$, we expect $\ell_{\rm mfp} \sim \hbar \beta$.
We always take the external sources to be slowly varying in spacetime, and vanishing at both spatial and temporal infinities.

\item[b.] Small amplitude expansion.  One takes the external sources to be small and considers small perturbations of dynamical variables $\chi_{r,a}$ around
equilibrium values.

\item[c.] $a$-field expansion. We expand the action $I_B$
in terms of the number of $a$-fields, i.e.
\be \label{aexp0}
I_B =  I^{(1)}_B +  I^{(2)}_B + \cdots
\ee
where $I^{(m)}_B$ contains altogether $m$ factors of $\phi_a, \chi_a$ and $c_a$. The expansion starts with $m=1$ due to~\eqref{keyp2}. From~\eqref{keyp3},  $I^{(m)}_B$ is pure imaginary for even $m$ and real for odd $m$.  The $a$-field expansion is motivated from the structure of generating functional $W[\phi_r, \phi_a]$.
As will be discussed in Sec.~\ref{sec:response},  the expansion of $W$ in $\phi_a$ gives rise to fluctuation functions of increasing orders. So if one is only interested in the fluctuation functions up to certain orders, one could truncate the expansion~\eqref{aexp0} to the appropriate order. In Sec.~\ref{sec:noise} we also show $\chi_a$ can be interpreted as noises. Thus  a-field expansion essentially corresponds to expansion in terms of noises. For this reason, we will also refer to it as noise expansion.

\een

\subsection{Plan for the rest of the paper}

In the next section, we review aspects of  generating functionals in the CTP formalism, which will play an important role in our discussions. Of particular importance is the discussion of the KMS conditions
at full nonlinear level as well as the constraints which the KMS conditions impose on response functions.

In Sec.~\ref{sec:hydro}, we explain how the standard formulation of hydrodynamics arises in our formulation, and aspects of our theory going beyond it. We first discuss how to recover the standard hydrodynamical equations of motion and then constraints on the equations of motion following from our symmetry principles.  In particular,  in addition to recovering all the currently known constraints, we will find a set of new constraints to which we refer as generalized Onsager conditions. We also discuss how to obtain the standard formulation of fluctuating hydrodynamics.

In the rest of the paper, we apply the formalism outlined in this introduction to two examples.
In Sec.~\ref{sec:diffusion}, we consider the hydrodynamics
associated with a conserved current~\eqref{1gent1}--\eqref{1newc}.
We discuss emergent supersymmetry in detail at quadratic and cubic level in the small amplitude expansion.
We work to all orders in derivatives. We give an explicit example in which the generalized Onsager conditions give new constraints at second derivative order at cubic level (details in Appendix~\ref{app:example}).
We also derive a minimal truncation of our theory which provides a path integral formulation for a variation of stochastic Kardar-Parisi-Zhang equation.

In Sec.~\ref{sec:charge}, we apply the formalism to full dissipative charged fluids. We write the action in a double expansion
of derivatives and $a$-fields.
We prove that it reproduces the standard formulation of hydrodynamics as its equations of motion.
We also use our formalism to derive the two-point functions of a neutral fluid, and
provide a path integral formulation of the relativistic stochastic Navier-Stokes equations.
Finally we show that a conserved entropy current arises at the ideal fluid level from an accidental symmetry.

We conclude in Sec.~\ref{sec:discussion} with future directions.
We have also included a number of technical appendices. In particular, in  Appendix~\ref{app:fdt} we discuss constraints from the KMS condition at general orders and prove a generalized Onsager relation. In Appendix~\ref{app:fdtar}, we show how the local KMS  condition leads to
the KMS condition for full correlation functions at tree-level for the vector model. %In Appendix~\ref{app:ward} we explore the implications of Ward identities from supersymmetry.
In Appendix~\ref{app:example} we give an explicit example in the vector theory which shows that local KMS counterpart of the
nonlinear Onsager relation gives new nontrivial constraints at second order in derivatives.
In Appendix~\ref{app:proof} we prove that at $O(a)$ level in the a-field expansion, the stress tensor and current can be solely expressed in terms of standard hydrodynamical variables.

\section{Generating functional for closed time path integrals} \label{sec:CTP}

Here we review aspects of the closed time path integral (CTP), or Schwinger-Keldysh formalism~(see e.g.~\cite{Chou:1984es,Niemi:1983nf,Wang:1998wg,Hubook}),
which will be used in this paper. At the end, we derive constraints on nonlinear response functions
from KMS conditions, which will play an important role later in constraining hydrodynamics. This discussion is new.

\subsection{Closed time path integrals}

\begin{figure}[!h]
\begin{center}
%$
%\begin{array}{cc}
\includegraphics[scale=1]{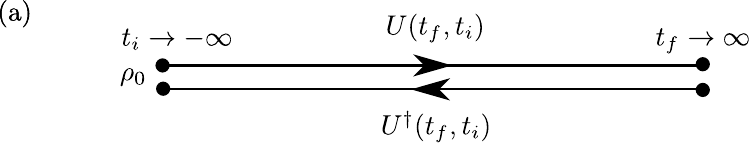} \quad
\includegraphics[scale=1]{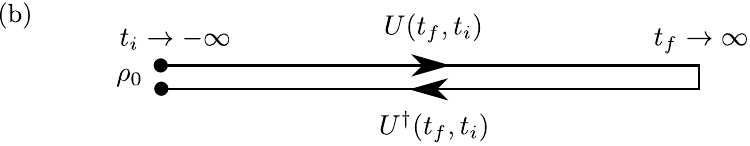}
%\end{array}
%$
\end{center}
\caption{(a) Evolution of a general initial density matrix $\rho_0$. (b) Closed time path contour from taking the trace.
Inserted operators should be path ordered as indicated by the arrows.
}
 \label{fig:SK}
\end{figure}

The evolution of a system with an initial state $\rho_0$ at some $t_i \to -\infty$ can be written as
\be
\rho (t) = U(t, t_i) \rho_0 U^\da (t, t_i),
\ee
where the evolution operator $U(t,t_i)$ can be expressed as a path integral from $t_i$ to $t$.
It then follows that $\rho(t_f)$ with $t_f \to \infty$ is described
by a path integral with two segments, one going forward in time from $-\infty$ to $+ \infty$ and one going backward in time from $+\infty$ to $-\infty$ (see Fig.~\ref{fig:SK}a),
\be
\vev{x''|\rho(t_f)|x'} = \int dx_0'' dx_0' \int_{x_1(t_i)=x_0''}^{x_1 (t_f) = x''} D x_1 \int_{x_2(t_i)=x_0'}^{x_2 (t_f) = x'} D x_2 \, e^{i S[x_1] - i S[x_2]} \, \vev{x_0''|\rho_0|x_0'}   \ .
\ee
For notational simplicity, we have written the above equation for the quantum mechanics of a single degree of freedom $x (t)$.

Setting $x'' = x' = x$ and integrating over $x$, we then find that
\be \label{heno}
\Tr \le(\rho_0 \sP \cdots \ri) \equiv \vev{\sP \cdots } = \int dx
\int_{x_1 (+\infty) = x_2 (+\infty) = x}Dx_1 Dx_2\,  e^{i S[x_1] - i S[x_2]} \, \cdots \, \vev{x_0''|\rho_0|x_0'},
%\sO_1 (t_1) \sO_2 (t_2) \cdots \sO_n (t_n)
\ee
where the path integrations on the right hand side are over arbitrary $x_{1,2} (t)$ with the only constraint
$x_1 (+\infty) = x_2 (+\infty) = x$ (see Fig.~\ref{fig:SK}b). In~\eqref{heno}  $\cdots$ denotes possible operator insertions,
and $\sP$ on the left hand side indicates that the inserted operators are path ordered:
operators inserted on the first (i.e. upper) segment are time-ordered, while those on the second (i.e. lower) segment are anti-time-ordered, and the operators on the second segment always lie to the left of those on the first segment.

It is often convenient to consider the generating functional
\be \label{pagr1}
Z [\phi_{1i}, \phi_{2i}] \equiv e^{W [\phi_{1i}, \phi_{2i}]} =  \Tr \le[\rho_0 \sP \exp \le(i \int d t \, (\sO_{1i} (t) \phi_{1i} (t) - \sO_{2i} (t) \phi_{2i} (t)) \ri)\ri],
\ee
where $i$ labels different operators, and the subscripts $1,2$ in $\sO_i$ denote whether the operators are inserted on the first  or second  segment of the contour (note $\sO_{1i}$ and $\sO_{2i}$ are the same operator), and $\phi_{1i}, \phi_{2i}$ are independent sources for the operator $\sO_i$ along each segment. The $-$ sign before terms with subscript $2$ arises from reversed  time integration.
Taking functional derivatives of $W$ gives path ordered connected correlation functions, for example
\bea
{1 \ov i^4} {\de^4 W \ov \de \phi_1 (t_1) \de \phi_2 (t_2)
\de \phi_1 (t_3)  \de \phi_2 (t_4)}
\biggr|_{\phi_{1} = \phi_2 =0}
 &=&  \vev{\sP \sO_{1} (t_1) \sO_{2} (t_2)  \sO_1 (t_3)  \sO_2 (t_4)}   \cr
&=& \vev{\tilde T (\sO (t_2) \sO (t_4)) T (\sO (t_1) \sO (t_3))},
\label{corrd}
\eea
where we have suppressed $i,j$ indices. In the second line, $T$ and $\tilde T$ denote time and anti-time ordering
respectively. In this notation, equation~\eqref{pagr1} can thus be written as
\be \label{pager1}
e^{W [\phi_{1i}, \phi_{2i}]} =  \Tr \le[\rho_0 \le(\tilde T e^{- i \int d t \,  \sO_{2i} (t) \phi_{2i} (t) } \ri) \le( T e^{i  \int d t \, \sO_{1i} (t) \phi_{1i} (t) } \ri) \ri]  \ .
\ee

We will take all operators $\sO_i$ under consideration to be Hermitian and bosonic.
$\phi_{1i}, \phi_{2i}$ are real. Taking the complex conjugate of~\eqref{pager1}, we then find that
\be \label{pro2}
W^* [\phi_{1i} , \phi_{2i}] = W[\phi_{2i}, \phi_{1i}] \ .
\ee

Equation~\eqref{pagr1} can also be written as
\be \label{pagr2}
e^{W [\phi_{1i}, \phi_{2i}]}  = \Tr\le[U (+\infty, -\infty; \phi_{1i}) \rho_0 U^\da (+\infty, -\infty; \phi_{2i})\ri],
\ee
where $U_1$ is the evolution operator for the system obtained from the original system under the deformation $\int dt \phi_{1i} \, \sO_i$, and similarly for $U_2$.
From~\eqref{pagr2}, we have
\be \label{pro1}
W [\phi_{i}, \phi_{i}] = 0, \qquad \phi_{1i} = \phi_{2i} = \phi_i  \ .
\ee

It is convenient to introduce the so-called $r-a$ variables with
\be \label{ra1}
\phi_{ri} = \ha (\phi_{1i} + \phi_{2i}), \quad \phi_{ai} = \phi_{1i} - \phi_{2i}, \quad
\sO_{ai} = \sO_{1i} - \sO_{2i}, \quad \sO_{ri} = \ha ( \sO_{1i} + \sO_{2i}),
\ee
for which~\eqref{pagr1} becomes
\be \label{ra2}
 e^{W[\phi_{ai} , \phi_{ri}] } = \Tr \le[\rho_0 \sP \exp \le(i \int d t \, (\phi_{ai} (t)  \sO_{ri} (t) +  \phi_{ri} (t) \sO_{ai} (t)) \ri)\ri] \ .
\ee
From~\eqref{ra2}, one obtains a set of correlation functions (in the absence of sources) with specific orderings
(suppressing $i,j$ indices for notational simplicity):
\be \label{dfun3}
G_{\al_1 \cdots \al_n} (t_1, \cdots t_n) \equiv {1 \ov i^{n_r}}
{\de^n W \ov \de \phi_{\bar \al_1} (t_1) \cdots
\de \phi_{\bar \al_n} (t_n)} \biggr|_{\phi_a = \phi_r =0} =  i^{n_a} \vev{\sP \sO_{\al_1} (t_1) \cdots \sO_{\al_n} (t_n)} \ ,
\ee
where $\al_1, \cdots , \al_n \in (a, r)$ and $\bar \al = r, a$ for $\al = a, r$. $n_{r,a}$ are the number of $r$ and $a$-index in $\{\al_1, \cdots, \al_n\}$ respectively ($n_a + n_r =n$). The $r-a$ representation~\eqref{ra1}--\eqref{dfun3} is convenient as~\eqref{dfun3} is directly related to (nonlinear) response and fluctuation
functions, which we will review momentarily.

Equations~\eqref{pro2}--\eqref{pro1} can also be written as
\be \label{pro3}
W [\phi_{ai}=0 , \phi_{ri}] = 0,
\ee
and
\be \label{pro4}
W^*[\phi_{ai} , \phi_{ri}] = W [ - \phi_{ai} , \phi_{ri}] \ .
\ee
Equation~\eqref{pro3} implies that
\be \label{3top}
G_{a\cdots a} = 0 \ .
\ee

\subsection{Nonlinear response functions} \label{sec:response}

In this subsection, for notational simplicity we will suppress $i,j$ indices on $\sO$ and $\phi$'s.
To understand the physical meaning of correlation functions introduced in~\eqref{dfun3}, let us first expand $W$ in terms of $\phi_{a}$'s:
\be \label{eem}
W [\phi_a , \phi_r]= i \int d t_1 \, D_r (t_1)  \phi_a (t_1) + {i^2 \ov 2!} \int d t_1 d t_2  \, D_{rr} (t_1, t_2)  \phi_a (t_1) \phi_a (t_2) + \cdots,
\ee
where
\be \label{dfun}
D_{r \cdots r} (t_1, \cdots,  t_n) =  {1 \ov i^n} {\de^n W \ov \de \phi_{a} (t_1) \cdots
\de \phi_{a} (t_n)} \biggr|_{\phi_a =0} = \vev{\sP \sO_r (t_1) \cdots \sO_r (t_n) }_{\phi_r} \ .
\ee
For $\phi_a =0$, we have $\phi_1 = \phi_2 = \phi_r \equiv \phi$.
Writing the last expression of~\eqref{dfun} explicitly in terms of orderings of $\sO$'s, we find that
\be \label{dfun1}
D_r (t) = \vev{\sO (t)}_{\phi} \ , \qquad D_{rr} (t_1, t_2)= \ha \vev{\{\sO (t_1), \sO(t_2)\}}_{\phi} \ , \quad \cdots
\ee
and $D_{r\cdots r} (t_1, \cdots, t_n)$ is the fully symmetric $n$-point fluctuation functions of $\sO$, in the presence of external source $\phi$.  They are  referred to as non-equilibrium fluctuation functions~\cite{bernard,peterson}~(see also~\cite{Wang:1998wg}).

One can further expand these non-equilibrium fluctuations functions in the external source $\phi (t)$, for example,
\begin{gather} \label{dr1}
D_r (t_1) = \vev{\sO}_\phi = G_r (t_1) +  \int d t_2 \, G_{ra} (t_1,t_2) \phi (t_2) + {1 \ov 2!} \int d t_2 d t_3 \, G_{raa} (t_1,t_2, t_3)  \phi (t_2) \phi (t_3) + \cdots \\
%\ee
%\be
D_{rr} (t_1, t_2)= \ha \vev{\{\sO (t_1), \sO(t_2)\}}_{\phi} = G_{rr} (t_1, t_2) + \int d t_{3} \, G_{rra} (t_1, t_2 , t_{3}) \phi (t_{3}) + \cdots
\label{drr1}
\end{gather}
%\be
%D_{r\cdots r} (t_1, \cdots t_n) = G_{r \cdots r} (t_1, \cdots t_n) + i \int d t_{n+1} \, G_{r\cdots ra} (t_1, \cdots t_n , t_{n+1}) \phi (t_{n+1})% + {1 \ov 2!} \int d t_3 d t_4 \, G_{rraa}  (t_1,t_2, t_3, t_4 ) \phi (t_3) \phi  (t_4)
%+ \cdots
%\ee
where $G_{\al_1 \cdots \al_n}$ were introduced in~\eqref{dfun3}. From~\eqref{dr1}, it follows that $G_r$ is the one-point function in the absence of source, and $G_{ra}, G_{raa}, \cdots$ are respectively linear, quadratic and high order response functions of $\sO$ to the external source.
Similarly, $G_{rr}$ is the symmetric two-point function in the absence of source, and $G_{rra}, G_{rraa}, \cdots$ are response functions for the second order fluctuations.
Indeed, writing the last expression of~\eqref{dfun3} explicitly in terms of orderings of $\sO$'s, one finds that
$G_{ra\cdots a}$ are the fully retarded $n$-point Green functions of~\cite{Lehmann:1957zz}, while
$G_{r \cdots r}$ is the symmetric $n$-point fluctuation function~\cite{bernard,peterson}. Other $G_{\al_1 \cdots \al_n}$
involve some combinations of symmetrizations and antisymmetrizations.

Note that, by definition, for hermitian operators, all of these functions are real in coordinate space.
At the level of two-point functions, one has
\be \label{yepi}
  G_{ra} (t_1, t_2) = G_R (t_1, t_2)  , \qquad   G_{ar} (t_1, t_2) = G_A (t_1, t_2)  , \qquad
 G_{rr} (t_1, t_2) = G_S  (t_1, t_2),
\ee
where $G_R, G_A$ and $G_S$ are retarded, advanced and symmetric Green functions respectively.
Explicit forms of various three-point functions are given in Appendix~\ref{app:functions}.

\subsection{Time reversed process and discrete symmetries} \label{sec:DS}

Let us now consider constraints on the connected generating functional $W$ when $\rho_0$ invariant under certain discrete symmetries.
We will now restore spatial coordinates using the notation $x = (t, \vx)$, and take spacetime dimension to be $d$.

Suppose that $\rho_0$ is invariant under parity $\sP$ or charge conjugation $\sC$, i.e.
\be
\sP \rho_0 \sP^\da = \rho_0, \quad {\rm or} \quad \sC \rho_0 \sC^\da = \rho_0.
\ee
Then, from~\eqref{pager1}
\bega
 \label{h1}
W[\phi_{1i}, \phi_{2i} ] = W[\phi_{1i}^P, \phi_{2i}^P] ,
 \qquad \phi^P_i (x) \equiv \eta_i^P \phi_{i} (\sP x), \\
W[\phi_{1i}, \phi_{2i} ] = W[\eta_i^C \phi_{1i}, \eta_i^C \phi_{2i}],
\label{h2}
\end{gather}
where we have taken
\be
\sP \sO_i (x) \sP^\da = \eta_i^P \sO_i (\sP x) , \qquad \sC \sO_i (x) \sC^\da = \eta_i^C \sO_i ( x)  \ .
\ee
For even spacetime dimensions, $\sP x$ changes the signs of all spatial directions, while for odd
dimensions, it changes the sign of a single spatial direction.

For time reversal, consider a process with $\rho_0$ the state at $t = +\infty$ with the same external perturbations:
\bln
e^{W_T [\phi_{1i}, \phi_{2i}]} &=  \Tr\le[U_2^\da (+\infty, - \infty; \phi_{2i}) \rho_0 U_1 (+\infty, -\infty; \phi_{1i})\ri]
 \cr
& =
 \Tr \le[\rho_0 \le(T e^{ i \int d t \,  \sO_{1i} (t) \phi_{1i} (t) } \ri) \le( \tilde T e^{- i  \int d t \, \sO_{2i} (t) \phi_{2i} (t) } \ri) \ri]  \ .
 \label{defwt}
\end{align}
It should be stressed that  $W_T$ is a definition and  we have not assumed time reversal symmetry.
At quadratic order in $\phi$'s, % from their respective definitions $W_T$ can be obtained from $W$. More explicitly, at quadratic level
we can write $W$ as
\be \label{1w}
W = i \int d^dx_1 d^dx_2 \, \le({i \ov 2} G_{ij} (x_1-x_2) \phi_{ai} (x_1) \phi_{aj} (x_2) + K_{ij} (x_1 - x_2) \phi_{ai} (x_1) \phi_{rj} (x_2) \ri), \
\ee
with symmetric, retarded and advanced Green functions given respectively by
\be
G^S_{ij} (x)= G_{ij} (x) = G_{ji} (-x) , \quad G^R_{ij} (x) = K_{ij} (x), \quad G^A_{ij} (x) = \bar K_{ij} (x)\equiv K_{ji} (-x) \ .
\ee
From~\eqref{defwt}, $W_T$ can be written as
\be \label{1wt}
W_T = i \int d^d x_1 d^d x_2 \, \le({i \ov 2} G_{ij} (x_1-x_2) \phi_{ai} (x_1) \phi_{aj} (x_2) +  \bar K_{ij} (x_1 - x_2) \phi_{ai} (x_1) \phi_{rj} (x_2) \ri) \ ,
\ee
but for higher point functions, $W_T$ can no longer be directly obtained from $W$.

Now let us suppose that $\rho_0$ is invariant under time-reversal symmetry, i.e.
\be
\sT \rho_0 \sT^\da = \rho_0,  \qquad \sT \sO (x) \sT^\da = \eta_i^T \sO(\sT x) , \quad \sT x \equiv (-t, \vx),
\ee
then from~\eqref{pager1} and~\eqref{defwt} we find (for real $\phi_i$'s)
\be \label{tins}
W[\phi_{1i}, \phi_{2i}] = W_T [\phi_{1i}^{T}, \phi_{2i}^{T}] ,  \qquad \phi_i^{T} (x) \equiv \eta_i^T \phi_i (\sT x)   \ .
\ee

For $\rho_0$ invariant under some products of $\sC, \sP, \sT$, the results can be readily obtained from~\eqref{h1}--\eqref{h2} and~\eqref{tins}. For example, suppose that $\rho_0$ is invariant under $\sP\sT$, i.e.
\be
\Th \rho_0 \Th^\da = \rho_0, \qquad \Th = {\cal P T},
\ee
then
\be \label{cpti1}
W [\phi_{1i}, \phi_{2i}] = W_T [ \phi_{1i}^{PT},  \phi_{2i}^{PT}], \qquad \phi_i^{PT} (x) \equiv \eta_i^{PT} \phi_i (-x),
\quad \eta_i^{PT} \equiv   \eta_i^P  \eta_i^T   \ .
\ee

From~\eqref{1w} and~\eqref{1wt}, for a system with $\sP \sT$ symmetry,~\eqref{cpti1} implies that
\be\label{1ptsy}
G_{ij} (x) = \eta_{i}^{PT}\eta_{j}^{PT} G_{ij} (-x), \qquad
K_{ij} (x) = \eta_{i}^{PT}\eta_{j}^{PT} K_{ji} (x) \ .
\ee
For higher point functions,~\eqref{cpti1} does not impose any direct constraints on $W$ itself, only relating $W$ to $W_T$.

\subsection{Thermal equilibrium and the KMS condition} \label{sec:fdt}

Let us now specialize to a thermal density matrix
\be \label{1emn}
\rho_0 = {1 \ov Z_0} e^{-\beta_0 (H - \mu_0 Q)}, \qquad Z_0 = \Tr e^{-\beta_0 (H - \mu_0 Q)} \ .
\ee
We will restrict to our discussion to Hermitian operators $\sO_i$ which commute with charge $Q$. This is satisfied by the stress tensor $T^{\mu \nu}$ and the current $J^\mu$ associated with $Q$ which are the main interests of this paper. Then $W$ satisfies the following KMS condition~\cite{kubo57, mart59,Kadanoff}:
\bln
\label{newfdt}
e^{W [\phi_{1i}, \phi_{2i}]} & = {1 \ov Z_0} \Tr \le[e^{-(\beta_0 -\th) \hat H} \le(\tilde T e^{- i \int \sO_{2i} \phi_{2i}} \ri) e^{(\beta_0 - \th) \hat H} e^{-\beta_0 \hat H}  e^{\th \hat H} \le( T e^{i  \int \sO_{1i} \phi_{1i}}  \ri) e^{- \th \hat H}  \ri] \cr
& = e^{W_T  [\phi_{1i} (t + i \th ), \phi_{2i} (t - i (\beta_0-\th)) ] },
\end{align}
for arbitrary $\th \in [0, \beta_0]$ where  $\hat H = H- \mu_0 Q$ and we have used that
\be
e^{-a \hat H} \le(\tilde T e^{ i \int \sO (t) \phi (t)} \ri) e^{a\hat H}
= \tilde T e^{i  \int \sO(t) \phi (t - i a)}
\ee
and~\eqref{defwt}. Similarly we have
\be
W_T [\phi_{1i}, \phi_{2i}] = W[ \phi_{1i} (t - i \th), \phi_{2i} (t + i (\beta_0 - \th))] \ .
\ee

At quadratic order in $\phi_i$'s, from~\eqref{1w}--\eqref{1wt}, equation~\eqref{newfdt} gives the standard fluctuation-dissipation theorem (FDT) for two-point functions:
\be \label{fdt4}
G_{ij} (k) %= - {i \ov 2} (K_{ij} - \bar K_{ij} ) {Q - \bar Q \ov Q + \bar Q}
= \ha \coth {\beta_0 \om \ov 2} \De_{ij} (k), \qquad
i \De_{ij} \equiv K_{ij} - \bar K_{ij} \ .
\ee

For higher point functions, $W_T$ cannot be expressed in terms of $W$, and
the KMS condition~\eqref{newfdt} by itself does not impose constraints on $W$ beyond quadratic order.
For a $ \sP \sT$ invariant Hamiltonian $H$,
$\rho_0$ is invariant under $\sP\sT$. Using~\eqref{cpti1}, we can then further write~\eqref{newfdt}  as
\bea \label{newfdt1}
W [\phi_{1i}, \phi_{2i}] &=&  W_T [\eta^{PT} \phi_{1i} (-x),  \eta^{PT}  \phi_{2i} (-x)] \cr
  &= & W [\eta^{PT} \phi_{1} (-t + i \th, -\vx) , \eta^{PT } \phi_{2} (- t - i( \beta_0- \th), -\vx )  ] \ .
\eea
For the stress tensor and  conserved currents, which are
our main interests of the paper, $\eta_i^{PT} =1$ for all components. Below we will
take $\eta_i^{PT} =1$.

For two point functions, with $\sP \sT$ symmetry in addition to~\eqref{fdt4} we also have~\eqref{1ptsy}, which in momentum space becomes
\be \label{1onsa}
G_{ij} (k)  = G_{ij} (-k) = G_{ij}^* (k) = G_{ji} (k)   , \qquad K_{ij} (k) = K_{ji} (k), %  \quad \Lra \quad  G = G^T = G^*, \quad K = K^T
\ee
the second of which are Onsager relations.
Recall that by definition, $G_{ij}$ is real in coordinate space and is Hermitian in momentum space.

At cubic level in $\phi$'s, let us write $W$ as
\be \label{lag3}
W = i \le[{1 \ov 3!} G_{ijk} \phi_{ai} \phi_{aj} \phi_{ak} + {i \ov 2} H_{ijk} \phi_{ai} \phi_{aj} \phi_{rk} +
\ha K_{ijk} \phi_{ai} \phi_{rj} \phi_{rk} \ri],
\ee
where we have used a simplified notation, e.g. the first term should be understood in momentum space as
\be
G_{ijk} \phi_{ai} \phi_{aj} \phi_{ak}
= \int d k_2 d k_3 \,
G_{ijk} (k_1, k_2, k_3) \phi_{ai} (k_1) \phi_{aj} (k_2)  \phi_{ak} (k_3) , \quad k_1 + k_2 + k_3 =0,
\ee
and similarly with others. Note that (suppressing $ijk$ indices)
\be
G = - G_{rrr} , \qquad  H = G_{rra}, \qquad K = G_{raa} \ .
\ee
By definition, the $G_{ijk} (k_1, k_2, k_3)$ are fully symmetric under simultaneous permutations of $i,j,k$ and the corresponding momenta, and
\begin{gather}
%G_{ijk} (k_2, k_3) = G_{jik} (-k_2 - k_3, k_3) = G_{ikj} (k_3, k_2) = G_{kji} (k_2, - k_2 - k_3) = \cdots  \\
H_{ijk} (k_1, k_2, k_3) = H_{jik} (k_2, k_1, k_3), \qquad K_{ijk} (k_1, k_2, k_3) = K_{ikj} (k_1, k_3, k_2) \ .
\end{gather}

To write the KMS condition for three-point functions, it is convenient  to introduce the following notation (suppressing all $i, j$ indices):
\be
H_3 \equiv G_{rra}, \quad H_2 \equiv G_{rar} , \quad H_1 \equiv G_{arr}, \quad
K_1 \equiv   G_{raa} , \quad K_2 \equiv  G_{ara}, \quad K_3 \equiv  G_{aar}\ .
\ee
Then~\eqref{newfdt1} applied to three-point level can be written in momentum space as~\cite{Wang:1998wg}
\bln \label{H1fdt}
H_1&=\frac i{2}(N_3+N_2)K_1^*-\frac i2 (N_2 K_3+N_3 K_2),\\
H_2&= \frac i{2}(N_3+N_1)K_2^*-\frac i2 (N_1 K_3+N_3 K_1),\\
H_3&= \frac i{2}(N_1+N_2)K_3^*-\frac i2 (N_1 K_2+N_2 K_1),\\
\label{H3fdt}
 G & =\frac 14\left(\left(K_1^*+K_2^*+K_3^*\right)+2 N_2 N_3 {\rm Re} \, K_1 + 2N_1 N_3 {\rm Re} \, K_2+2N_1 N_2 {\rm Re} \, K_3\right),
\end{align}
where we have introduced
\be
 N_a=\coth\left(\frac{\beta \om_a}2\right) , \quad a=1,2,3\ .
 \ee

Expressions of~\eqref{newfdt1} in terms of correlation functions  at general orders  are reviewed in Appendix~\ref{app:fdt}.

\subsection{The classical statistical limit} \label{sec:CSL}

Let us now consider the classical limit of the generating functional~\eqref{ra2}
for a density matrix $\rho_0$ which has a classical statistical mechanics description.

With $\hbar$ restored, each term in~\eqref{eem} and~\eqref{dr1}--\eqref{drr1} should have
a factor $\hbar^{-n}$ with $n$ equal to the number of $\phi_{r,a}$ factors.
As defined, the symmetric Green functions~\eqref{dfun} should all have a well defined $\hbar \to 0$ limit, and after taking the limit, they describe classical statistical fluctuations.  $G_{r \cdots r  a \cdots a} $ with $n_a$ $a$-indices should have the limiting behavior
\be
G_{r \cdots r  a \cdots a} \to \hbar^{n_a}  G_{r \cdots r a \cdots a}^{\rm (cl)}, \qquad \hbar \to 0,
\ee
as it has $n_a$ commutators. $G_{r \cdots r a \cdots a}^{\rm (cl)}$ is defined exactly as $G_{r \cdots r a \cdots a}$, but with all commutators replaced by Poisson brackets. From now on, to simplify notation, we will suppress the subscript ``cl'' and use the same notation
to denote the quantum and classical correlation functions.
Thus, for $W[\phi_a, \phi_r]$ to have a well-defined limit, the sources $\phi_a, \phi_r$ should scale as
\be
\phi_a \to \hbar \phi_a , \qquad \phi_r \to \phi , \qquad \hbar \to 0  \ .
\ee

Let us now look at the $\hbar \to 0$ limit of the KMS conditions~\eqref{newfdt1}.
With $\hbar$ restored, $\beta_0$ in all expressions should be replaced by $\beta_0 \hbar$.
At the level of two-point functions, equation~\eqref{fdt4} then becomes
\be
G_{ij} = {1 \ov \beta_0 \om} {\rm Im} \De_{ij}
\ .
\ee
At cubic level, given $G \sim O(\hbar^0), H \sim O(\hbar), K \sim O(\hbar^2)$, equations~\eqref{H1fdt} and~\eqref{H3fdt} become
\bega \label{1hh}
H_1 = - {i \ov \beta \om_2 \om_3} \le( \om_1K_1^* + \om_2 K_2 + \om_3 K_3 \ri), \\
%{\rm Im} G = 0, \qquad {\rm Re}
G ={2 \ov \beta^2 \om_1 \om_2 \om_3} \le(\om_1 {\rm Re} K_1+
\om_2 {\rm Re} K_2 + \om_3 {\rm Re} K_3 \ri), \
\label{2hh}
\end{gather}
and $H_2, H_3$ can be obtained from~\eqref{1hh} by permutations.

\subsection{Constraints on response functions from KMS conditions} \label{sec:icon}

The KMS conditions~\eqref{newfdt1} not only relate various nonlinear response
and fluctuation functions, they also imply conditions on correlation functions themselves.
For example, at two point function level,~\eqref{fdt4}, regularity of  $G_{ij}$ in the limit $\om \to 0$
requires  that
\be
{\rm Im} \De_{ij} \to  0, \quad \om \to 0 \ .
\ee
Similarly, in~\eqref{H1fdt}--\eqref{H3fdt}, regularity of $H_{1,2,3}$ and $G$
when taking some combinations of $\om_{1,2,3}$ to zero also imposes constraints
on $K_{1,2,3}$ in various zero frequency limits. The complete set of conditions are given in equations~\eqref{con1}--\eqref{con3} of Appendix~\ref{app:fdt}.

Of particular interest to us are consistency conditions involving only
response functions $G_{ra \cdots a}$, which will play an important role in our discussion of hydrodynamics. For general $n$-point response functions, let us denote
\be \label{defkk}
K_1 = G_{ra \cdots a}, \qquad K_2 = G_{ara \cdots a} , \qquad \cdots  \qquad
K_n = G_{a \cdots a r}  \ .
\ee
We can show that when taking any $n-2$ frequencies to zero, e.g.
\be \label{allc2}
K_1 = K_2^* , \qquad \om_3, \om_4, \cdots , \om_n \to 0  \ .
\ee
From equation~\eqref{allc2} and permutations of it, it then follows that
\be \label{allc}
 K_1 = K_2 = \cdots = K_n \equiv K, \quad {\rm Im} \, K = 0, \qquad {\rm all} \; \om_i \to 0   \ .
\ee
Except for two-point functions, equations~\eqref{allc2}--\eqref{allc} for general $n$ appear to be new. We prove~\eqref{allc2} in Appendix~\ref{app:fdtproof}.
Equations~\eqref{allc} have simple physical interpretations: the first equation says that in the
stationary limit, there is no retardation effect, while the second equation says that there is no dissipation.

For two-point functions, denoting $K \equiv K_1$, then $K_2 = K^\da$, equation~\eqref{allc2} reduces to
\be
K_{ij} (\om, \vk) = K_{ji} (\om, \vk) ,
\ee
i.e. the familiar Onsager relations.  From now on we will refer to~\eqref{allc2} as generalized  Onsager relations.

It appears to us~\eqref{allc2} and~\eqref{allc} are the only relations involving response functions alone. If one leaves more than two frequencies nonzero,
then the KMS relations will necessary involve functions with more than one $r$-indices, as in $n=3$ relations~\eqref{H1fdt}--\eqref{H3fdt}.

Equations~\eqref{allc2}--\eqref{allc} can be written in a compact way in terms of one-point function~\eqref{dr1} in the presence of sources. For this purpose, it is convenient to define
\bln
&\sG_{i_1 i_2} (x_1, x_2; \phi_i (\vx)] =  {\de \vev{\sO_{i_1} (x_1)}_\phi \ov \de \phi_{i_2} (x_2)} \biggr|_{S} \cr
& =  K_{i_1 i_2} (x_1, x_2) +  \int d^d x_3 \, K_{i_1 i_2 i_3} (x_1,x_2, x_3)  \phi_{i_3} (\vx_3)\cr
& + \ha
\int d^d x_3  d^d x_4 \, K_{i_1 i_2 i_3 i_4} (x_1,x_2, x_3, x_4)  \phi_{i_3} (\vx_3)  \phi_{i_4} (\vx_4) +
 \cdots,
\end{align}
where again $K \equiv K_1$, and the subscript $S$ in the first line denotes the procedure that after taking the differentiation one should set all sources to be time-independent. The notation $\sG (\cdots ]$ highlights that it is a function of $x_1, x_2$, but a functional of $\phi_i (\vx)$. In the second line, $\phi (\vx)$ indicates that
the sources only have spatial dependence. Then~\eqref{allc2} can be written as
\be \label{1con1}
\sG_{i j} (x, y; \phi_i (\vx)] = \sG_{j i} (-y, -x; \phi_i (-\vx) ] ,
\ee
or in momentum space
\be
 \sG_{i j} (k_1, k_2; \phi_i (\vk)] = \sG_{j i} (-k_2, -k_1; \phi_i (-\vk)] =
\sG_{j i}^* (k_2, k_1; \phi_i (-\vk)] \ .
\ee

Now look at the first equation of~\eqref{allc}, which implies that in the stationary limit there exists some functional $\tilde W [\phi_i (\vx)]$ defined on the {\it spatial} part of the full spacetime, from which
\be \label{1con2}
 \vev{\sO_{i} (\om =0, \vx)}_\phi  = {1 \ov i} {\de \tilde W [\phi (\vx)] \ov \de \phi_{i} (\vx)}  \ .
 \ee
The above equation implies that for stationary sources to first order in $\phi_a$, the generating functional~\eqref{eem} can be written in a ``factorized'' form:
\be \label{1fac}
W [\phi_r, \phi_a] =i  \int d^{d-1} \vx \, \vev{\sO_i (\om=0, \vx)}_\phi  \, \phi_{ai} (\vx) + \cdots =
i \tilde W [\phi_1] - i \tilde W [\phi_2] + \cdots \ .
\ee

The second equation of~\eqref{allc} is the statement that  $K_{i_1 \cdots i_n} (\vk_1, \cdots \vk_n) $
are real in momentum space. By definition, $K$'s are real in coordinate space. That they are also real in momentum space implies that
\be
K_{i_1 \cdots i_n} (\vk_1, \cdots \vk_n)= K_{i_1 \cdots i_n} (-\vk_1, \cdots , - \vk_n) = {\rm real}
\ee
which in turn implies that
\be
\tilde W [\phi (\vx) ] = \tilde W [\phi (-\vx)]  \ .
\ee

\section{Relations with standard formulations} \label{sec:hydro}

In this section we first explain how the standard hydrodynamical equations of motion arise in our framework.
Then we consider constraints on hydrodynamical equations of motion following from our symmetry principles outlined in the introduction. In particular,  the prescription~\cite{Banerjee:2012iz,Jensen:2012jh} that in a stationary background the stress tensor and current should be obtainable from a stationary partition function will arise as a subset of our conditions. We will find a set of new constraints to which we refer as generalized Onsager conditions.

Finally we discuss how to recover the standard formulation of fluctuating hydrodynamics and aspects of our theory going beyond it.

\subsection{Recovering  hydrodynamical equations of motion}
\label{sec:fluc}

Let us first explain how the standard hydrodynamical equations of motion arise in our formulation.
To illustrate the basic idea, we again use the same simplified notation of~\eqref{qft1}.
Since we are interested in the equations of motion (i.e. in the thermodynamical limit of Sec.~\ref{sec:hbar}), it is enough to consider the bosonic theory,
with all ghost dependence ignored.

Recall from Sec.~\ref{sec:eom} that the equations of motion for the dynamical variables $\chi_{a,r}$ correspond to the conservation of $J_{a,r}$, which we can schematically write as\footnote{Equations of motion for $\tau_{r,a}$ do not have this structure. They can be solved algebraically and do not affect the argument below.  }
\bega
\label{j0eq}
\p J_r % =  \p J_r^{(0)} + \p J_r^{(1)} + \cdots
= 0
, \qquad
\p J_a %=  \p J_a^{(1)} + \p J_a^{(2)} + \cdots
= 0 \ .
\end{gather}
Let us now expand the bosonic action $I$ in terms of the number of $a$-fields, as discussed around~\eqref{aexp0},
\be \label{aexp1}
I =  I^{(1)} +  I^{(2)} + \cdots,
\ee
where $I^{(m)}$ contains altogether $m$ factors of $\phi_a$ and $\chi_a$. From~\eqref{j11},  the current operators $J_{a,r}$ can be similarly expanded as
\bega \label{jexp}
J_r = J_r^{(0)} + J_r^{(1)} + \cdots , \qquad J_a = J_a^{(1)} + J_a^{(2)} + \cdots,   %\\
\end{gather}
where $m$ in the superscript $(m)$ again denotes the number of $a$-fields in each expression.
Note that $J_a$ starts with $m=1$, i.e. $J_a |_{\phi_a =0, \chi_a =0} =0$, and $J_r^{(0)}$ only depends on
the lowest order action $I^{(1)}$.

With~\eqref{jexp}, the  equations of motion~\eqref{j0eq} also have the expansion
\bega
\label{j1eo}
\p J_r  =  \p J_r^{(0)} + \p J_r^{(1)} + \cdots  = 0 , \\
\p J_a =  \p J_a^{(1)} + \p J_a^{(2)} + \cdots  = 0 \ .
\label{j2eo}
\end{gather}

To make connection with the standard hydrodynamical equations, let us now take the background fields of the two segments of CTP to be the same,  i.e.
\be
\phi_1 = \phi_2 = \phi_r = \phi, \qquad \phi_a =0 ,
\ee
or in terms of our original fields,
\be \label{yuen}
g_{1\mu \nu} = g_{2 \mu\nu} = g_{\mu \nu}, \qquad A_{1 \mu} = A_{2 \mu} = A_\mu \ .
\ee
With $\phi_a =0$,  as already discussed after~\eqref{class}, the equations of motion give that
\be \label{j3eo}
\chi_a^{(\rm cl)} =0  \quad \to \quad \chi_1 = \chi_2 = \chi_r \equiv \chi \ .
\ee
 In terms of our original dynamical variables, one then has
\be  \label{yuen1}
X^\mu_1 = X^\mu_2 = X^\mu, \qquad \vp_1 = \vp_2 = \vp  \ .
\ee

With $\phi_a = \chi_a =0$, $J_a$ vanishes identically and all terms in $J_r$ except for $J^{(0)}$ vanish. Thus,
\be \label{eun}
 J_1 = J_2 = J_r = J^{(0)}_r,  %\equiv J
\ee
and the remaining equations of motion are
\be \label{j4eo}
%\p J =
\p J^{(0)}_r = 0 \  .
\ee
In terms of original variables, equation~\eqref{eun} corresponds to
\be \label{1hyd}
\hat T^{\mu \nu}_1 = \hat T^{\mu \nu}_2  = (\hat T_r^{\mu \nu})^{(0)} \equiv \hat T^{\mu \nu}_{\rm hydro} ,\qquad
\hat J_1 = \hat J_2 = (\hat J_r^{\mu})^{(0)} \equiv  \hat  J^{\mu}_{\rm hydro}
\ee
and~\eqref{j4eo} to
\be \label{rehyd}
 \nab_{\mu}   \hat  J^\mu_{\rm hydro} =0, \qquad
 \nab_{\nu}  \hat  T^{\nu \mu}_{\rm hydro} - F^{\mu}{_\nu}  \hat  J^\nu_{\rm hydro} = 0 \ .
 \ee
Furthermore, one can show from the symmetry requirements~\eqref{sdiff}--\eqref{daug}, as the zeroth order terms in the $a$-field expansion of currents, $ \hat  T^{\mu \nu}_{\rm hydro}$
and $ \hat  J^{\mu}_{\rm hydro}$ can be expressed solely in terms of the velocity field~\eqref{umu}, local chemical potential~\eqref{mub} and local temperature field~\eqref{tay0} (which we will prove explicitly in
in Sec.~\ref{sec:stress} and Appendix~\ref{app:proof}). Equations~\eqref{rehyd} then reproduce the standard hydrodynamical equations.

To summarize, {\it the standard hydrodynamical equations of motion correspond to the zeroth
order approximation in the $a$-field expansion in the thermodynamical limit.}
%In other words it only concerns $I^{(1)}$ in the $a$-field expansion~\eqref{aexp0} of the action.

\subsection{Constraints on hydrodynamics} \label{sec:consh}

For $\rho_0$ given by the thermal ensemble~\eqref{emn}, we also need to impose the local KMS conditions
on the source action $I_s$~\eqref{keyp4}. As far as the hydrodynamical equations of motion~\eqref{rehyd}
are concerned, we only need to look at constraints on $I^{(1)}_s$, which encode the contact contributions to all of the response functions.

In the standard formulation of hydrodynamics one needs to impose constraints from the local second law of thermodynamics, existence of stationary equilibrium, and the Onsager relations. In our formulation, these constraints are fully taken care of by the local KMS conditions~\eqref{keyp4}. At an abstract level, this is a consequence of the facts that: (i)  the local KMS conditions ensure that the full KMS conditions are satisfied in the thermodynamical limit; (ii) the full KMS conditions are known to imply the local second law (see e.g.~\cite{sewell}) as well as existence of stationary equilibrium; (iii) time reversal symmetry is encoded in our
formulation of local KMS conditions. In fact, from the discussion of Sec.~\ref{sec:icon}, local KMS conditions include not only  the Onsager relations for linear responses, but also give full nonlinear generalizations.

More explicitly,  restricted to $I^{(1)}_s$, the local KMS conditions give the following three types of constraints:

\ben
\item[(a)] Relations between coefficients in $I^{(1)}_s$ and higher order terms in $a$-expansion. For example, at first derivative order,~\eqref{fdt4} relates transport coefficients such as shear, bulk viscosities and conductivity in $I^{(1)}_s$ to coefficients in $I^{(2)}_s$ (FDT relations). From~\eqref{keyp3} $I^{(2)}$, terms in the action are pure imaginary and their coefficients should satisfy certain non-negativity conditions in order for the path integral to be well defined. Altogether, this implies the non-negativity of various transport coefficients.
As we shall see in Sec.~\ref{sec:trans}, while this works out easily for the shear viscosity, for conductivity and bulk viscosity it is highly nontrivial. At first derivative order,  the non-negativity of shear, bulk viscosities and conductivity are all one gets. These are also the inequality constraints from the non-negative divergence of the entropy current. In fact it has been argued recently~\cite{Bhattacharyya:2013lha,Bhattacharyya:2014bha} these are the only inequality constraints from the entropy current
to all orders in derivatives. It is conceivable, in our context at higher derivative orders
the well-definedness of the integration measure combined with  FDT relations may give additional
inequality relations, thus predicting new relations going beyond those from the entropy current.

\item[(b)] When all sources in $I^{(1)}_s$ are taken to be time-independent, $I^{(1)}_s$ should satisfy~\eqref{allc}.  From~\eqref{1fac}, this means that for stationary sources we can write $I^{(1)}_s$
in a factorized form
\be
I_s^{(1)} [g_{1}, A_1; g_2, A_2 ] = \tilde W [g_1, A_1] - \tilde W [g_2, A_2]
\ee
where $\tilde W [g, A]$ is a local functional of stationary metric $g_{\mu\nu} (\vx)$ and gauge field $A_\mu (\vx)$ on the {\it spatial} manifold. Note that for stationary backgrounds, the dynamical modes will not be excited and thus
$I_s^{(1)}$ is the full contribution to the leading generating functional $W_{\rm tree}^{(1)}$ in the $a$-field expansion in the thermodynamical limit. We thus have derived the prescription~\cite{Banerjee:2012iz,Jensen:2012jh}  that in a stationary background the stress tensor and current should be derivable from a partition function. In~\cite{Bhattacharyya:2013lha,Bhattacharyya:2014bha} it has also been shown that this requirement is equivalent to equality-type constraints from the entropy current. Now this coincidence becomes completely natural.

\item[(c)] For time dependent sources, we have an additional set of constraints following from the generalized Onsager relations~\eqref{1con1} on $I^{(1)}_s$ coefficients. In the next section (and Appendix~\ref{app:example}), we will see that they lead to new constraints in the hydrodynamics of a single current
starting at second order in derivative expansion. For a full charged fluid including the stress tensor, these new constraints will also start operating at the second derivative order, but we will not work
them out explicitly in this paper.

\een

\subsection{Recovering stochastic hydrodynamics}  \label{sec:noise}

Now we show how to recover the standard formulation of fluctuating hydrodynamics~\cite{landau1,LL}.
For this purpose, consider the first two terms in the $a$-field expansion~\eqref{aexp1}:
\be
I = I^{(1)} + I^{(2)}    \ .
\ee
From our discussion of Sec.~\ref{sec:fluc} we can write $I^{(1)}$ as
\be
I^{(1)} =  \chi_a \p J_r^{(0)} \ ,
\ee
which gives the equations of motion~\eqref{j4eo} when varied with respect to $\chi_a$.
$I^{(2)}$ can be schematically written as
\be
 I^{(2)}  =  {i \ov 2} \chi_a G (\p , \chi_r) \chi_a  \ ,
 \ee
where $G$ is a local differential operator depending on $\chi_r$. Now, expanding
$G (\p, \chi_r)$ in powers of $\chi_r$,
\be
 G (\p , \chi_r) = G_0 (\p) + O(\chi_r)\ ,
\ee
where now $G_0$ is a local differential operator with no dependence on dynamical variables.
Keeping only the $G_0$ term in $I^{(2)}$, we can write the action  schematically as
\be \label{truc}
I = \chi_a \p J_r^{(0)} + {i \ov 2} \chi_a G_0 \chi_a   \ .
\ee
Note that we are not doing any $\chi_r$ expansion in $I^{(1)}$.

Now consider a Legendre transformation of the second term of~\eqref{truc}, i.e. introducing $\xi  = - {\p I_{aa} \ov \p \chi_a}$
%= -iK_1 \vp_a$
to rewrite $I_{aa} = {i \ov 2} \chi_a G_0 \chi_a   $ as
\be \label{ltr1}
I_{aa} = - \chi_a \xi + \tilde I_{aa} [\xi] , \quad {\rm with} \quad
\tilde I_{aa} = {i \ov 2} \xi {1 \ov G_0} \xi.
\ee
$I$ can then be written as
\be \label{zero9}
 I = {i \ov 2} \xi  {1 \ov G_0} \xi  + \chi_a \le(\p J_r^{(0)} - \xi \ri)  \ .
\ee
The path integral then becomes
\be
e^W = \int D \xi D \chi_r D \chi_a \, e^{i I}
= \int D \xi D \chi_r \, \de \le(\p J_r^{(0)}- \xi \ri) \, e^{- \ha \int d^d x \, \xi G^{-1}_0 \xi},
\ee
i.e. $\chi_a$ is now a Lagrange multiplier, whose integration gives the stochastic diffusion equation
\be\label{stoeom1}
\p J_r^{(0)} = \xi  \ ,
 % \quad \psi \equiv \p_0 \vp_r , \quad
%\eta = {1 \ov \chi} \xi
\ee
where $\xi$ is a stochastic force with local Gaussian distribution:
\be \label{stoeom2}
\vev{\xi} = 0, \qquad \vev{\xi (x) \xi (0)} = G_0  \de^{(d)} (x) \ .
\ee
Equations~\eqref{stoeom1}--\eqref{stoeom2} recover the standard formulation of fluctuating hydrodynamics~\cite{landau1,LL}.\footnote{Of course, at this stage our discussion is rather schematic. Explicit expressions
can be found in Sec.~\ref{sec:kpz} and Sec.~\ref{sec:stochastic}.}
We see that $\chi_a$ is the conjugate variable for the noises, and thus the expansion in $a$-fields
may be considered as an expansion in noises.

The above discussion  makes clear  the aspects of our formulation that go beyond the traditional formulation of fluctuating hydrodynamics: (i) In addition to the $G_0$ term, the full $I^{(2)}$ also includes interactions between dynamical variables and the noises. (ii) $I^{(n)}$ with $n \geq 3$ includes interactions among noises and higher order interactions among noises and dynamical variables.
(iii) Beyond~\eqref{zero9}, dynamical variables can fluctuate on their own and are not constrained by fluctuations of noises as in~\eqref{stoeom1}. Furthermore, once we include  interactions between $\chi_r$ and $\chi_a$
in $I^{(2)}$, it is no longer convenient to perform the Legendre transform~\eqref{ltr1} from $\chi_a$ to $\xi$ which will result in a non-local and non-polynomial action.
It is more sensible to simply work with $\chi_a$.

From the renormalization group perspective, the effective theory we are writing down
is  defined at a cutoff scale $\Lam$, below which hydrodynamics is defined.\footnote{For example, for a strongly coupled theory, $\Lam$ is of order temperature.} If one is interested in physics at some energy scale $E \ll \Lam$, then one should further integrate out hydrodynamical degrees of freedom with energies $\om \in (E, \Lam)$. It may happen for certain situations that the neglected interactions in~\eqref{zero9} are all irrelevant.  In such a case, the standard stochastic formulation~\eqref{stoeom1}--\eqref{stoeom2} is already adequate for obtaining the leading physics at energies $E \ll \Lam$.

\subsection{Correlation functions}

We conclude the discussion of this section by making some comments on correlation functions.

Let us use $(J_r^{(0)})_{\rm cl}$ to denote the expression obtained by evaluating $J_r^{(0)}$ on the solution to the equations of motion. Then expanding $(J_r^{(0)})_{\rm cl}$  in $\phi_r$
from~\eqref{dr1}, one obtains the full set of nonlinear response functions $G_{ra}, G_{raa}, \cdots$ in the thermodynamical limit.
This constitutes the standard hydrodynamical approach to response functions~\cite{Kadanoff} (see also~\cite{Kovtun:2012rj} for a recent review).

In the thermodynamical limit,  we can go beyond the standard formulation by turning on $\phi_a \neq 0$.
Then both equations~\eqref{j1eo}--\eqref{j2eo} are nontrivial.  Solving these equations to obtain $(J_a^{(n)})_{\rm cl}, (J_r^{(n)})_{\rm cl}$ and expanding them in $\phi_a$ and $\phi_r$, we can
% with $n \geq 1$  perturbatively in $\phi_a$.
%The solutions to higher order terms $J_a^{(n)}, J_r^{(n)}$ with $n \geq 1$ in the $a$-field expansion
now obtain the full set of nonlinear fluctuation and response functions of Sec.~\ref{sec:response}
in thermodynamical limit.  Note that beyond the leading order term $J_r^{(0)}$, $J_{a,r}^{(n)}$ with $n \geq 1$ cannot be expressed solely in terms of velocity-type variables $u^\mu (\sig)$, $ \mu (\sig), T(\sig)$. Instead, the more fundamental fluid field variables, $X^\mu_s$ and $\vp_s$, must be used.

Beyond the thermodynamical limit, we also need to include loop corrections from statistical or quantum fluctuations. Recall the expansion in~$\heff$ discussed in Sec.~\ref{sec:hbar}, which we copy here for convenience:
\be
W  [\phi_r, \phi_a]  = {1 \ov \heff} W_{\rm tree} + W_1 + \heff W_2 + \cdots  \ .
\ee
Corrections from $W_1, W_2, \cdots$ will give rise to phenomena such as long time tails, as well as running  transport coefficients with scales, and so on (see e.g.~\cite{Kovtun:2011np,Kovtun:2012rj} for recent discussions).
Such fluctuation effects may be particularly important near classical and quantum phase transitions and in non-equilibrium situations. %Our formulation provides a systematic framework for dealing with these issues.
%We will leave the study of loop effects for future investigations.

\section{A baby example: stochastic diffusion} \label{sec:diffusion}

As a baby example of the general formalism introduced earlier, we consider the hydrodynamical action
associated with a conserved current discussed in~\eqref{1gent1}--\eqref{1newc},
which we copy here for convenience
\be \label{newc1}
e^{ W[A_{1\mu} , A_{2 \mu}]} =\Tr \le(\rho_0 e^{i \int d^d x \, A_{1\mu} J_1^\mu -  i \int d^d x \, A_{2\mu} J_2^\mu} \ri)
= \int D \vp_1 D \vp_2 \, e^{i I [B_{1\mu},  B_{2 \mu}]}, \
\ee
with
\be\label{Bvar1}
B_{1 \mu}  \equiv A_{1 \mu} + \p_\mu \vp_1, \qquad B_{2 \mu} \equiv A_{2 \mu} + \p_\mu \vp_2 \ .
\ee
%The relevant physics here is diffusion and we will derive a low energy effective action for the diffusion modes $\vp_{1,2}$.
This theory applies to situations where $J^\mu$ either decouples from the stress tensor (as for example for a particle-hole symmetric neutral fluid) or the coupling of $J^\mu$ to the stress tensor is  small enough to be neglectable. % can  be considered as a special limit of the full hydrodynamical theory, %of the stress tensor and current,
%where one ignores the backreaction of the current dynamics to that of the stress tensor.
 In the stress tensor sector one takes the equilibrium solution $X_1^\mu = X^\mu_2
=x^a \de_a^\mu, \; \tau =0$ with
the metric backgrounds $g_{1 \mu \nu}  = g_{2 \mu \nu} = \eta_{\mu \nu}$. Thus in this case the fluid and physical spacetimes coincide. We will take $\rho_0$ to be the thermal ensemble~\eqref{emn}.

It is convenient to introduce the $r-a$ variables,
\be
B_{a \mu} = B_{1 \mu} - B_{2 \mu} = A_{a\mu} + \p_\mu \vp_a  , \qquad B_{r \mu} = \ha (B_{1 \mu} + B_{2 \mu})
= A_{r \mu} + \p_\mu \vp_r \ .
\ee
The local action $I[B_r, B_a]$ should satisfy symmetry conditions 1-8 outlined in the introduction.
%~\eqref{keyp2}--\eqref{keyp4},
In particular, in this case equations~\eqref{sdiff}--\eqref{tdiff} simply reduce to  rotational symmetries in spatial directions. From~\eqref{daug}, it should also be invariant under
\be \label{1daug}
 B_{ri} \to B_{ri} - \p_i \lam (x^i) \ .
\ee
Writing
\be
I = \int d^d x \, \sL,
\ee
we will expand $\sL$  in powers of $B_{r,a}$.

\subsection{Quadratic order}

\subsubsection{The quadratic action}

At quadratic order in $B_{r,a}$, the most general bosonic $\sL$ consistent with rotational symmetries,~\eqref{keyp3} and~\eqref{keyp2}
can be written as
\begin{gather}
\sL = {i \ov 2} a B_{a0}^2 + {i \ov 2} b  B_{ai}^2 + {i \ov 2} c (\p_i B_{ai})^2
+i f B_{a0} (\p_i B_{ai}) + g B_{a0} B_{r0}
+ h B_{a0} \p_i \p_0 B_{ri}  \cr
+ u \p_i B_{ai} B_{r0} + v B_{ai} \p_0 B_{ri} + {w \ov 2}  F_{aij}  F_{rij},
\label{genex}
\end{gather}
where the coefficients $a,b,c, \cdots$ should be understood as real scalar (under spatial rotations) local differential operators  constructed out of $\p_t$ and $\p_i$, and act on the second factor of a term. For example
\be
a B_{a0}^2 \equiv B_{a0} a( \p_t, \p_i) B_{a0} =  B_{a0} (-k_\mu) a (k) B_{a0} (k_\mu), \quad k_\mu = (-\om, \vk),
\ee
where in the second equality we have also written the expression in momentum space.
All of the coefficients can be expanded in the number of derivatives, for example, in momentum space ($q = |\vk|$),
\begin{gather} \label{eiep}
a (k)= a_{00} + a_{20} \om^2 + a_{02} q^2 + \cdots , \qquad
b (k)= b_{00} +  b_{20} \om^2 + b_{02} q^2 + \cdots , \cr
g (k) = g_{00} + i g_{10} \om + g_{20} \om^2 + g_{02} q^2 + \cdots,
\end{gather}
and so on. Note that there is no term with odd powers of $\om$ in the expansions of $a, b, c$ as these
correspond to total derivatives. Thus $a, b, c$ are real in momentum space.
Other coefficients can have odd powers in $\om$ and are complex in momentum space with, e.g.
\be
g(-k) = g^*(k) , \quad h(-k) = h^*(k) ,  \quad \cdots \ .
\ee
In coordinate space $g^*$ is the operator obtained from $g$ by integration by parts
i.e. $g^* (\p_t , \p_i) = g (-\p_t, -\p_i)$.  In the last term of~\eqref{genex}, $F_{ij} = \p_i A_j - \p_j A_i$ and  is independent of $\vp_s$.

Due to~\eqref{keyp3}, the $aa$ terms in~\eqref{genex} are pure imaginary, and thus are real
in the exponent of the path integral~\eqref{newc1}. This implies that the coefficients of the
leading terms in the derivative expansion must be non-negative, for example,
\be \label{1pso}
a_{00} \geq 0, \qquad b_{00} \geq 0 \ .
\ee

Equation~\eqref{genex} applies to general dimensions and is parity invariant. For
a specific dimension, say $d=3$,  one can write down additional parity-breaking terms using fully antisymmetric $\ep$-symbol. %see Appendix~\ref{app:parity} for a discussion.

We still need to impose the local KMS condition~\eqref{keyp4}, which at quadratic level amounts to
imposing~\eqref{fdt4} on the source action obtained by setting dynamical fields $\vp_{r,a}$ to zero in~\eqref{genex}. The source action is the same as~\eqref{genex} with $B_{r\mu}$ and $B_{a\mu}$ replaced by $A_{r \mu}$ and $A_{a\mu}$.
From~\eqref{genex} we can read
\bega
G_{00} = a, \quad G_{ij} =  b \de_{ij}  + c q_i q_j = \tilde b \de_{ij} - c q^2 P^T_{ij}, \quad
G_{0i} = i q_i f , \quad G_{i0} = - i q_i f^*,  \\
K_{00} = g , \quad K_{0i} = q_i \om h, \quad K_{i0} = - i q_i u, \quad K_{ij} = - i \om v \de_{ij} + w q^2 P^T_{ij},  \\
\bar K_{00} = g^*,  \quad \bar K_{0i} = i  q_i u^*    \quad \bar K_{i0} =  q_i \om h^*, \quad \bar K_{ij} =  i \om v^* \de_{ij} + w^* q^2 P^T_{ij}.
\end{gather}
where we have introduced
\be
\tilde b = b + c q^2 , \qquad P^T_{ij} =  \de_{ij} - {q_i q_j \ov q^2} \ .
\ee
Applying~\eqref{fdt4} we then have
\begin{gather} \label{ci1}
a = -{i \ov 2} \coth {\beta \om \ov 2} (g - g^*), \\
\label{ci2}
\tilde b =  -{ \om \ov 2} \coth {\beta \om \ov 2}  (v + v^*) , \qquad
 c =  {i \ov 2} \coth {\beta \om \ov 2} (w-w^*),  \\
 f = -\ha  \coth {\beta \om \ov 2} (\om h  - i u^*).
 \label{ci3}
\end{gather}
In particular, in~\eqref{ci3},  since the left hand side is regular as $\om \to 0$, we need $u$ to contain at least one power of $\om$, i.e.
\be \label{stac}
u_{00} = u_{02} = u_{04} = \cdots = 0, \
\ee
where various coefficients in the expansion of $u$ are defined as in~\eqref{eiep}.
Further imposing $\sP\sT$ symmetry on the source action, i.e. requiring $G$ and $K$ to be symmetric (Onsager relations), we have additional constraints:
\be \label{ptc1}
f = - f^* , \qquad  \om h = - i u \ .
\ee
The second equation above automatically implies~\eqref{stac}, and one can check that equations~\eqref{allc} are also automatically satisfied. Equation~\eqref{ci3} can now be written as
\be
 f = {i \ov 2}  \coth {\beta \om \ov 2} (u + u^*) = - { \om \ov 2}  \coth {\beta \om \ov 2} (h - h^*) \ .
 \ee

%We note by passing that we have been able to obtain an explicit off-shell action vector action
%from holographic duality, where the coefficients are indeed consistent with~\eqref{ci1}--\eqref{stac}~\cite{cgl}.

\subsubsection{Off-shell currents and constitutive relations}
%\subsubsection{Existence of stationary equilibrium solutions}

From~\eqref{genex}, we find the corresponding off-shell currents
\begin{gather} \label{ofc1}
\hat J^0_a = g^*  B_{a0} + u^* \p_i  B_{ai}, \qquad
\hat J^i_a = h^* \p_i \p_0 B_{a0} - v^*  \p_0 B_{ai}  + w^* \p_j F_{aij}, \\
\label{ofc2}
\hat J^0_r = i a   B_{a0} + i f \p_i B_{ai} + g B_{r0} + h  \p_i \p_0 B_{ri}, \\
\hat J^i_r = i b  B_{ai} - i c \p_i \p_j B_{ja} - i f^* \p_i B_{a0} - u \p_i  B_{r0} + v \p_0 B_{ri} + w  \p_j  F_{rij}  \ .
\label{ofc3}
\end{gather}
The equations of motion for $\vp_r$ and $\vp_a$ correspond to the conservation of $\hat J_a^\mu$ and $\hat J_r^\mu$ respectively.  To leading order in the $a$-field expansion, i.e. setting all the
$a$-fields to zero (and dropping $r$-subscripts), we have
\begin{gather} \label{1hatj}
\hat J^0 =  P_0 \mu  - h \p_i E_i ,
, \quad
\hat J^i =  - P_z \p_i \mu - v E_i
+ w  \p_j  F_{ij}  , \\
 \quad P_0 \equiv g  + h  \p_i^2 , \qquad P_z \equiv u - v , \qquad E_i = -\p_0 A_i + \p_i A_0,
 \label{2hatj}
\end{gather}
where from~\eqref{mub} $\mu = B_0 = A_0 + \p_0 \vp$ is the chemical potential. That at leading order in the $a$-field expansion $\hat J^\mu$ can be expressed solely in terms of $\mu$ to all derivative orders is a consequence of fluid gauge symmetry~\eqref{daug}.
In fact, one can immediately see that this works  at full nonlinear level, as the fluid gauge symmetry means that $B_{ri}$ can only appear either with a time derivative $\p_0 B_{ri} =-  E_i + \p_i \mu$ or through $F_{rij} = \p_i B_{rj} - \p_j B_{ri}$. It is also clear from~\eqref{ofc1}--\eqref{ofc3} that at higher orders in the $a$-field expansion,
 $\hat J_{r,a}^\mu$ cannot be expressed in terms of $\mu_{r,a}$ alone, and the more fundamental $\vp_{a}$ has to be used.

It can also be readily checked from conservation of~\eqref{1hatj} that
equation~\eqref{stac} is equivalent to the existence of a stationary equilibrium for a stationary background field $A_\mu$.

Finally, let us expand~\eqref{1hatj} in derivatives, at the leading order
\be \label{1ccu}
\hat J^0 =  \chi \mu + \cdots  , \qquad
\hat J^i =  \sig (E_i - \p_i \mu) + \cdots,
\ee
from which we can identify
\be \label{1sig}
\chi =  g_{00}, \qquad \sig = - v_{00}, \
\ee
as  charge susceptibility and conductivity respectively. From~\eqref{ci2}, $v_{00}$ is related
to $b_{00}$  as
\be \label{fd0}
b_{00} =- {2 \ov \beta} v_{00}  \ .
\ee
From~\eqref{1pso}, we thus conclude that
\be\label{poss0}
\sig \geq 0 \ .
\ee

\subsubsection{BRST invariance and supersymmetry} \label{sec:vsusy}

Let us now set $A_{a\mu} =0$ in~\eqref{genex} and introduce ghost partners $c_{a,r}$ for $\phi_{a,r}$.
 Here the BRST transformation~\eqref{brst0} becomes
 \be \label{brst1}
\de \vp_r = \ep c_r, \qquad \de c_a = \ep \vp_a  \ .
\ee
From the discussion of~\eqref{se}--\eqref{see} we can readily write down the corresponding
BRST invariant  Lagrangian density $\sL_B$ as
\begin{gather}
\sL_B = g \p_0 \vp_a B_{r0} + h  \p_0 \vp_a \p_i \p_0 B_{ri} + u \p_i^2 \vp_a B_{r0} + v \p_i \vp_a \p_0 B_{ri} - c_a K \p_0 c_r  + {i \ov 2} \vp_a G \vp_a,
\label{vend}
\end{gather}
where (with $P_0, P_z$ introduced in~\eqref{2hatj})
 \be
 K = - P_0 \p_0 + P_z \p_i^2 ,  %- g \p_0 - h \p_0 \p_i^2 + (u -v) \p_i^2 ,
 \qquad  G = - a \p_0^2 - \tilde b \p_i^2 -2  f \p_0 \p_i^2
  \ .
 \ee
Note that the ghost action is uniquely determined and the currents $\hat J^\mu_{a,r}$ are not modified.

Further setting $A_{r\mu} =0$ in~\eqref{vend}, we obtain the Lagrangian density for dynamical fields in the absence of external fields:
\be \label{tne}
\sL_{\rm tot} = \vp_a K \p_0 \vp_r - c_a K \p_0 c_r  +  {i \ov 2} \vp_a G \vp_a  \ .
\ee
One can now verify that  if the local KMS conditions~\eqref{ci1}--\eqref{ci3} are satisfied, in addition to~\eqref{brst1},~\eqref{tne} is also invariant under the following fermonic transformation ($\bar \ep$ is a constant Grassman number):
\be \label{brst3}
\bar \de \vp_r =  c_a \bar \ep , \qquad
\bar \de c_r =  (\vp_a  + \Lam  \vp_r ) \bar \ep, \qquad
 \bar \de \vp_a = - \Lam c_a  \bar \ep,
\ee
where
\be
\Lam = 2 \tanh {i \beta \p_0 \ov 2} \ .
\ee
In other words, for~\eqref{tne} to be invariant under~\eqref{brst3}, $G$ and $K$ should satisfy
\be
(K + K^*) \p_0 = {i \ov 2} \Lam (G + G^*)
\ee
which follow from~\eqref{ci1}--\eqref{ci3}.

It can readily be checked that $\de$ and $\bar \de$ satisfy the following ``supersymmetric'' (SUSY) algebra:
\be \label{1alg}
\de^2 = 0, \qquad \bar \de^2 = 0, \qquad [\de , \bar \de ] = \bar \ep \ep \Lam  \ .
\ee
This is not the usual SUSY algebra, as $\Lam$ involves an infinite number of derivatives.

With all background fields set to zero, the currents have the form
\begin{gather} \label{1ofc1}
\hat J^0_a = (g^*  \p_0 + u^* \p_i^2 ) \vp_a , \qquad
\hat J^i_a = (h^* \p_i \p_0^2 - v^*  \p_0 \p_i  ) \vp_{a}  ,  \\
\label{1ofc2}
\hat J^0_r = (i a   \p_0  + i f \p_i^2 ) \vp_{a}  + P_0 \p_0  \vp_r , \qquad
\hat J^i_r = (i \tilde b  \p_i  - i f^* \p_i \p_0 ) \vp_{a}   -P_z \p_i \p_0 \vp_{r},
%\label{1ofc3}
\end{gather}
which can be readily checked to satisfy the same transformations as $\vp_{r,a}, c_{r,a}$, i.e.
\bega \label{1nal}
\de J_r^\mu = \ep \xi_r^\mu, \quad \bar \de J_r^\mu = \xi_a^\mu \bar \ep  , \quad
\de \xi_a^\mu = \ep J_a^\mu , \quad \bar \de \xi_r^\mu = (J_a^\mu +\Lam J_r^\mu) \bar \ep, \quad  \bar
\de J_a^\mu = - \Lam \xi_a^\mu \bar \ep, \
%\label{12a}
\end{gather}
with $\xi^\mu_{r,a}$ given by
\bega
\xi_r^0= P_0 \p_0 c_r, \qquad \xi_r^i = - P_z \p_i \p_0 c_r, \cr
\xi_a^0 = (P_0 \p_0 - i (a \p_0 + f \p_i^2) \Lam) c_a , \quad
\xi_a^i = - (P_z \p_0 + i (\tilde b - f^* \p_0 ) \Lam ) \p_i  c_a   \ .
\end{gather}
Again, the local KMS conditions~\eqref{ci1}--\eqref{ci3} are crucial.

\subsubsection{The full generating functional}

For the quadratic action~\eqref{genex}, the path integrals~\eqref{newc1} can be evaluated exactly  by solving the equations of motion for $\vp_{r,a}$. The ghost part does not
contribute at quadratic order as it gives an overall constant (which cancels the determinant from the bosonic part). We can directly verify that the FDT~\eqref{fdt4} for the full correlation functions are satisfied given the local KMS conditions~\eqref{ci1}--\eqref{ci3}, although this is a special case of the general argument given in Appendix~\ref{app:fdtar}. We now restore the background fields $A_{r \mu}, A_{a \mu}$.

To evaluate~\eqref{newc1}, it is convenient to work in momentum space. Taking $k_\mu \equiv (k_0, k_z, k_\al) =  (- \om, q, \vec 0)$, one can readily see that $\vp_{r,a}$ only couples to $A_\parallel \equiv (A_0, A_z)$, and  $B_{r\al} = A_{r \al}, \, B_{a\al} = A_{a\al}$. We can then directly read
from~\eqref{genex} the generating functional for
$ A_{r \al}, A_{a \al}$ as
\begin{gather} \label{transG}
W[A_{r \al}, A_{a \al}]  =  i \int {d^d k \ov (2 \pi)^d} \le[{i \ov 2} b  A_{a\al }^2 + v A_{a\al } \p_0 A_{r\al } + {w }  F_{a z \al }  F_{r z \al }  \ri] \ .
\end{gather}
By comparing with~\eqref{yepi}, we find that the corresponding components of the retarded and symmetric correlation functions in momentum space are
\be \label{trns}
G^S_{\al \al} = b (\om, q^2)   , \qquad G^R_{\al \al} = - i \om v (\om, q^2)  + q^2 w (\om, q^2)   \ .
\ee
The FDT relation~\eqref{fdt4} requires  that
\be
b = -\ha \coth {\beta \om \ov 2} \le(\om (v + v^*) + i q^2 (w-w^*) \ri),
\ee
which is satisfied as result of~\eqref{ci2}.

Integrating out $\vp_{r,a}$ leads to a nonlocal
 generating functional for $A_r^{\parallel}, A_a^{\parallel}$,
 \be \label{longG}
 W[A_r^{\parallel}, A_a^{\parallel}] = i \int {d^d k \ov (2 \pi)^d} \, \le[E_a^*  \Pi^L E_r +  {i \ov 2} E_a^* G^L  E_a \ri],
 \ee
where
\be \label{Edef}
E_{a}  (\om, q) \equiv q A_{a0} (\om, q) + \om A_{az} (\om, q) , \quad E_{r} \equiv q A_{r0} + \om A_{rz} , \quad
E_{a,r} (-\om, - q) = -E_{a,r}^*(\om, q),
\ee
and
\begin{gather} \label{logex}
 \Pi^L ={g \hat D- u  \ov - i \om + \hat D q^2},
 \quad G^L = {a q^2 D D^*  - q^2 ( f D +  f^* D^*) + \tilde b  \ov (- i \om + \hat D q^2)(i \om + \hat D^* q^2) } ,
 \quad  \hat D \equiv {P_z \ov P_0} \ .
 \end{gather}
As desired, there is no $rr$-type term in~\eqref{longG}.  That $A_\parallel$ appears only through the combinations in $E_{a,r}$ is a consequence
of the gauge invariance of $W$.
The nonlocality is reflected in the presence of a diffusion pole in $\Pi^L$ and $G^L$. $\hat D$ can be considered as
a diffusion function, which has also been discussed recently in~\cite{Bu:2015ame} as well it holographic calculation.

From~\eqref{yepi}, we can read various components of the symmetric and retarded Green functions
\be \label{logrn0}
G_R^{00} = q^2 \Pi^L, \quad G^{0z}_R = \om q \Pi^L, \quad
G^{zz}_R = \om^2 \Pi^L, \quad G_S^{00} = q^2 G^L, \quad G^{0z}_S = \om q G^L , \quad
G^{zz}_S =\om^2 G^L, \
\ee
and the FDT relation~\eqref{fdt4} requires that
\be \label{logfdt}
G^L =\coth{\beta \om \ov 2} \, {\rm Im} \Pi^L   \ .
\ee
One can readily check from~\eqref{logex} that given~\eqref{ci1}--\eqref{ci3},~\eqref{logfdt} is indeed satisfied.

Keeping the lowest order terms in~\eqref{logex} in derivative expansion of various quantities we find
\be
\Pi^L %= - {v_{00} g_{00}  \ov - i g_{00} \om  - v_{00} q^2  }
= {\sig \ov - i \om + q^2 D}, \qquad G^L  %= {  b_{00} g_{00}^2 \ov g_{00}^2 \om^2 + v_{00}^2 q^4 }
  =
{2 T \sig \ov \om^2 + D^2 q^4},
\ee
where we have used~\eqref{1sig}--\eqref{fd0}, and $D$, which is the leading term of $\hat D$, is given by
\be\label{oimp}
 D = -{ v_{00} \ov g_{00}} = {\sig \ov \chi}   \ .
 \ee
We see that the form of the diffusion constant $D$ is consistent with the Einstein relations.
Note that $\chi$ should be non-negative for a stable {\it equilibrium} state. Given~\eqref{poss0}, we then find that
$D$ is non-negative for a stable equilibrium state, and the pole of retarded Green functions~\eqref{logrn0} indeed lies in the lower half $\om$-plane.

Note that the full generating functional given in~\eqref{transG} and~\eqref{longG} automatically satisfies time-reversal invariance (i.e. Onsager relations)  without imposing conditions~\eqref{ptc1}.
This is an accident due to the simplicity of the system under consideration. This is no longer the case
when including parity breaking terms or the stress tensor.

\subsection{Cubic order} \label{sec:cubic}

\subsubsection{The cubic action}

Let us now consider the bosonic action $I$ of~\eqref{newc1} at cubic order. We can write the corresponding Lagrangian as
\be \label{13r}
 \mathcal L_{3b}=\frac 1{3!}G^{\mu\nu\rho}B_{a\mu}B_{a\nu}B_{a\rho}+\frac i{2}H^{\mu\nu\rho}B_{a\mu}B_{a\nu}B_{r\rho}+\frac 12 K^{\mu\nu\rho}B_{a\mu}B_{r\nu}B_{r\rho},
 \ee
where $G, H, K$ are real {\it local} differential operators acting on various fields. For example, the first term can be  understood in momentum space as
\be
G^{\mu\nu\rho}B_{a\mu}B_{a\nu}B_{a\rho}=\int dk_1dk_2 dk_3\, \delta(k_1+k_2+k_3)G^{\mu\nu\rho}(k_1, k_2,k_3)B_{a\mu}(k_1)B_{a\nu}(k_2)B_{a\rho}(k_3),
\ee
where $G^{\mu\nu\rho}(k_1, k_2,k_3)$ can be expressed as a power series of $k_{1,2,3}$. By definition,
$G^{\mu\nu\rho}(k_1, k_2,k_3)$ is fully symmetric under simultaneous exchanges of subscripts $\mu, \nu ,\rho$ and $k_{1,2,3}$. Similarly,
\be
H^{\mu \nu \rho} (k_1, k_2, k_3) = H^{\nu \mu \rho} (k_2, k_1, k_3), \qquad
K^{\mu \nu \rho} (k_1, k_2, k_3) = K^{\mu \rho \nu} (k_1, k_3, k_2) \ .
\ee
$G, H, K$ should be such that $\sL_3$ is rotationally invariant and satisfies~\eqref{daug}.
It is possible to write~\eqref{13r} more explicitly as in~\eqref{genex} to make these properties manifest, but the expression becomes quite long and we will not do it here.

Imposing local KMS conditions amounts to requiring that $G, H, K$  satisfy~\eqref{H1fdt}--\eqref{H3fdt}.
$H$ in~\eqref{13r} corresponds to $H_3$,  $K$ corresponds to $K_1$, and the other are obtained by permutations. For example,
\be
(H_1)^{\mu \nu \rho} (k_1, k_2, k_3) \equiv H^{\rho \nu \mu} (k_3, k_2 , k_1) , \qquad (K_2)^{\mu \nu \rho} (k_1, k_2 , k_3) \equiv K^{\nu \mu \rho}  (k_2, k_1 , k_3)
\ee
and similarly with the others. %One could again verify that existence of a stationary equilibrium for a stationary external field is implied by~\eqref{allc}.

As an illustration of implications of the local KMS conditions on~\eqref{13r}, we consider a truncation of it in Appendix~\ref{app:example}. In particular, we see that the generalized Onsager relations~\eqref{1con1} lead to nontrivial relations on the transport coefficients at second order in derivative expansions at nonlinear level.

Setting the external fields to zero, we find the action for dynamical modes:
\be \label{1lb}
i \sL _{3b}= {\sG \ov 6} \vp_a^3 + {i \ov 2} \sH \vp_a^2  \vp_r + {\sK \ov 2} \vp_a \vp_r^2,
\ee
where (note the $i$ factor on left hand side of~\eqref{1lb})
\be
\sG(k_1, k_2,k_3)= G^{\mu \nu \rho} k_\mu k_\nu k_\rho,
\label{1sg}
\ee
and similarly with $\sH$ and $\sK$. It is clear that $\sG$ inherits the symmetry properties of $G$ and is
fully symmetric under exchanges of $k_{1,2,3}$. Similarly  $\sH$ is symmetric under exchange of $k_1, k_2$ and
$\sK$ symmetric under exchange of $k_2, k_3$.
Furthermore, it can be readily checked that $\sG, \sH, \sK$ satisfy~\eqref{H1fdt}--\eqref{H3fdt}
as a result of $G,H,K$ satisfying these relations. Again $\sH$ and $\sK$ in~\eqref{1lb} should be understood as $\sH_3$ and $\sK_1$ respectively, and
\bega
\sH (k_3, k_2 , k_1) \equiv \sH_1 (k_1, k_2, k_3), \qquad \sH (k_1, k_3 , k_2) \equiv \sH_2  (k_1, k_2, k_3) , \\
\sK (k_3, k_2 , k_1) \equiv \sK_3  (k_1, k_2, k_3), \quad \sK (k_2, k_1 , k_3) \equiv \sK_2 (k_1, k_2, k_3) \ .
\end{gather}
Also note that due to~\eqref{daug}
\be \label{gey}
\sH_\al \propto \om_\al, \qquad \sK_\al \propto {\om_1 \om_2 \om_3  \ov \om_\al} , \qquad \al=1,2,3 \ .
\ee

\subsubsection{BRST invariance and supersymmetry}

Setting $A_{a \mu}$ to zero, and applying~\eqref{se}--\eqref{see} to~\eqref{13r} we can obtain an BRST invariant action by adding to~\eqref{13r}
the following fermionic action
\bega \label{14r}
\sL_{3f} = -{i \ov 4} H_{\mu\nu\rho} (\p_\mu c_a \p_\nu \vp_a + \p_\mu \vp_a \p_\nu c_a)
\p_\rho c_r -  f  c_a  \vp_a  c_r -  K_{\mu\nu\rho} \p_\mu c_a B_{r \nu} \p_\rho c_r \ .
\end{gather}
 As noted in~\eqref{ambi}, the BRST invariant action is not unique (beginning at cubic order). In~\eqref{1l}, this non-uniqueness is parameterized by the term with coefficient $f (k_1, k_2, k_3)$
which has the symmetry  properties
\be
f(k_1, k_2, k_3) = - f (k_2, k_1, k_3) \ .
\ee
The full BRST invariant action in the absence of sources of can then be written as
\be \label{1l}
i \sL_B = {\sG \ov 6} \vp_a^3 + {i \ov 2} \sH \vp_a^2 \vp_r + {\sK \ov 2} \vp_a \vp_r^2
- {i \ov 2} \sH c_a \vp_a c_r - i f c_a \vp_a c_r - \sK c_a \vp_r c_r  \ .
\ee
Following our earlier notations, below we will denote $f$ as $f_3$, and similarly introduce
\be
 f_1 (k_1, k_2 , k_3) \equiv f (k_3, k_2 , k_1) , \qquad  f_2 (k_1, k_2 , k_3) \equiv f (k_3, k_1 , k_2)\ .
\ee

As already mentioned in Sec.~\ref{sec:susy}, the fermionic transformation~\eqref{brst3} cannot remain
a symmetry at nonlinear orders due to higher derivative nature of $\Lam$. For example, were~\eqref{brst3}
a symmetry of our cubic Lagrangian, then from~\eqref{1alg}, $\Lam$ would also be a symmetry. However, this is not the case, as
\be
\Lam_1 + \Lam_2 + \Lam_3 \neq 0 \quad {\rm for} \quad
\om_1 + \om_2 + \om_3 =0 ,
\ee
where $\Lam_i \equiv 2 \tanh {\beta_0 \om_i \ov 2}, \; i=1,2,3$. There is a basic contradiction in~\eqref{1alg}:
while the left hand side is a derivation by definition, the right hand side is not.

We will now show that in the $\hbar \to 0$ limit (i.e. the classical statistical limit discussed in Sec.~\ref{sec:CSL} and Sec.~\ref{sec:hbar}), in which
\be \label{susa}
\Lam = i \beta_0 \p_t, \qquad [\de, \bar \de] =i  \bar \ep \ep \beta_0 \p_t ,
\ee
 the local KMS conditions satisfied by $\sG, \sH, \sK$
ensure that~\eqref{1l} is supersymmetric.  In particular, supersymmetry fixes uniquely the undetermined local operator $f$ in~\eqref{1l}
in terms of other quantities.

As discussed in Sec.~\ref{sec:CSL} and Sec.~\ref{sec:hbar}, in the $\hbar \to 0$ limit, various quantities in~\eqref{1l} should scale as
\be
\sG \to \sG, \quad \sH \to \hbar \sH, \quad \sK \to \hbar^2 \sK , \quad f \to \hbar f , \quad
(c_a , \vp_a) \to \hbar (c_a , \vp_a) , \quad c_r , \vp_r \to c_r, \vp_r,
\ee
and the local KMS conditions in this limit are given by~\eqref{1hh}--\eqref{2hh}, which we copy here for convenience:
\bega \label{1gg}
\sH_3 = - {i \ov \beta \om_1 \om_2} \le( \om_1 \sK_1 + \om_2 \sK_2 + \om_3 \sK_3^* \ri), \\
%{\rm Im} \sG = 0, \qquad {\rm Re}
\sG ={2 \ov \beta^2 \om_1 \om_2 \om_3} \le(\om_1 {\rm Re} \sK_1+
\om_2 {\rm Re} \sK_2 + \om_3 {\rm Re} \sK_3 \ri) \ .
\label{2gg}
\end{gather}
%Note that these expressions are regular in the $\om_i \to 0$ limits due to~\eqref{gey}.

Under~\eqref{brst3}, we find
\be
i \bar \de \sL_3 =  C_1 \vp_a^2 c_a \bar \ep + C_2 \vp_a c_a \vp_r \bar \ep + C_3 c_a \vp_r^2 \bar \ep
+ C_4 c_a^2 c_r \bar \ep,
\ee
with
\bega \label{1c1}
C_1 = -{\sG \ov 2} \Lam_3 + {i \ov 2} \sH_3 - {i\ov 4} (\sH_1 + \sH_2) - {i \ov 2} (f_1 + f_2), \\
C_2 = - i \sH_3 \Lam_2 + \sK_1 - \sK_2 - {i \ov 2} \sH_3 \Lam_3 + i f_3 \Lam_3,  \\
C_3 = - \ha \sK_1 (\Lam_1 + \Lam_2 + \Lam_3), \\
C_4 = {i \ov 4} \sH_3 (\Lam_1 - \Lam_2) - {i \ov 2} f_3 (\Lam_1 + \Lam_2)
+ \ha (\sK_1 - \sK_2)   \ .
\end{gather}
In the  $\hbar \to 0$ limit, $C_3$ and the symmetric part of $C_2$ are automatically zero, while the antisymmetric
part of $C_2$ is equivalent to $C_4$. Setting $C_4=0$, we can solve for $f$:
\be \label{1f3}
f_3 = {1 \ov \beta \om_3} \le(i (\sK_1 - \sK_2) - \ha \beta \sH_3 (\om_1 - \om_2) \ri) \ .
\ee
Note that $f_3$ is regular as $\om_3 \to 0$ due to~\eqref{gey}. Thus, $f_3$ is a well-defined {\it local} differential operator.
Plugging~\eqref{1f3} into~\eqref{1c1} we find that
\be \label{2f3}
C_1 = {1 \ov \beta \om_1 \om_2}
\le[-{\sG \ov 2} \beta^2 \om_1 \om_2 \om_3 + {i \beta \ov 2} \le(\om_1 \om_2 \sH_3
+ \om_1 \om_3 \sH_2 + \om_2 \om_3 \sH_1 \ri) - \ha \le(\om_1 \sK_1 + \om_2 \sK_2 + \om_3 \sK_3 \ri) \ri].
\ee
Now one can readily check from~\eqref{1gg}--\eqref{2gg} that $C_1 =0$.

\subsubsection{Multiplet of currents}

Now let us look at the $\hat J_{r,a}^\mu$ in the absence of background fields.
From~\eqref{13r} and~\eqref{14r}, we find
\be \label{2j}
J_a^\mu = {i \ov 2} (H_1)^{\mu\nu\rho} \p_\nu \vp_a \p_\rho \vp_a + (K_2)^{\mu\nu\rho}
(\p_\nu \vp_a \p_\rho \vp_r - \p_\nu c_a \p_\rho c_r ), \
\ee
while expanding~\eqref{13r} to first order in $A_{a \mu}$, we find
\be \label{1j}
J_r^\mu = \ha G^{\mu\nu\rho} \p_{\nu} \vp_a \p_{\rho} \vp_a + i H^{\mu\nu\rho} \p_\nu \vp_a \p_\rho \vp_r
+ \ha K^{\mu\nu\rho} \p_\nu \vp_r  \p_\rho \vp_r    \ .
\ee
From the discussion around~\eqref{1newib},  there is freedom to add ghost terms to~\eqref{1j} of the form $ R^{\mu \nu \rho} \p_\nu c_a \p_\rho c_r$, with $R^{\mu \nu \rho} $ a local differential operator. We thus now have
\be \label{3j}
J_r^\mu = \ha G^{\mu\nu\rho} \p_{\nu} \vp_a \p_{\rho} \vp_a + i H^{\mu\nu\rho} \p_\nu \vp_a \p_\rho \vp_r
+ \ha K^{\mu\nu\rho} \p_\nu \vp_r  \p_\rho \vp_r + R^{\mu \nu \rho} \p_\nu c_a \p_\rho c_r \ .
\ee
We now show that requiring that $J_a^\mu$ and $J_r^\mu $ satisfy the $\hbar \to 0$ limit of the transformations~\eqref{1nal}, i.e.
\bega \label{1a}
\de J_r^\mu = \ep \xi_r^\mu, \quad \bar \de J_r^\mu = \xi_a^\mu \bar \ep  , \quad %\\
\de \xi_a^\mu = \ep J_a^\mu , \quad \bar \de \xi_r^\mu = (J_a^\mu + i \beta \p_0 J_r^\mu) \bar \ep, \quad  \bar \de J_a^\mu = - i \beta \p_0 \xi_a^\mu \bar \ep \
%\label{2a}
\end{gather}
uniquely fixes $R$. Note that the first two equations of~\eqref{1a} should be viewed as the definition for $\xi_{r,a}^\mu$, while the last two equations follow from~\eqref{susa} once the thrid equation is satisfied. So we only need to check the third equation of~\eqref{1a}.

From~\eqref{3j}, we have
\bega
\xi_r^\mu = i H^{\mu\nu\rho} \p_\nu \vp_a \p_\rho c_r
+ K^{\mu\nu\rho} \p_\nu \vp_r  \p_\rho c_r + R^{\mu \nu \rho} \p_\nu \vp_a \p_\rho c_r ,  \\
\xi_a^\mu  = (- \beta \om_2 G + i H_2  + R)^{\mu\nu\rho} \p_{\nu}  c_a \p_{\rho} \vp_a
+ (- i \om_2 \beta H_3 + K_1 + \beta \om_3 R )^{\mu\nu\rho}  \p_\nu c_a \p_\rho \vp_r,  \
\end{gather}
where in the second equation for notational simplicities we have used a mixed coordinate and momentum
representation. Now imposing the third equation of~\eqref{1a}, we find
\bega \label{11hh}
{i \ov 2} H_1 = \ha  \beta \om_1 G + {i \ov 2} (H_2 + H_3) + R_s  , \\
K_2 =  - i \om_2 \beta H_3 + K_1 + \beta \om_3 R, \
\label{12hh}
\end{gather}
where
\be
R_s^{\mu \nu \rho}  = \ha (R^{\mu \nu \rho} + R^{\mu \rho \nu }  ) \ .
\ee
One can now verify that equation~\eqref{11hh} is equivalent to the symmetric part (in terms of the last two indices) of~\eqref{12hh}, if~\eqref{2f3} vanishes. Thus we have a consistent set of equations.  $R$ can now be solved as
\be \label{1rr}
R =  \frac 1{\beta\omega_3}\left(K_2-K_1+i\omega_2\beta H_3\right),
\ee
%{1 \ov \beta \om_1}
%(K_3^* - K_2),  \
%where we have again used the local KMS conditions.
Note that $R$ is local as due to~\eqref{daug}, $H_3, K_1, K_2$ should all be proportional to $\om_3$.

To summarize, both the invariance of the action~\eqref{14r} under the supersymmetric transformation~\eqref{brst3} and
the existence of supermultiplet structure~\eqref{1a} can be attributed to the vanishing of equation~\eqref{2f3}. Now one can readily check that the combination of~\eqref{1gg} and~\eqref{2gg} which gives~\eqref{2f3} precisely coincides with~\eqref{ggg1} for $n=3$.  Thus we conclude that in the current context, it is the local part of~\eqref{ggg1} (i.e. this KMS condition applied to $I_s$) that is responsible for the emergence of supersymmetry. As we already discussed in the paragraph after~\eqref{12a}, supersymmetry in turn ensures that~\eqref{ggg1} is satisfied for full correlation functions at all loop orders.

\subsection{A minimal model for stochastic diffusion}  \label{sec:kpz}

Let us now combine the quadratic and cubic actions and truncate them to the lowest nontrivial order in derivative expansions. From~\eqref{1hh}--\eqref{2hh}, the local KMS conditions imply that
coefficients of $O(a)$ terms with $n$ derivatives are related to those of $O(a^2)$ terms with $n-1$ derivatives, and those of $O(a^3)$ terms with $n-2$ derivatives. Thus
at lowest order in the derivative expansion, we will keep the first derivative in $O(a)$ terms, zero derivatives in $O(a^2)$ terms, and drop $O(a^3)$ terms.

\subsubsection{Linear stochastic diffusion}

In~\eqref{genex}, keeping zero derivative terms in $O(a^2)$ terms and first derivative terms in $O(a)$ terms, we find
\be \label{1yue}
\sL_2 = %{i \ov 2} a_0 B_{a0}^2 +
i \sig T  B_{ai}^2 + \chi  B_{a0} B_{r0}
 - \sig B_{ai} \p_0 B_{ri}  + c_a (\chi \p_0 - \sig \p_i^2 ) \p_0 c_r,
\ee
where we have used~\eqref{1sig}--\eqref{fd0}. In~\eqref{1yue}, we have dropped a zeroth derivative $O(a^2)$ term  $a_{00} B_{a0}^2$ and a first derivative $O(a)$ term $g_{10}  \p_0 B_{a0} B_{r0}$.  The $g_{10}$
term is subleading compared to the term with coefficient $\chi$. The $a_{00}$ term is dropped since
it is related to $g_{10}$ by the local KMS conditions:
\be
a_{00} = {2 \ov \beta} g_{10}  \ .
\ee
In counting the relevance of terms we always drop terms which are related by local KMS conditions together.
At this order, the off-shell currents are
\bega \label{11c}
\hat J_r^0 = \chi \p_0 \vp_r , \qquad \hat  J_r^i = 2 i \sig T \p_i \vp_a  - \sig \p_0 \p_i \vp_r, \\
\hat  J_a^0 = \chi \p_0 \vp_a , \qquad \hat  J_a^i = \sig \p_i \p_0 \vp_a  \ .
\end{gather}
Turning off the external fields, we get~\eqref{tne}, with
\be \label{1po}
K =  \chi(-  \p_0 +D  \p_i^2),  %+\cdots ,
\qquad G = % -a_{00} \p_0^2
-  2 \sig T \p_i^2 \ .%+ \cdots
\ee

Now following the procedure outlined in~\eqref{ltr1}--\eqref{stoeom1} we obtain
the stochastic diffusion equation
\be
\le( - \p_0 + D \p_i^2 \ri)n = \xi ,  \
  \quad n \equiv \p_0 \vp_r , \quad
%\eta = {1 \ov \chi} \xi
\ee
where the noise force $\xi$ is the Legendre conjugate of $\vp_a$ and has a local Gaussian distribution given by
\be
\vev{\xi} = 0, \qquad \vev{\xi (x) \xi (0)} = - 2 T \sig \p_i^2 % (a_{00} \p_0^2 +2 T \sig \p_i^2 % + \cdots)
\de^{(d)} (x)  \ .
\ee
%We thus see that $\vp_a$ is the conjugate variable for the stochastic force.

\subsubsection{Action for a variation of stochastic Kardar-Parisi-Zhang equation}

At  cubic level, in~\eqref{13r} we keep first derivative terms in $K$, zero derivative terms in $H$, and
drop all $G$ terms. Then, after imposing local KMS conditions (see Appendix~\ref{app:example}), we find
\be \label{1yh}
\sL_{3b} =  %{i \ba \ov 2} B_{a0}^2 B_{r0}    +
{i \sig_1 T } B_{ai}^2   B_{r0}
+{\chi_1 \ov 2}  B_{a0} B_{r0}^2  - %{\ta_1 \ov 2}  \p_0 B_{a0} B_{r0}^2  +
 \sig_1  B_{ai}  B_{r0} \p_0  B_{ri}
\ee
where we have dropped $\p_0 B_{a0} B_{r0}^2$ and $B_{a0}^2 B_{r0}$. The former is subleading compared to
$B_{a0} B_{r0}^2$ while the latter is related to the former by local KMS conditions. Now setting the background fields to zero, and combining~\eqref{1yh} with the cubic fermionic action~\eqref{14r} and the quadratic action~\eqref{1yue}, we obtain the full action
\bega
\sL =  i T \sig  (\p_i \vp_{a})^2 + \chi  \p_0 \vp_a  \p_0 \vp_r - \sig \p_i \vp_a  \p_0 \p_i \vp_r
 +  c_a  (\chi \p_0 - \sig \p_i^2 ) \p_0 c_r
\cr
+  i T  \sig_1  \p_i \vp_{a} \p_i (\vp_a + i \beta \p_0 \vp_r)  \p_0 \vp_r
 - i T \sig_1 ( \p_i c_a \p_i \vp_a  \p_0 c_r + (\p_0 c_a \p_i \vp_a - \p_i c_a \p_0 \vp_a) \p_i c_r )
 \cr
  - \sig_1 \p_i^2 c_a  \p_0 \vp_r \p_0 c_r
    + {\chi_1 \ov 2} \p_0 \vp_a  \p_0 \vp_r \p_0 \vp_r
 - \chi_1 \p_0 c_a \p_0 \vp_r \p_0 c_r ,
   \label{1fi}
\end{gather}
where we have used~\eqref{1f3}, which gives
\be \label{1f}
f =  - {T \sig_1} (\om_1 k_2 - \om_2 k_1) \cdot k_3  \ .
\ee
The off-shell currents are
 \bega
 \hat J_a^0=  \chi \p_0 \vp_a + {i T \sig_1} (\p_i \vp_a)^2  +\chi_1 (\p_0 \vp_a \p_0 \vp_r   - \p_0 c_a \p_0 c_r)
 - \sig_1 (\p_i \vp_a  \p_0 \p_i \vp_r  - \p_i c_a \p_0 \p_i c_r) \cr
\hat J_a^i =  \sig \p_i \p_0 \vp_a + \sig_1 \p_0 (\p_i \vp_a  \p_0 \vp_r -  \p_i c_a  \p_0 c_r ) ,
\label{1fc}
\end{gather}
and
\bega
\hat J_r^0 = \chi \p_0 \vp_r + {\chi_1 \ov 2} (\p_0 \vp_r)^2 + i T \sig_1 \p_i c_a \p_i c_r
 ,\cr
\hat J_r^i =
 2 i \sig T \p_i \vp_a  - \sig \p_0 \p_i \vp_r  + 2 i T \sig_1  \p_i \vp_a \p_0 \vp_r -\sig_1 \p_0 \vp_r \p_0 \p_i \vp_r
 - i T \sig_1 (\p_0 c_a \p_i c_r + \p_i c_a \p_0c_r),
 \label{2fc}
 \end{gather}
 where we have used~\eqref{1rr}. The Lagrangian~\eqref{1fi} is invariant under~\eqref{brst1} and~\eqref{brst3}, with $\Lam$ given by~\eqref{susa}. The currents satisfy~\eqref{1a}.

For~\eqref{1fi},  as in the quadratic case, one can again consider the Legendre transform
$\sL_{aa} = - \vp_a \xi + \tilde \sL_{aa} [\xi, \vp_r]$. The equation of motion then obtained from varying $\vp_a$ has the form
\be \label{1kpz}
(\p_0 - D \p_i^2) n + \ha \le(\lam_1 \p_0 - {\lam} \p_i^2 \ri) n^2 = \xi,
\ee
with $\lam_1 = {\chi_1 \ov \chi^2}$ and $\lam = {\sig_1 \ov \chi^2}$. Equation~\eqref{1kpz} resembles the Kardar-Parisi-Zhang (KPZ) equation~\cite{kpz}. {Note that with nonlinear terms such as $(\p_i \vp_a)^2 \p_0 \vp_r$,
$\tilde \sL_{aa}$ now contains interactions between $\vp_r$ and $\xi$. In fact, $\tilde \sL_{aa}$ is neither local nor polynomial, thus it no longer makes sense to
replace $\vp_a$ by $\xi$ via a Legendre transform.
It could still happen that nonlinear terms such as $(\p_i \vp_a)^2 \p_0 \vp_r$ turn out to be irrelevant when going further into the IR, in which case the very low energy physics would still be governed by~\eqref{1kpz}, with $\xi$
a local Gaussian noise. We will leave understanding the renormalization group flow of~\eqref{1fi} for future work. }

Finally we should emphasize that in our framework, the forms of the action~\eqref{1fi} and the equation~\eqref{1kpz} are completely determined by symmetries, with no other freedom.

\section{Effective field theory for general charged fluids} \label{sec:charge}

In this section, we proceed to write down the bosonic part of the hydrodynamical action for a charged
fluid.
% The full theory with ghosts will be left for future.

%It becomes much more tedious
%Including ghosts becomes much more complicated  than the vector story we discussed in last section and will be left
%for future investigation.

\subsection{Preparations}

\subsubsection{Organization of variables} \label{sec:var}
%\subsection{A convenient decomposition}

We first introduce a convenient set of variables which will make imposing~\eqref{sdiff}--\eqref{tdiff} and~\eqref{daug}
 convenient. Below, if not written explicitly, it should always be understood
that the CTP indices $s=1,2$ are suppressed.  In particular, any equation without explicit CTP
indices should be understood as a relation between variables pertaining to one segment of the CTP contour,
and altogether there are two copies of the equations.

Given the identification of the velocity field~\eqref{umu} and the form of the symmetries~\eqref{sdiff}--\eqref{tdiff}, it is convenient to
decompose the matrix $\p_a X^\mu$ in~\eqref{hdef1} as
\begin{gather}  \label{tbe0}
{\p X^\mu \ov \p \sig^0} \equiv b u^\mu, \quad u^\mu u_{ \mu} = -1, \quad
u_{ \mu} = g_{\mu \nu}  u^\nu ,\quad
{\p X^\mu \ov \p \sig^i} \equiv - v_i  b u^\mu + \lam_i{^\mu}, \quad u_\mu \lam_i{^\mu} = 0, \
\end{gather}
and conversely,
\begin{gather}
b = \sqrt{- \p_0 X^\mu  g_{\mu \nu} \p_0 X^\nu }, \quad
u^\mu = {1 \ov b} \p_0 X^\mu,  \quad
v_i = {1 \ov b^2} g_{\mu \nu} \p_0 X^\mu \p_i X^\nu, \quad
\lam_i{^\mu} = \p_i X^\mu + \p_0 X^\mu v_i  \ .
\end{gather}
$h_{ab}$ in~\eqref{hdef1} can then be written as
\be \label{1deco}
h_{ab} d\sig^a d\sig^b =  - b^2  \le( d\sig^0 -  v_i d\sig^i \ri)^2 + a_{ij} d\sig^i d\sig^j,
\ee
where
\be
a_{ij} \equiv \lam_i{^\mu} \lam_{j}{^\nu} g_{\mu \nu},
\ee
and we will denote its inverse as $a^{ij}$. The inverse transformation can be written as
\be \label{tbe1}
{\p \sig^i \ov \p X^\mu} =  \lam^i{_\mu} \equiv  g_{\mu \nu} a^{ij} \lam_j{^\nu}
, \qquad {\p \sig^0 \ov \p X^\mu} = -{1 \ov b} u_\mu +  v_i \lam^i_\mu  \ .
\ee
It can be readily checked that
\be
\lam^i{_\mu} \lam_j{^\mu} = \de_j^i  , \qquad
\lam^{i\mu} \lam_i{^\nu}  = \De^{\mu \nu} \equiv g^{\mu \nu} + u^\mu u^\nu \ .
\ee
The various quantities $b, u^\mu, v_i, \lam_i{^\mu}$ are not arbitrary. Following their definitions from ${\p X^\mu \ov \p \sig^a}$ and ${\p \sig^a \ov \p X^\mu}$, they satisfy various integrability conditions, which are given in Appendix~\ref{app:inter}.

%For notational convenience we will often suppress the CTP indices $s=1,2$  below.
%We now express various components of $h_{ab}$ and $B_a$ in terms of hydrodynamic modes $\tau, X^\mu, \vp$.
%Given the form of the symmetries~\eqref{sdiff}--\eqref{tdiff},
Similar to~\eqref{1deco} we can
decompose  $B_a$ as
\begin{gather}
 %\label{nep0}
%h_{ab} d\sig^a d\sig^b =  - E^2  \le( d\sig^0 -  v_i d\sig^i \ri)^2 + %e^{2  \eta}
%\al_{ij} d\sig^i d\sig^j, \\ %= - \ell_a \ell_b + \De_{ab}
B_a d\sig^a = \mu b (d\sig^0 - v_i d\sig^i) + \fb_i d\sig^i  \ .
\label{bdec}
\end{gather}
with
\be \label{mhat}
%E = e^{- \tau} b ,  \quad \al_{ij} = %e^{-2 \tau} \lam_i{^\mu} \lam_{j \mu}\equiv
%e^{-2 \tau} a_{ij}  , \quad
 \mu =  u^\mu A_{\mu} + D_0 \vp,  %u^\mu B_\mu ,
\qquad \fb_i =   \lam_{i}{^\mu}  A_{\mu} + D_{i} \vp,   %\lam_i{^\mu} B_\mu , \qquad
%\hat A_\mu = A_\mu - \p_\mu \vp
\ee
where the local chemical potential $\mu$ was introduced before in~\eqref{mub} and we have also introduced
``covariant'' derivatives:
\be \label{cobe}
D_0 \equiv   {1 \ov b} \p_0 , \qquad  D_{i} \equiv \p_i + v_{i} \p_0 \ .
\ee
Also note that
\be \label{onlam}
\Lam = \le|\det {\p X \ov \p \sig} \ri| = {\sqrt{a} b \ov \sqrt{-g}} %= {\sqrt{\al} \ov \sqrt{-g}} E  e^{d \tau}
 \ .
\ee

Under spatial diffeomorphisms~\eqref{sdiff}, $b, \mu$  transform as scalars, $\fb_i, v_i$ as vectors and  $a_{ij}$ as a  symmetric tensor.
Under time diffeomorphisms~\eqref{tdiff}, $a_{ij}, \mu, \fb_i$ transform as scalars while
\begin{gather}
\label{nei}
b' (\sig^0, \sig^i)  =\p_0 f b (f(\sig), \sig^i), \qquad
v_i' (\sig^0, \sig^i) = {1 \ov \p_0 f} \le(v_i (f(\sig^0, \sig^i), \sig^i) -  \p_i f \ri)
\ .
% \\
%B_0' (\sig^0, \sig^i) = \p_0 f B_0 (f(\sig), \sig^i), \qquad B_i' (\sig^0, \sig^i) = B_i (f(\sig^0, \sig^i), \sig^i) +  \p_i f B_0 (f(\sig), \sig^i) \ .
\end{gather}
$\vp, \tau$ transform as scalars under both diffeomorphisms.

Now for $r-a$ variables, we introduce $\mu_{r,a}, v_{ai}, v_{ri}, \fb_{ai}, \fb_{ri}$
as usual (see~\eqref{ra1}), while for $b, a_{ij}$ it is convenient to introduce
instead the following definitions
\begin{gather}  \label{1nr}
%\tau_a = \tau_1 - \tau_2, \quad
 %\tau_r = \ha (\tau_1 + \tau_2), \quad v_{ai} = v_{1i} - v_{2i} , \quad  v_{ri} = \ha  (v_{1i} + v_{2i}) , \\
 E_r = %\ha (E_1 + E_2) =
 \ha \le( b_1 +b_2 \ri) ,\qquad E_a %= \log (E_2^{-1} E_1)
 =   \log \le(b_2^{-1} b_1 \ri) , \\
 a_{r ij} = \ha  (a_{1ij} + a_{2ij}), \quad %=\ha  e^{-2\tau}(a_{1ij} + a_{2ij})
 %\label{0nr}
% = e^{- 2 \tau} a_{rij} ,  %\ha \le(e^{-2\tau_1} a_{1ij} + e^{- 2\tau_2} a_{2 ij} \ri),
%\\
\chi_a %= \ha \log \det (\al_2^{-1} \al_1)
= \ha \log \det (a_2^{-1} a_1), \quad  % -(d-1) \tau_a + \ha \tr \log \le(a_2^{-1} a_1 \ri), \quad
\Xi =  \log \le(\hat a_2^{-1} \hat a_{1}\ri),
% \hmu_{a} = e^{\tau_1} u^\mu_1 B_{1\mu} - e^{\tau_2} u^\mu_2 B_{2\mu},   \quad
 % \hmu_{r} = \ha \le(e^{\tau_1} u^\mu_1 B_{1\mu} + e^{\tau_2} u^\mu_2 B_{2\mu} \ri), \\
% \fb_{ai} =  \lam_{1i}{^\mu} B_{1\mu} -  \lam_{2i}{^\mu} B_{2\mu}, \quad
 % \fb_{ri} =  \ha \le(\lam_{1i}{^\mu} B_{1\mu} +  \lam_{2i}{^\mu} B_{2\mu} \ri)\ .
\label{1na}
\end{gather}
where $\hat a_{1,2}$ denotes the unit determinant part of $a_{1,2}$ and thus $\Xi$ is traceless. Under~\eqref{sdiff}, $a_r$ transforms as tensor, $E_{r,a}, \chi_a,  \mu_{a}, \mu_{r},\tau$ as scalars, $ v_{ai}, v_{ri}, \fb_{ai}, \fb_{ri}$ as vectors, while $\Xi$ transform as
\be
 \Xi' (\sig^0, \sig'^i) = Q^{-1} \Xi (\sig^0, \sig^i (\sig')) Q, \qquad Q^i{_j} = {\p \sig^i \ov \p \sig'^j}  \ .
\ee
Under~\eqref{tdiff}, $a_r, \chi_a ,\Xi, E_a, \tau, \mu_{a}, \mu_{r}, \fb_{ai}, \fb_{ri}$ transform as a scalar while
\begin{gather}
E_r' (\sig^0, \sig^i) = \p_0 f E_r (f(\sig^0, \sig^i), \sig^i) , \quad %B_{a0,r0}' (\sig^0, \sig^i) =  \p_0 f B_{a0,r0} (f(\sig^0, \sig^i), \sig^i) \\
v_{ai}' (\sig^0, \sig^i) = {1 \ov \p_0 f} v_{ai} (f(\sig^0, \sig^i), \sig^i),
\\
v_{ri}' (\sig^0, \sig^i) = {1 \ov \p_0 f} \le(v_{ri} (f(\sig^0, \sig^i), \sig^i) -  \p_i f \ri),
%, \\
%B_{ai,ri} = B_{ai,ri} (f(\sig^0, \sig^i), \sig^i) +  \p_i f B_{a0,r0} (f(\sig^0, \sig^i), \sig^i)\ .
\end{gather}
which motivates us to further introduce
\begin{gather}
%b_{a0} = {1 \ov E_r} B_{a0} , \quad b_{r0} = {1 \ov E_r} B_{r0} , \quad
V_{ai} = E_r v_{ai}  , \quad  V_{ri} = E_r v_{ri}   \ .
%\\b_{ai} = B_{ai} + v_{ri} B_{a0},  \qquad b_{ri} = B_{ri} + v_{ri} B_{r0}
\end{gather}
Now $V_{ai}$ transforms as a scalar while $V_{ri}$ as
\be
 V_{ri}' (\sig^0, \sig^i) = V_{ri} (f(\sig^0, \sig^i), \sig^i)  - \p_i f E_r   \ .
 \ee

Finally under~\eqref{daug}, $\fb_{ai}$ is invariant while $\fb_{ri}$ transforms as
\be \label{guhe}
\fb_{ri} \to \fb_{ri}' = \fb_{ri} - \p_i \lam (\sig^i) \ .
\ee

\subsubsection{Covariant derivatives} \label{sec:cord}

Consider $\phi$ and $\phi_i$, which are a scalar and vector respectively under spatial diffeomorphisms~\eqref{sdiff}, and
are scalars under  time diffeomorphisms~\eqref{tdiff}.   We would like to construct a covariant spatial derivative
$D_i=\partial_i+\cdots$ such that:
\ben

\item   $D_i\phi$ and $D_i\phi_j$ are tensors with respect to~\eqref{sdiff}.

\item It is compatible with $a_{rij}$, i.e.
\be
D_i a_{rjk}=0  \ .
\ee

\item $D_i\phi$ and $D_i\phi_j$ remain scalars under~\eqref{tdiff}.
\een
The action of $D_i$ on higher rank and upper index tensors can be obtained using the Leibniz rule.
Here and below, unless otherwise noted, all the indices are raised and lowered by $a_r$.

It can be readily verified the following definitions satisfy the above conditions
\bega \label{des}
D_i \phi =  \p_i \phi+ v_{ri} \p_0 \phi \equiv d_i \phi, \\
D_i\phi_j = d_i\phi_j -\tilde\Gamma^k_{ij}\phi_k,
\label{des1}
\end{gather}
where $d_i \equiv \p_i + v_{ri} \p_0$ and %$ \tilde\Gamma^k_{ij}$ is defined as
\be \label{omm}
\tilde\Gamma^i_{jk} \equiv \ha a^{il}_r \left(d_j a_{rkl}+ d_k a_{rjl}
-d_l a_{rjk}\right)=\Gamma^i_{jk}+\frac{1}{2}a^{il}_r \left(v_{rj}\partial_0a_{rkl} + v_{rk}\partial_0a_{rjl}
-v_{rl}\partial_0a_{rjk}\right)
\ee
with $\Gamma^i_{jk}$ the Christoffel symbol corresponding to $a_r$.

For the time derivative, one can check for a scalar $\phi$ under~\eqref{tdiff},
\be \label{urm}
D_0 \phi \equiv {1 \ov E_r} \p_0 \phi
\ee
is a scalar.

One should be careful to note that the $D_0, D_i$ introduced here are different from those in~\eqref{cobe}.
$E, v_i$ in~\eqref{cobe} should be understood to have subscripts $s=1,2$ and there are two copies of them.
The $D_0, D_i$ introduced here in a sense correspond to the $r$-version of the derivatives there.

$E_r$ and $V_{ri}$ do not transform as a scalar under~\eqref{tdiff}.  We can construct a combined object
\be \label{der}
D_i E_r \equiv {1 \ov E_r} \le(\p_i E_r  + \p_0 V_{ri} \ri)
\ee
which transforms under~\eqref{tdiff} as a scalar and under~\eqref{sdiff} as a vector. %It does not appear possible to construct a nice geometric quantity out of time derivatives of $E_r$ or spatial derivatives of $V_{ri}$.

While $\fb_{ri}$ is not gauge invariant~\eqref{guhe}, at first derivative order the gauge invariant forms are
\be
\sB_{ij} = D_i \fb_{rj} - D_j \fb_{ri}  , \qquad D_0 \fb_{ri} = {1 \ov E_r} \p_0 \fb_{ri},
\ee
which are scalars under~\eqref{tdiff} and are tensors under~\eqref{sdiff}.

Finally, we note the identity
\be
D_i \phi^i +  \phi^i D_i E_r
= \frac 1{\sqrt{a_r} E_r}\left(\partial_i(\sqrt{a_r} E_r\phi^i)+\partial_0(\sqrt{a_r}\phi^i V_{ri})\right),
\ee
which allows us to do integration by part under the integrals:
\be
\int d^d \sig \, \sqrt{a_r} E_r \, D_i \phi^i  = - \int  d^d \sig \, \sqrt{a_r} E_r \, \phi^i D_i E_r \ .
\ee

\subsubsection{Torsion and curvature} \label{sec:curb}

Now consider the commutator of $D_i$ acting on a scalar:
\be\label{1tor}
[D_i , D_j ] \phi \equiv \ft_{ij} D_0\phi, \qquad \ft_{ij} = E_r (d_i v_{rj} - d_j v_{ri}) ,
  \ee
where we used $\tilde \Gamma^k_{[ij]}=0$. Clearly the torsion $\ft_{ij}$
has good transformation properties under both time and spatial diffeomorphisms as the left hand side of~\eqref{1tor} does. Similarly, we can introduce  the
``Riemann tensor'' $\tilde R^k_{\ lij}$ by
\be\label{1defR}
[D_i,D_j]\phi_k=\tilde R_{ijk}{^l}\phi_l+ \ft_{ij} D_0\phi_k \
\ee
with
\be \label{1riem}
\tilde R_{ijk}{^l} = d_j \tilde\Gamma^l_{ik} - d_i \tilde\Gamma^l_{jk}  + \tilde\Gamma^m_{ki}\tilde\Gamma^l_{jm} - \tilde\Gamma^m_{kj}\tilde\Gamma^l_{im} \ .
\ee
One can check that we still have
\be
\tilde R_{ijk}{^l} + \tilde R_{kij}{^l} + \tilde R_{jki}{^l} =0,
\ee
but due to the extra term on the right hand side of~\eqref{1defR},
\bega \label{2defR}
\tilde R_{ijkl} + \tilde R_{ijlk} = - \ft_{ij} D_0 a_{rkl}.
\end{gather}
As a result, there are two ``Ricci tensors'':
\be
\tilde R_{ik}^1 =  \tilde R_{ijk}{^j}, \qquad \tilde R_{ik}^2 =  \tilde R_{ij}{^j}{_k},
\ee
neither of which is symmetric. It is convenient to consider
\be
W_{ik} = \tilde R_{ik}^1 + \tilde R_{ik}^2 = - \ft_{ij} a_r^{jl} D_0 a_{r kl} ,
\qquad S_{ik} = \ha \le(\tilde R_{ik}^1 - \tilde R_{ik}^2 \ri) \ ,
\ee
where the second equality of the first equation follows from~\eqref{2defR}.
Also note that
\be
\tilde R_{[ij]}^1=\ha [d_i, d_j] \log\sqrt a_r=\ha \ft_{ij} D_0\log \sqrt{a_r} \ .
\ee
Finally one can check that there is no new invariant from $[D_0, D_i]$.

\subsection{The bosonic  action}  \label{sec:action}

\subsubsection{General structure}

We are now ready to write down the bosonic part of the hydrodynamical action,
\be
 I [h_1, B_1; h_2, B_2;\tau] = I [\Phi_r,\Phi_a]
\ee
with
\be \label{1fie}
\Phi_r = \{a_r,  E_r, \tau, \mu_r, v_{ri},  \fb_{ri} \}, \qquad \Phi_a = \{E_a , \chi_a, \Xi, \mu_a, V_{ai},  \fb_{ai} \},
\ee
which is  invariant under~\eqref{sdiff}--\eqref{tdiff} and~\eqref{daug}, and satisfies conditions~\eqref{keyp3} and~\eqref{keyp2}. Constraints from the local KMS condition~\eqref{keyp4}
will be discussed later in Sec.~\ref{sec:kmsc}.
 Note that there is no separate dependence on $\vp$ in $ I$ other than that contained in $\mu$ and $\fb_i$.
 %\HL{ For a conformal system, one removes $\tau_a$ and $\tau_r$ from the list,
%with all $\tau$-dependence implicit in other variables.}

From~\eqref{keyp3},
\be \label{refl1}
 I^*[\Phi_r, \Phi_a] = -  I [\Phi_r, - \Phi_a],
\ee
and equation~\eqref{keyp2} implies that
\be \label{aim0}
I [\Phi_r, \Phi_a=0]  = 0 \ .
\ee
From~\eqref{aim0}, we cannot use any negative power of $\Xi$. In particular, while we start with
two spatial metrics $a_1$ and $a_2$, only $a_r$ can serve as a metric to raise and lower indices
in constructing the action.
   We can write the action as
\be  \label{intm}
\int d^d \sig \, \sqrt{a_r} E_r \, \sL [\Phi_r, \Phi_a]  ,
\ee
where $\sL$ is a function of $\Phi$'s and their derivatives, and should be a scalar under~\eqref{sdiff}--\eqref{tdiff}.

We will write $\sL$ as a double expansion in terms of the number of $a$-type fields in~\eqref{1fie}, and the number of derivatives.\footnote{Due to nonlinear relations in~\eqref{1nr}--\eqref{1na}, this $a$-field expansion is slightly different from that outlined in Sec.~\ref{sec:hbar} and Sec.~\ref{sec:fluc}, but qualitatively the same.}
%The number of $a$-type fields corresponds to the order of fluctuations as discussed in Sec.~\ref{sec:CTP}.
More explicitly,
\be \label{aexp}
\sL =  \sL^{(1)} +  \sL^{(2)} + \cdots,
\ee
where $ \sL^{(m)}$ contains $m$ factors of $\Phi_a$'s. From~\eqref{refl1},  $\sL^{(m)}$ is pure imaginary for  even $m$ and real for odd $m$. Each  $\sL^{(m)}$ can then be further expanded in
the number of derivatives.

%\subsection{Leading order in fluctuation expansion}

Let us first consider terms with only a single factor of $\Phi_a$. By using the covariant derivatives of Sec.~\ref{sec:cord}, we find  to first order in derivatives the most general Lagrangian density
can be written as
\bega \label{kee}
\sL^{(1)} = - f_1 E_a + f_2  \chi_a + f_3  \nu_a
-  { \eta  \ov 2}  \Xi^{ij} D_0 a_{rij}   -   \lam_{1}  V_{a}^i D_i E_r
- \lam_2 \fc^i_a \hat D_0 \fb_{ri} + \lam_{12} V^i_a \hat D_0 \fb_{ri}
\cr  + \lam_{21}  \fc_{a}^i D_i E_r +
\lam_{5}  %\al_r^{ij}
 D_i \tau V^i_{a} + \lam_{6}  %\al_r^{ij}
 D_i \mu_r V_{a}^i
+ \lam_{7} % \al_r^{ij}
 D_i \tau \fc_{a}^i + \lam_{8} D_i \mu_r \fc_{a}^i  + \cdots,
\end{gather}
where $\Xi^{ij} \equiv  \Xi^i{_k} a^{kj}_r$ is symmetric and traceless, and for later convenience\footnote{With these choices the coefficients of various terms of the stress tensor and current, e.g. those in~\eqref{1p},~\eqref{01p},~\eqref{11p}, simplify.} we introduce
\be \label{odef}
\nu_a = \mu_a + E_a \mu_r , \quad \hmu = \mu_r \beta (\sig), \quad  \fc_{ai} = \fb_{ai} - \mu_r V_{ai} , \quad
\hat D_0 \fb_{ri} \equiv D_0 \fb_{ri} - \mu_r D_i E_r \
\ee
where the local inverse temperature $\beta (\sig)$ was introduced in~\eqref{tay0}.

 In~\eqref{kee}, $\eta$ and $\lam$'s are all real functions of $\mu_r$ and $\tau$.
$f_{1,2,3}$ can be further expanded in derivatives as
\bea
f_1 &=&  \ep_0 + f_{11} D_0 \tau +  f_{12}  D_0 \le(\log \sqrt{\det a_r} \ri) +  f_{13}  \beta^{-1} (\sig) D_0 \hmu  %\tr (\al_r^{-1} D_0 \al_r)
+ {\rm higher \, derivatives} , \\
f_2 &=&  p_0 + f_{21} D_0 \tau -  f_{22} D_0 \le(\log \sqrt{\det a_r} \ri) + f_{23} \beta^{-1} (\sig) D_0 \hmu
+ {\rm higher \, derivatives} , \\
f_3 &=&  n_0 + f_{31} D_0 \tau +  f_{32} D_0 \le(\log \sqrt{\det a_r} \ri) - f_{33} \beta^{-1} (\sig) D_0 \hmu
+ {\rm higher \, derivatives} ,
\eea
with all coefficients $f_{11}, f_{12}, \cdots$ real functions of $\mu_r$ and $\tau$.
Note that $a_{rij}$ was introduced in~\eqref{1na}. Various signs are chosen for later convenience.

At $O(a^2)$, to zeroth order in derivatives, we have
\bea
- i \sL_0^{(2)} &=& s_{11}E^2_a+s_{22}\chi^2_a+s_{33}\nu^2_a + 2 s_{12} E_a \chi_a
 + 2 s_{13} E_a \nu_a  \cr
&&+ 2 s_{23}\chi_a \nu_a
+ r \tr  \Xi^2 +r_{11} V_a^i V_{ai} + 2 r_{12} V_a^i\fc_{ai} + r_{22} \fc_a^i \fc_{ai}
 \label{aa0},
\eea
where again all coefficients are real and are functions of $\mu_r, \tau$.

It is straightforward to write down terms at higher order in the $a$-field expansion or
with more derivatives, but the number of terms increases quickly. For the rest of this section, we will focus on analyzing~\eqref{kee}--\eqref{aa0}.

As usual, one has the freedom of making field redefinitions
\be
\chi \to \chi + \de \chi  \quad \Longrightarrow \quad I \to  I + \int d^d \sig \,{\de  I \ov \de \chi} \de \chi,
\ee
where $\chi$ collectively denotes all dynamical variables and $\de \chi$ involves derivatives of $\chi$. Equivalently,
we could set to zero all terms in the action which are proportional to the equations of motion at lower derivative order.

\subsection{Stress tensor and current operators} \label{sec:stress}

We now consider the stress tensor and current operators following from the action written above.
%In particular, we will show that at the lowest order in the expansion of $a$-fields, these operators recover the standard
%constitutive relations.

\subsubsection{General discussion}

The stress tensor and current operators are defined in~\eqref{defst} by varying the action with respect to $g_{s\mu \nu} (x), A_{s \mu} (x)$.  Since both the action $I$ and $g_{s\mu \nu} (x), A_{s \mu} (x)$ are invariant under~\eqref{sdiff}--\eqref{tdiff} and~\eqref{daug}, by definition $\hat T_{s}^{\mu \nu}$ and $\hat J_s^\mu$ are also invariant.
As emphasized below~\eqref{defst},  $x$ denotes the spacetime location at which $\hat T_s^{\mu \nu}, \hat J_s^\mu$ ($s=1,2$) are evaluated and should be distinguished from either $\sig$ or $X$.
Given the dependence of the action on $g_s$ and $A_s$ is of the form
\be
I = \int d^d \sig \, \tilde \sL [g_{s\mu \nu} (X(\sig)), A_{s \mu} (X(\sig))], \qquad \tilde \sL = \sqrt{a_r} E_r \sL,
\ee
the stress tensor has the structure
\be \label{ien}
\ha \sqrt{-g_s} T_s^{\mu \nu} (x) = \int d^d \sig \, \de^{(d)} (x - X_s (\sig)) \, {\de\tilde \sL \ov \de g_{s \mu \nu} (X_s (\sig))},
\ee
and similarly for  the current. Note that since $X^\mu_s (\sig)$ are dynamical variables, in the full ``quantum'' theory
defined by the path integral~\eqref{qft0}, the delta function $\de^{(d)} (x - X_s (\sig))$ on the right hand side of~\eqref{ien}
is a quantum operator and should be understood as
\be
\de^{(d)} (x - X_s (\sig)) = \int {d^d k \ov (2 \pi)^d} e^{i k \cdot (x - X_s (\sig))} \ .
\ee

At the level of equations of motion, one can solve the delta function to find $\sig_s (x) = X^{-1}_s (x)$ and evaluate the integrals of~\eqref{ien}.
For example, the stress tensor for the first segment can be written as
\bega
 \sqrt{-g_1} |\Lam_1| \hat T^{\mu \nu}_1 (x) =  \le(\mu \le({\de I \ov \de \mu_a} + \ha  {\de I \ov \de \mu_r}\ri) - {\de I \ov \de E_a}
- {b \ov 2} {\de I \ov \de E_r} \ri) u^\mu u^\nu +  {\de I \ov \de \chi_a} \De^{\mu \nu} +
{\de I \ov \de a_{rij}}  \lam_i{^\mu} \lam_j{^\nu} \cr
+ 2 {\de I \ov \de \Xi^i{_j}} \le(\lam^{i (\mu} \lam_j{^{\nu)}} - {\De^{\mu \nu} \ov d-1} \de_i^j + \cdots \ri) +
2  \le( \mu \le({\de I \ov \de \fb_{ai}} + \ha  {\de I \ov \de \fb_{ri}}\ri) + {1 \ov b} \le({\de I \ov \de v_{ai}}
+\ha {\de I \ov \de v_{ri}} \ri) \ri) u^{(\mu} \lam_i{^{\nu)}},
\label{stress1}
\end{gather}
where $\Lam$ was introduced in~\eqref{onlam} and we have suppressed the subscript $1$ (all variables without an explicit subscript should be understood as with index $1$). In obtaining~\eqref{stress1}, we have used expressions in Appendix~\ref{app:vary}, and it should be understood that  the right hand side is evaluated at $\sig_1 (x) = X^{-1}_1 (x)$.
Similarly, from variation of $A_{1 \mu}$ we find that
 \be \label{cur1}
 \sqrt{-g_1} |\Lam_1| \hat J^\mu_1 =  \le({\de I \ov \de \mu_a} + \ha  {\de I \ov \de \mu_r}\ri) u^\mu +
  \le({\de I \ov \de \fb_{ai}} + \ha  {\de I \ov \de \fb_{ri}}\ri) \lam_i{^\mu}   \ .
 \ee
$\hat T^{\mu \nu}_2$ and $\hat J^\mu_2$ can be obtained from~\eqref{stress1}--\eqref{cur1}
by switching the signs of the terms involving derivatives with respect to the $a$-fields.

We can expand~\eqref{stress1}--\eqref{cur1} in  the number of $a$-fields.
At zeroth order, as we discuss below and in more detail in Appendix~\ref{app:proof}, as a consequence of symmetries~\eqref{sdiff}--\eqref{tdiff} and~\eqref{daug}, the stress tensor and current can be expressed solely in terms of velocity-type variables $u^\mu, \hmu, \tau$ and their derivatives to all derivative orders.

%\subsection{Structure of stress tensor and current at higher orders}

Going beyond zeroth order in the $a$-field expansion,
other dependence on $X^\mu_{1,2}$ will be involved. For example, at $O(a)$, the following quantities~(which are invariant under~\eqref{sdiff}--\eqref{tdiff} and~\eqref{daug}):
\be
\lambda_{1i}^{\ \mu}\lambda_{2j}^{\ \nu}a_r^{ij},\qquad
a_r^{ij}v_{ai}b_r\lambda_{rj}^\mu,\qquad
a_r^{ij}\lambda_{ri}^{\ \mu}\mathfrak b_{aj},
\ee
will contribute to the stress tensor. These quantities cannot be written in terms of the velocity or chemical potential.

%\subsection{Leading order in $a$-expansion}

%Given the expansion in terms of the number of $a$-fields of $\sL$ in~\eqref{aexp}, the corresponding off-shell stress tensor and
%current operators will also have such an expansion.

\subsubsection{Lowest order in $a$-field expansion}

Let us now look at the stress tensor and current at leading order in the $a$-field expansion, where
we can take
\bega \label{yyy}
g_1 = g_2 = g, \quad A_1 = A_2 = A , \quad X_1^\mu = X_2^\mu = X^\mu, \quad
 \vp_1  = \vp_2  = \vp, %\quad \tau_1 = \tau_2 = \tau,
 \cr
\mu_1 =\mu_2 =  \mu, \qquad \sig_1^a (x) = \sig_2^a (x) \equiv \sig^a (x) = X^{-1} (x),  \qquad
X^\mu (\sig^a (x)) = x^\mu,
\end{gather}
and then
\be \label{2hyd}
\hat T^{\mu \nu}_1 = \hat T^{\mu \nu}_2  = (\hat T_r^{\mu \nu})^{(0)} \equiv \hat T^{\mu \nu}_{\rm hydro} ,\qquad
\hat J_1 = \hat J_2 = (\hat J_r^{\mu})^{(0)} \equiv  \hat  J^{\mu}_{\rm hydro} \ .
\ee
Setting all the $a$-fields to zero in~\eqref{stress1}--\eqref{cur1} and dropping the $r$-indices, we find that they can be written  as
\be\label{1gens}
\hat T^{\mu \nu}_{\rm hydro} = \ep u^\mu u^\nu + p \De^{\mu \nu} + %\De^{\mu \nu, \rho \sig}
2  u^{(\mu} q^{\nu)} + \Sig^{\mu \nu} ,
%\De^{\nu) \rho} q_\rho ,
\qquad \hat J^\mu_{\rm hydro} = n u^\mu + %\De^{\mu \nu}
j^\mu,
\ee
where
\bega \label{1der}
\ep = \mu {\de \sL \ov \de \mu_a} - {\de \sL \ov \de E_a} , \qquad p =   {\de \sL \ov \de \chi_a} , \qquad  \Sig^{\mu \nu} = 2   \lam^{i (\mu} \lam_j{^{\nu)}} {\de \sL \ov \de \Xi^i{_j}}  , \\
q^\mu =    \lam_i{^\mu} \le( \mu {\de \sL \ov \de \fb_{ai}} + {1 \ov E} {\de \sL \ov \de v_{ai}}  \ri), \qquad
n =  {\de \sL \ov \de \mu_a} , \qquad j^\mu =   \lam_i{^\mu}   {\de \sL \ov \de \fb_{ai}}\ .
\label{2der}
\end{gather}
It should be understood in~\eqref{1der}--\eqref{2der} that after taking the derivative, one should set all the $a$-fields to zero. In Appendix~\ref{app:proof}, we show that all quantities of~\eqref{1der}--\eqref{2der} can be expressed in terms of standard hydrodynamical variables.

Applying~\eqref{1der}--\eqref{2der} to~\eqref{kee}, we find to first derivative order
\be \label{1fird}
\ep = \ep_0 + h_\ep , \qquad p= p_0 + h_p, \qquad \Sig^{\mu \nu} = - \eta \sig^{\mu \nu}, \qquad
n = n_0 + h_n,
\ee
with
\bega
 \label{1p}
 h_\epsilon = %e^{-\tau} \hmu h_n
f_{11} \p\tau +  f_{12}  \th +  f_{13} e^{-\tau} \p \hmu , \; h_p =  f_{21} \p \tau - f_{22} \th +  f_{23} e^{-\tau} \p \hmu  , \;
h_n  = f_{31} \p\tau + f_{32}   \th -   f_{33} e^{-\tau}  \p \hmu  \\
 \p \equiv u^\mu \nab_\mu, \quad \th\equiv \nab_\mu u^\mu, \quad \sig^{\mu \nu} \equiv \De^{\mu \lam} \De^{\nu \rho} \le(\nab_\lam u_\rho + \nab_\rho u_\lam - {2 \ov d-1} g_{\lam \rho}
\nab_\al u^\al \ri),
\label{qcur4}
\end{gather}
and
 \bega\label{01p}
 j^\mu =\lam_{21} \p u^\mu
 -\lam_2  \le( \De^{\mu \nu} \p_\nu  \mu + u_\lam  F^{\lam \mu} \ri)+ \lam_7 \De^{\mu \nu} \p_\nu  \tau +\lam_{8} \De^{\mu \nu} \p_\nu  \mu \\
 q^\mu  =- \lam_1 \p u^\mu
  +\lam_{12}   \le( \De^{\mu \nu} \p_\nu  \mu + u_\lam  F^{\lam \mu} \ri)+ \lam_5 \De^{\mu \nu} \p_\nu  \tau +\lam_{6} \De^{\mu \nu} \p_\nu  \mu \ .
 \label{11p}
\end{gather}

As advertised in Sec.~\ref{sec:fluc}, equations~\eqref{1gens} and~\eqref{1fird} are precisely the standard constitutive relations for  $\hat T^{\mu \nu}$ and $\hat J^\mu$ to first derivative order in a general frame (before one imposes entropy current constraints). In particular, $ \epsilon_0, p_0, n_0$ are  the local energy, pressure and charge densities in the ideal fluid limit, with $h_\epsilon, h_p , h_n$
their respective first order derivative corrections. $\eta$ is the shear viscosity.
%Recall also from~\eqref{mhat} and~\eqref{tay0} that
%\be
%\hmu (\sig)= {\mu (\sig) \ov T (\sig)} T_0 \ .
%\ee
We should emphasize that~\eqref{1fird}--\eqref{11p} are not yet the final form of the stress tensor and current, as we have not imposed the local KMS conditions in~\eqref{kee}.
In particular, at this stage, the energy density $\ep_0$, pressure $p_0$,  and charge density $n_0$
are completely independent. There are no relations among them. In the next subsection, we will
discuss how thermodynamical relations emerge, along with other constraints on~\eqref{kee}.

%To close, note that  $\tau$-dependences in~\eqref{qcur1}--\eqref{allpara} % let us consider a neutral conformal fluid, for which $\hat \mu =0$ and all coefficients in~\eqref{kee} become constants.
%The $\tau$ dependence of a quantity then
% give local temperature scalings of each quantity expected on dimensional ground, e.g.
% \be
 % \ep_0 , p_0  \propto T^d, \qquad n_0 \propto T^{d-1}, \qquad \eta \propto T^{d-1} ,\quad \cdots  \ .
%  \ee

\subsection{Formulation in the physical spacetime} \label{sec:forph}

The formulation of Sec.~\ref{sec:action} is convenient for  writing down an action invariant under various fluid space diffeomorphisms.
The resulting action is defined in the fluid spacetime.
Here we discuss how to rewrite the action in the physical spacetime, which is more convenient for many questions.

For this purpose, consider
\be \label{1ax}
X_a = X_1 (\sig) - X_2(\sig),  \quad X_1 (\sig) = X(\sig) + \ha X_a (\sig), \quad X_2 (\sig) = X(\sig) - \ha X_a(\sig) \ .
\ee
We now invert $X^\mu (\sig^a)$ to obtain $\sig^a (X^\mu)$, and treat $\sig^a (X)$ as dynamical
variables. Other dynamical variables $X_a^\mu (\sig), \vp_{r,a} (\sig), \tau (\sig)$ %$X_a^\mu (x) = X_a^\mu (x (X)), \vp_{r,a} (x (X)), \tau_{r,a} (x(X))$
are now all considered as functions of $X^\mu$ through $\sig^a (X)$. Since $X^\mu$ are now simply the coordinates for the physical spacetime, there is no need to distinguish them from $x^\mu$.
 Thus the dynamical variables are now
$\sig^a (x), X_a^\mu (x), \vp_{r,a} (x), \tau (x)$. Below we will drop all $r$-subscripts.
%The background fields can be written as $g_{1 \mu \nu} (X_1) = g_{1 \mu \nu} \le(x + \ha X^a (x)\ri)$ $g_{2 \mu \nu} (X_2) = g_{2 \mu \nu} \le(x - \ha X^a (x) \ri)$ and similarly with $A_1$ and $A_2$.

Now let us consider the actions~\eqref{kee} and~\eqref{aa0} expressed in these variables. For simplicity, we will put all background fields to zero (except that corresponding to the chemical potential at infinity), i.e.
\be
g_{1 \mu \nu} = g_{2 \mu \nu} = \eta_{\mu \nu}, \qquad
A_{1 \mu} = A_{2 \mu} = \mu_0 \de_{\mu}^0 \ .
\ee
So below all contractions between $\mu, \nu, \cdots$ indices are through $\eta_{\mu \nu}$. Using
$\sig^a (x)$ we can define a velocity field as in~\eqref{tbe0}:
\be
u^\mu = {1 \ov b} {\p x^\mu \ov \p \sig^0} , \qquad b^2 = - \eta_{\mu \nu} {\p x^\mu \ov \p \sig^0}  {\p x^\nu \ov \p \sig^0}, \
\ee
which can also be written as
\be \label{1vel}
u^\mu = {1 \ov \sqrt{-j^2}} j^\mu, \quad j^2 \equiv j^\mu j_\mu , \quad
j^\mu = \ep^{\mu \mu_1 \cdots \mu_{d-1}} {\p \sig^{1} \ov \p x^{\mu_1}} \cdots {\p \sig^{d-1} \ov \p x^{\mu_{d-1}}}  \ .
\ee
Note that in the form of~\eqref{1vel}, $\sig^0$ is not needed to define $u^\mu$. Various quantities defined earlier can be straightforwardly
converted into the new variables. For example, to first order in $X_a,  \vp_a$, we have
\be
u_1^\mu = u^\mu + {1 \ov 2} \De^{\mu \nu} \p X_{a\nu}, %\quad
%\hmu_{a} = e^{\tau} \p \vp_a + \hmu  u^\mu \p X_{a\mu} ,
\quad \nu_a = \p \vp_a,
 \quad
  \mu =  u^0 A_0 + \p   \vp  \ . %\equiv \hmu \ .
\ee

Expanded in $X_a^\mu, \vp_a$, the action can be written as
\be \label{1nex}
I = \tilde I^{(1)} + \tilde I^{(2)} + \tilde I^{(3)} + \cdots \ .
\ee
Note that since the $\Phi_a$ defined in Sec.~\ref{sec:action} depend nonlinearly on dynamical variables,
the expansion~\eqref{1nex} does not coincide with~\eqref{aexp}. For example, $\sL^{(1)}$ in~\eqref{aexp} also contributes to $\tilde I^{(3)}, \tilde I^{(5)}, \cdots$.  But note $\tilde I^{(1)}$ is  determined solely
from $\sL^{(1)}$ and $\tilde I^{(2)}$ solely from $\sL^{(2)}$. We then find from~\eqref{kee}
\bega \label{kee2}
\tilde I^{(1)} = \int d^d x \,  \le[  \hat T^{\mu \nu}_{\rm hydro}  \p_\mu X_{a \nu}
+ \hat J^\mu_{\rm hydro} \p_\mu \vp_a \ri].  \
\end{gather}
This form of~\eqref{kee2} is of course expected since, as we discussed in Sec.~\ref{sec:eom}, the equations of motion for $X_a^\mu$ and $\vp_a$ simply correspond to the conservation of the stress tensor and current respectively. For this reason, we expect~\eqref{kee2} to apply to all derivative orders. Equation~\eqref{kee2} was
considered recently in~\cite{Kovtun:2014hpa} from exponentiating the hydrodynamical equations of motion.

At $O(a^2)$, from~\eqref{aa0} we find
\be
\begin{split}
\tilde I^{(2)}_0=& i \int d^dx \, \left[ r \eta^{\mu\rho}\eta^{\nu\sigma}(2 \partial_{<\mu}X_{a\nu>})(2 \partial_{<\rho} X_{a\sigma>} ) +r_{11}\Delta^{\mu\rho}  (2 u^\nu \partial_{(\mu}X_{a\nu)}) (2 u^\sigma \partial_{(\rho}X_{a\sigma)}) \right. \\
 & + r_{22} \Delta^{\mu\nu} \p_\mu \vp_a  \p_\nu \vp_a   + 2 r_{12}  \Delta^{\mu\rho} (2 u^\nu\partial_{(\mu}X_{a\nu)})  \p_\rho \vp_a  \\
& +  s_{11}(u^\mu \partial X_{a\mu})^2 + s_{22} (\Delta^{\mu\nu}\partial_\mu X_{a\nu})^2 + s_{33} (\p \vp_a)^2   \\
& -2 s_{12} \Delta^{\mu\nu}\partial_\mu X_{a\nu} u^\rho\partial X_{a\rho} + 2 s_{23} ( \p \vp_a) \Delta^{\mu\nu}\partial_\mu X_{a\nu}  -  2 s_{13}  u^\mu\partial X_{a\mu} ( \p \vp_a)
 \bigg]  \ .
\end{split}
\label{1qua}
\ee
In the above equations,
 the angular brackets denote the symmetric transverse traceless part of a tensor, i.e. for an arbitrary two-index tensor $C_{\mu \nu}$
\be
C_{< \mu \nu>} \equiv \De_{\mu \rho} \De_{\nu \lam} \le(C^{(\rho \lam)} - {1 \ov d-1} \De^{\rho \lam}
\De_{\al \beta} C^{\al \beta} \ri) \ .
\ee
We also follow the standard convention of using square brackets and parentheses to denote antisymmetrization and symmetrization respectively, i.e.
\be
C_{(\mu \nu)} = \ha (C_{\mu \nu} + C_{\nu \mu}), \qquad C_{[\mu \nu]} = \ha (C_{\mu \nu} - C_{\nu \mu}) \ .
\ee

Note that in both~\eqref{kee2} and~\eqref{1qua}, $\sig^0$ has dropped out, which is a consequence of the time diffeomorphism~\eqref{tdiff}. In fact, we expect $\sig^0$ to completely decouple to all orders.

For a neutral fluid, we find that
\be
\begin{split}
I^{(2)}_0=&\int d^dx\;  \bigg[ r \eta^{\mu\rho}\eta^{\nu\sigma}(2 \partial_{<\mu}X_{a\nu>})(2\partial_{<\rho} X_{a\sigma>})  +r_{11}\Delta^{\mu\rho}  (2u^\nu \partial_{(\mu}X_{a\nu)}) (2u^\sigma \partial_{(\rho}X_{a\sigma)}) \\
&+ s_{11}(u^\rho\partial X_{a\rho})^2 + s_{22} (\Delta^{\mu\nu}\partial_\mu X_{a\nu})^2 -  2 s_{12} (\Delta^{\mu\nu}\partial_\mu X_{a\nu})(u^\rho\partial X_{a\rho}) \bigg] \  .
\end{split}
\ee

Equation~\eqref{1qua} contains three quadratic forms: one each for the tensor, vector, and scalar
sectors. Since $\tilde I^{(2)}$ is pure imaginary, for the path integral to be well defined the
three quadratic forms should be separately non-negative, which implies that
\be \label{1non}
r \geq 0,
\ee
$r_{11}, r_{12}, r_{22}$ should be such that
\be \label{2non}
%C_v =
r_{11} x^2 +2  r_{12} xy + r_{22} y^2 \geq 0
\ee
for any real $x,y$, and $ s_{11},  s_{22},   s_{12},   s_{23},  s_{13},  s_{33}$ should be such that
\be  \label{3non}
%C_s =
 s_{11}x^2  +  s_{22} y^2 +  s_{33}z^2  -2 s_{12}xy +  2 s_{23} yz  -  2  s_{13}  xz \geq 0
\ee
for any real $x,y,z$.

For a neutral fluid, we  then have
\be \label{1neu}
r \geq 0, \qquad r_{11} \geq 0, \qquad  s_{11}x^2 + s_{22} y^2 - 2 s_{12} xy \geq 0 \ .
\ee

\subsection{The source action} \label{sec:kmsc}

We now discuss how to impose the local KMS conditions
on the actions~\eqref{intm}.

For this purpose, we first need to obtain the corresponding action for sources only.
Recall that from the prescription of Sec.~\ref{sec:thermal} we should first set all dynamical
fields to zero. Here we have a complication regarding what should be the appropriate ``background'' values for $\tau$. We propose the following prescription:
%how to deal with
%corresponding to~\eqref{kee}. Compared with the baby model
%of a conserved current of Sec.~\ref{sec:diffusion}, here we have a complication regarding how to deal with the $\tau_{1,2}$ modes. We will use the

\ben

\item  Set
\be \label{bakv}
X^\mu_{1,2} = \sig^a \de_a^\mu , \qquad \vp_{1,2} =0, % \qquad \tau_1 = \tau_2 = \tau_0
\ee
and then
\bega \label{1hdef1}
h_{s a b} (\sig) = g_{s\mu \nu} (x) \de_a^\mu \de_b^\nu ,%= e^{- 2 \tau_s} (\eta_{\mu \nu} + \lam_{s\mu \nu}) ,
\qquad B_{s a} (\sig) = A_{s\mu} (x) \de_a^\mu  \ .
\end{gather}
Now $\sig^a= \de^a_\mu x^\mu$ spans the physical spacetime and we will simply use $x^\mu$. % while $\lam_{s\mu \nu}, A_{s\mu}$ are physical external fields.
By definition, the resulting action obtained,
$I [g_s, A_s,\tau] $, is only invariant under (i) time diffeomorphisms, (ii) spatial diffeomorphisms, (iii) time-independent gauge transformations, of the {\it physical} spacetime.

\item  Recall that
\be
e^{-\tau} = {T_{\rm prop} \ov T_0},
\ee
where $T_{\rm prop}$ denotes the local proper temperature in the {\it fluid space}. In the absence of dynamics, it is natural to identify
\be
T_{\rm prop} = {T_0 \ov \sqrt{- g_{00}}},
\ee
which then motivates us to set for $\tau$ the following background value
\be \label{1yen}
 \tau =  \ha \log (- g_{r00}) \ .
\ee
%\HL{Note that this identification breaks time diffeomorphisms from $t \to f (t, x^i)$ to
%\be \label{fixt}
%t \to t + f (x^i)
%\ee
%as $\tau (x)$  is a scalar field, i.e. it transforms as a scalar under time diffeomorphisms, but %$g_{r00}$ does not transform as a scalar field under time diffeomorphisms.}

\een
The resulting  action $I_{s} [g_1, A_2; g_2, A_2]$ is then the one on which we will impose the local KMS conditions~\eqref{keyp4}.

\subsection{Constraints on constitutive relations
from local KMS conditions} \label{sec:cont}

As outlined in Sec.~\ref{sec:consh}, the local KMS conditions include
relations between coefficients of $\sL^{(1)}_s$ and those of $\sL^{(2)}_s$,
which will give rise to the non-negativity of various transport coefficients, as well as consistency conditions~\eqref{allc2}--\eqref{allc},
which concern only $\sL_s^{(1)}$ and give rise to constraints on constitutive relations.
In this subsection, we focus on $\sL^{(1)}_s$ and consider the latter type of constraints.

Imposing~\eqref{bakv} and~\eqref{1yen} amounts to setting in~\eqref{1gens}
\be
\tau = \log b = \ha \log (- g_{00}) , \qquad \mu = {A_0 \ov b} , \qquad u^\mu = {1 \ov b} (1, \vec 0) , \qquad b = \sqrt{- g_{00}}\ .
\ee

Let us now discuss~\eqref{allc} and~\eqref{allc2} in turn.

\subsubsection{Spatial partition function condition}

Following the discussion~\eqref{1con2}--\eqref{1fac}, equation~\eqref{allc} says that $\hat T_{\rm hydro}^{\mu \nu}$ and $\hat J_{\rm hydro}^{\mu}$ in a stationary background should be obtainable  from a partition function defined on the spatial manifold. This is precisely the prescription recently analyzed in detail in~\cite{Banerjee:2012iz,Jensen:2012jh}.

At zeroth order in derivatives, we have
\be \label{vare2}
\hat T^{\mu \nu}_{\rm hydro} = (\ep_0 + p_0) u^\mu u^\nu + p_0 g^{\mu \nu} , \qquad \hat J^\mu_{\rm hydro}  = n_0 u^\mu , \qquad
\ee
 where $\epsilon_0 = \ep_0 (\log b, A_0/b) $, and similarly with $p_0$ and $n_0$. For them to be obtainable from a single functional, we need to impose the integrability conditions
 \bega
 {\de (\sqrt{-g} \hat  T^{\mu \nu}_{\rm hydro} ) \ov \de g_{\lam \rho} }=  {\de (\sqrt{-g} \hat  T^{\lam \rho}_{\rm hydro} )\ov \de g_{\mu \nu} },  \qquad \ha {\de (\sqrt{-g} \hat  T^{\mu \nu}_{\rm hydro} ) \ov \de  A_\lam }=  {\de (\sqrt{-g} \hat J^\lam_{\rm hydro} )\ov \de g_{\mu \nu} }, \quad \cdots
%  \cr \quad    {\de (\sqrt{-g} J^{\mu}_{\rm hydro} ) \ov \de A_\nu }=  {\de (\sqrt{-g} J^\nu_{\rm hydro} )\ov \de A_{\mu } }
 \end{gather}
which lead to the thermodynamical relations
\be \label{thermo}
\ep_0 + p_0 - \mu n_0 = - {\p p_0 \ov \p \tau} , \qquad
n_0 = {\p p_0 \ov \p \mu},
\ee
and the functional from which they can be derived is simply $\int d^{d-1} \vx \, \sqrt{-g} \, p_0 (\log b, A_0/b)$ as one would have anticipated. It is also convenient to introduce
\be \label{1ent}
\hat s_0 = \ep_0+ p_0 - \mu n_0
\ee
which at the ideal fluid level corresponds to the local entropy density time local temperature.

%\subsubsection{First order in derivative}

At first order in derivatives, with time-independent sources we find that\footnote{Note that $\p u_i = \p_i \log b$.}
\bega
h_\ep = h_p = h_n = q^0 = j^0 = \sig_{\mu \nu} = 0,   \cr
j_i = (\lam_{21} + \mu \lam_2 + \lam_7) {\p_i  b \ov b} + \lam_8 \p_i \le({A_0 \ov b} \ri) , \quad
q_i  = (  \lam_5 -\lam_1 - \mu \lam_{12}  ) {\p_i  b \ov b} + (\lam_6 + \mu \lam_8)  \p_i \le({A_0 \ov b} \ri)  ,
\end{gather}
but with rotational symmetry, there cannot be any first order derivative term in a partition function
in general dimensions\footnote{With some specific dimensions, one may be able to construct
first derivative terms using the $\epsilon$ tensor. We will consider such terms elsewhere.}
and thus we need
%\be
%c_u + c_\tau = 0, \qquad c_\mu = c_F e^{-\tau}  , \qquad e_u + e_\tau =0, \qquad e_\mu = e_F e^{-\tau},  \
%\ee
%which in turn leads to
\be \label{1con}
 \lam_{5}  = \lam_{1} + \mu \lam_{12} , \qquad \lam_{7} = - \lam_{21 } - \mu \lam_2 , \qquad  \lam_{6}   =\lam_{8} =0
 \ee
 which gives (recall $\hmu$ was introduced in~\eqref{odef})
  \bega
 j^\mu =\lam_{21} (\p u^\mu - \De^{\mu \nu} \p_\nu  \tau)
 -\lam_2 (e^{-\tau} \De^{\mu \nu} \p_\nu  \hmu +  u_\lam  F^{\lam \mu} ) , %-   \lam_{21 }  \De^{\mu \nu} \p_\nu  \tau
 \\
 q^\mu  =- \lam_{1}  (\p u^\mu - \De^{\mu \nu} \p_\nu  \tau)
  +\lam_{12}  (e^{-\tau} \De^{\mu \nu} \p_\nu  \hmu +  u_\lam  F^{\lam \mu} )  \ .
\end{gather}

To consider the implications of~\eqref{1con} for the constitutive relations for the stress tensor and
current, let us consider the frame-independent vector %~\cite{Kovtun:2012rj}
\be \label{oii}
\ell_\mu \equiv j_\mu - {n \ov \ep + p}  q_\mu  \ .
%= \sig \le(F_{\mu \nu} u^\nu  - e^{-\tau} \De_\mu{^\nu} \p_\nu \hmu \ri)
\ee
Before imposing~\eqref{1con}, upon using the thermodynamical relations~\eqref{thermo}and the zero-derivative order equations of motion, $\ell_\mu$ has the form\footnote{
Note from zeroth order equations of motion and thermodynamic relations~\eqref{thermo} we have
$\p u^\mu - \De^{\mu \nu} \p_\nu \tau = -{n_0 \ov \ep_0 + p_0} \le(u_\lam F^{\lam \mu}  + e^{-\tau} \De^{\mu\nu} \p_\nu \hmu \ri)$.}
\be \label{0se}
\ell_\mu = c_1  F_{\mu \nu} u^\nu + c_2  \De_\mu{^\nu} \p_\nu \tau
+ c_3  \De_\mu{^\nu} \p_\nu \hmu,
\ee
where $c_1, c_2, c_3$ are independent functions of $\tau, \mu$. With~\eqref{1con}, we find that
\be \label{1se}
\ell_\mu
= \sig \le(F_{\mu \nu} u^\nu  - e^{-\tau} \De_\mu{^\nu} \p_\nu \hmu \ri),
\ee
with conductivity $\sig$ given by
\be \label{1xon}
\sig = \lam_2 + (\lam_{12} + \lam_{21}) {n_0 \ov \ep_0 + p_0} + \lam_1 \le({n_0 \ov \ep_0 + p_0}\ri)^2
%{1 \ov (\ep_0 + p_0)^2 } \le(\lam_1 n^2_0 + \lam_2 \hat s_0^2 + (\lam_{12} + \lam_{21}) \hat s_0 n_0  \ri)
%+\frac{n}{\ep+p} \left(e^{-\tau} \hat\mu \left(e_F - \frac{n}{\ep+p} e_u \right) \right)
 \ .
\ee

Comparing with~\eqref{0se}, we see that the thermal conductivity is determined from conductivity in the usual way and the $c_2$ term vanishes. In the conventional formulation, both of these relations follow from entropy current constraints.
%where $s = {\p p \ov \p T}$ is the entropy density.

The bulk viscosity $\zeta$ can be obtained by examining the other
frame-independent quantity %~\cite{Kovtun:2012rj}
\be \label{1ka}
h_p - {{\p p_0 \ov \p\tau} {\p n_0 \ov \p \mu} - {\p p_0 \ov \p \mu} {\p n_0 \ov \p \tau} \ov {\p \ep_0 \ov \p \tau}
 {\p n_0 \ov \p \mu} -  {\p n_0 \ov \p \tau} {\p \ep_0 \ov \p \mu}}
  h_\ep - {{\p p_0 \ov \p\mu} {\p \ep_0 \ov \p \tau} - {\p p_0 \ov \p \tau} {\p \ep_0 \ov \p \mu} \ov {\p \ep_0 \ov \p \tau}
 {\p n_0 \ov \p \mu} -  {\p n_0 \ov \p \tau} {\p \ep_0 \ov \p \mu}}  h_n  = - \ze \th , \
 \ee
where one needs to use the zeroth derivative order equations of motion to obtain the right hand side.

One can also check that the reality condition in~\eqref{allc2} does not appear to impose any additional constraints at these orders.

\subsubsection{Generalized Onsager relations}

Let us now consider the implications of the generalized Onsager relations~\eqref{allc2} and~\eqref{1con1}. The nonlinear source action for~\eqref{kee} can be written as
\be\begin{split} I^{(1)}_1=&\int \sqrt ab\bigg[\left(f_{11}\frac 1b\partial_0 b\right.+ f_{13}\frac 1b\partial_0 A_0+f_{12}\partial_0\log\sqrt a\bigg)\frac{g_{a00}}{2b^3}\\
&+\left(f_{21}\frac 1b\partial_0 b+ f_{23}\frac 1b\partial_0 A_0- f_{22}\partial_0\log\sqrt a\right)\frac 1{2b} a_{aij}a^{ij}\\
&+\left(f_{31}\frac 1b\partial_0 b- f_{33}\frac 1b\partial_0 A_0+f_{32}\partial_0\log\sqrt a\right)\frac{A_{a0}}{b^2}\\
&-\frac{\eta}{2b}\left(a_{aik}-\frac {a_{alj}a^{lj}}{d-1}a_{ik}\right)a^{km}a^{in}\partial_0 a_{mn}+\lambda_{12}  v_{ai}a^{ij}\partial_0 ( A_i+v_iA_0)\\
&- \lambda_2 b^{-1} (A_{ai}+A_{a0}v_i) a^{ij}\partial_0 ( A_i+v_iA_0)\\
&-\lambda_1 b v_{ai} a^{ij}\partial_0 v_j+\lambda_{21}  (A_{ai}+A_{a0}v_i)a^{ij}\partial_0 v_j
\bigg] ,
\end{split}
\label{1souc}
\ee
where we have used the decomposition~\eqref{1deco} and
\be
 g_{a00}=g_{100}-g_{200},\quad a_{aij}=a_{1ij}-a_{2ij},\quad v_{ai}=v_{1i}-v_{2i} \ .
 \ee
Applying~\eqref{1con1} to~\eqref{1souc}, we then find that,
\be\label{11ons}
\lambda_{12} = \lambda_{21}, \qquad
- f_{13}  = f_{31}  ,\qquad
f_{23} = f_{32}, \qquad
- f_{12}  = f_{21} \ .
\ee

%with
%\be
%\tilde r_{11} = r_{11} + r_{22} \mu^2 - 2 r_{12} \mu , \qquad \tilde r_{12} = r_{12} - \mu r_{22}
%\ee
%Relations
%\be
%\tilde \lam_1 =  \lam_1 + \mu^2 \lam_2 + (\lam_{12} + \lam_{21}) \mu , \qquad \tilde \lam_{12} = \lam_{12} + \lam_2 \mu , \quad
%\tilde \lam_{21} = \lam_{21} + \mu \lam_2 , \quad \tilde \lam_5 = \lam_5  - \mu \lam_7 , \quad
%\tilde \lam_{6}   = \lam_6 - \mu \lam_8
%\ee

Note that all the relations above can be obtained from the Onsager relations at linearized
level. So to first derivative order, nonlinear generalizations do not yield new relations.

\subsection{Non-equilibrium fluctuation-dissipation relations}  \label{sec:fullfdt}

Now let us consider the relations between coefficients of $I^{(1)}$ and $I^{(2)}$ which follow from the local KMS conditions. We find the source action by following the procedure
outlined in Sec.~\ref{sec:kmsc}, which gives
\be\begin{split} -iI^{(2)}_0=&\int \sqrt ab\bigg[s_{11}\left(\frac{g_{a00}}{2b^2}\right)^2+s_{22}\left(\frac{a_{aij} a^{ij}}2\right)^2+s_{33}\frac{A_{a0}^2}{b^2}-2 s_{12}\frac{g_{a00}}{2b^2}\frac{a_{aij}a^{ij}}2\\
& -2 s_{13}\frac{g_{a00}}{2b^2}\frac{A_{a0}}{b}+ 2s_{23}\frac{a_{aij}a^{ij}}2 \frac{A_{a0}}b+r \tr\left( a^{ij}a_{ajk}-\frac 1{d-1}a^{kl}a_{akl}\delta^i_j\right)^2\\
& +r_{11} b^2 (v_{ai})^2+2 r_{12} b a^{ij}v_{ai}\left(A_{aj}+v_jA_{a0}\right) +r_{22}\left(A_{ai}+v_iA_{a0}\right)^2
\bigg] \ .
\end{split}
\ee
Imposing \eqref{keyp4}, we find the following relations:
\be
r =\frac{\eta}{2} T(\sig), \quad
r_{11}=\lambda_1 T(\sig),\quad
r_{12}= -\frac{\lambda_{12}+\lambda_{21}}{2}  T(\sig) = - \lam_{12} T(\sig),\quad
r_{22}=\lambda_{2} T(\sig) \ ,
\label{1tv}
\ee
and
\begin{align}
\label{1sc}
 s_{11} & = {f_{11} T(\sig)} , \qquad
  s_{12} = { f_{12}  T(\sig)}  , \qquad
   s_{13} =  { f_{13}  T(\sig)}, \qquad
\\
s_{22}
&=  f_{22}T(\sig), \qquad
s_{23}
=-\frac{f_{32}+f_{23}}{2} T(\sig) = - { f_{23}T(\sig)},\qquad
 s_{33}
=f_{33} T(\sig) \ .
\label{2sc}
\end{align}
We stress that all relations above are for arbitrary  $\tau (\sig^a)$ and $\mu (\sig^a)$ (i.e. arbitrary local
temperature and local chemical potential) and thus are valid for far-from-equilibrium situations.

\subsection{Non-negativity of transport coefficients}  \label{sec:trans}

We now show that the conductivity $\sig$, shear viscosity $\eta$, and
bulk viscosity $\ze$ are non-negative. The shear viscosity
$\eta$ is non-negative following from the first equation of~\eqref{1tv}
and~\eqref{1non}.

With~\eqref{11ons} and~\eqref{1tv} the conductivity~\eqref{1xon}  becomes
\bea
\sig = \beta (\sig) \le(r_{22} - 2 r_{12} {n_0 \ov \ep_0 + p_0} + r_{11} \le({n_0 \ov \ep_0 + p_0}\ri)^2 \ri) \ .
\eea
%\bea
%\sig & = &  {1 \ov (\ep_0 + p_0)^2 } \le(\lam_1 n^2_0 + \lam_2 \hat s_0^2 + 2 \lam_{12}  \hat  s_0 n_0  \ri)  \cr
%& = &  {\beta (x)  \ov (\ep_0 + p_0)^2}  \le[r_{11} n^2_0 + r_{22} \hat  s^2_0  - 2r_{12} \hat  s_0 n_0
%\ri],
%\eea
whose  non-negativity of $\sig$ then follows from~\eqref{2non}.

From~\eqref{1ka},  using zeroth order equations of motion and thermodynamical relations~\eqref{thermo}
we find after some manipulations the bulk viscosity $\ze$ can be written as
% (\HL{clean up below, note
%that $b$ and $a$ are no longer introduced.})
%\be
%- \ze = \Lam_p - {{\p p_0 \ov \p\tau} {\p n_0 \ov \p \mu} - {\p p_0 \ov \p \mu} {\p n_0 \ov \p \tau} \ov {\p \ep_0 \ov \p \tau}
 %{\p n_0 \ov \p \mu} -  {\p n_0 \ov \p \tau} {\p \ep_0 \ov \p \mu}}
 % \Lam_\ep - {{\p p_0 \ov \p\mu} {\p \ep_0 \ov \p \tau} - {\p p_0 \ov \p \tau} {\p \ep_0 \ov \p \mu} \ov {\p \ep_0 \ov \p \tau}
 %{\p n_0 \ov \p \mu} -  {\p n_0 \ov \p \tau} {\p \ep_0 \ov \p \mu}}  \Lam_n,
 %\ee
%where
%\bega
%\Lam_p = b_\th - \bma b_\mu & b_\tau \ema \bma {\p \ep_0 \ov \p \hmu} & {\p \ep_0 \ov \p \tau} \cr
%{\p n_0 \ov \p \hmu} & {\p n_0  \ov \p \tau} \ema^{-1} \bma \ep_0 + p_0 \cr n_0 \ema , \cr
%\Lam_n = d_\th - \bma d_\mu & d_\tau \ema \bma {\p \ep_0 \ov \p \hmu} & {\p \ep_0 \ov \p \tau} \cr
%{\p n_0 \ov \p \hmu} & {\p n_0  \ov \p \tau} \ema^{-1} \bma \ep_0 + p_0 \cr n_0 \ema , \cr
%\Lam_\ep =
% a_\th - \bma a_\mu & a_\tau \ema \bma {\p \ep_0 \ov \p \hmu} & {\p \ep_0 \ov \p \tau} \cr
%{\p n_0 \ov \p \hmu} & {\p n_0  \ov \p \tau} \ema^{-1} \bma \ep_0 + p_0 \cr n_0 \ema  \ .
%\end{gather}
%After some manipulations, we can write $\ze$ as
\be \label{2ka}
\ze =  {1 \ov M^2_2} (f_{11} M_1^2 + f_{22} M_2^2 + f_{33} M_3^2 -
2 f_{23} M_2 M_3 - 2 f_{12} M_1 M_2 - 2 f_{13} M_3 M_1),
\ee
with
\be
M_1 = - (\ep_0+p_0) \p_\mu n_0 + n_0 \p_\mu \ep_0, \; \; M_2 = {\p n_0 \ov \p \tau} {\p \ep_0 \ov \p \mu} -  {\p \ep_0 \ov \p \tau}
 {\p n_0 \ov \p \mu}   , \; \; M_3 = \hat s_0 \p_\mu \ep_0 + n_0 \p_\tau \ep_0 \ .
 \ee
% and
%\bega
%s_1= f_{11} ,  \quad s_2 =  f_{22}, \quad s_3=  f_{33},\quad
%s_{23} = 2 f_{23}, \quad  s_{12} = - 2 f_{12}
%, \quad s_{13} =  2  f_{13} \ .
%\end{gather}
Now using~\eqref{1sc}--\eqref{2sc}, we find~\eqref{2ka} can be written as
\be
\ze =  {\beta (x) \ov M^2_2} (s_{11} M_1^2 +  s_{22} M_2^2 +  s_{33} M_3^2 +
2s_{23} M_2 M_3 - 2s_{12} M_1M_2 - 2 s_{13} M_2 M_3),
\ee
which is non-negative from~\eqref{3non}.

For a neutral fluid, the corresponding expression is
\bea \label{1bul}
 \ze &=&   {1 \ov (\p_\tau \ep_0)^2} \le[f_{22} (\p_\tau \ep_0)^2 - 2 f_{12} (\ep_0 + p_0) \p_\tau \ep_0
 + f_{11}  (\ep_0 + p_0)^2 \ri] \\
& = &  { \beta (x)  \ov (\p_\tau \ep_0)^2} \le[ s_{22} (\p_\tau \ep_0)^2 +s_{11} (\ep_0 + p_0)^2
 - 2 s_{12} (\ep_0 + p_0) \p_\tau \ep_0 \ri],
\eea
which is again non-negative from~\eqref{1neu}.
%where $\tilde f_{222}, \tilde f_{211},  \tf_{212}$ are obtained from the corresponding quantities in~\eqref{1varp} by setting
%$\hmu =0$.

\subsection{Full action to $O(a^2)$ in physical spacetime}  \label{sec:stochastic}

Let us now collect~\eqref{kee2} and~\eqref{1qua}, and all the relations on the coefficients found in Sec.~\ref{sec:cont} and Sec.~\ref{sec:fullfdt}. % which give us the full action
%to order $O(a^2)$ in the absence of external fields:
We have up to order $O(a^2)$
\be \label{xfull}
I = \tilde I^{(1)} + \tilde I^{(2)} + \cdots
\ee
where
\bega \label{4kee}
\tilde I^{(1)} = \int d^d x \,  \le[\hat T^{\mu \nu}_{\rm hydro}  \p_\mu X_{a \nu}
+ \hat J^\mu_{\rm hydro} \p_\mu \vp_a \ri],  \
\end{gather}
and to first derivative order
\be\label{11gens}
\hat T^{\mu \nu}_{\rm hydro} = (\ep_0 + h_\ep) u^\mu u^\nu + (p_0 + h_p) \De^{\mu \nu} +
2  u^{(\mu} q^{\nu)} - \eta \sig^{\mu \nu} ,
\quad \hat J^\mu_{\rm hydro} = (n_0 + h_n) u^\mu + %\De^{\mu \nu}
j^\mu,
\ee
with (using~\eqref{1con} and~\eqref{11ons})
\bega
 h_p=- f_{22}\theta - f_{12} \partial\tau+f_{23} e^{-\tau} \partial\hat\mu, \\
h_n=f_{23}\theta - f_{13} \partial\tau - f_{33}  e^{-\tau} \partial\hat\mu,
 \\
h_\epsilon =   f_{12} \th + f_{11} \p\tau +  f_{13} e^{-\tau} \p \hmu ,  \\
 j^\mu =  \lam_{12}  (\p u^\mu - \De^{\mu \nu} \p_\nu  \tau) - \lam_2(e^{-\tau} \De^{\mu \nu} \p_\nu  \hmu +  u_\lam  F^{\lam \mu} ) \\
 q^\mu =     - \lam_{1} (\p u^\mu -  \De^{\mu \nu} \p_\nu \tau) +   \lam_{12} ( e^{-\tau} \De^{\mu \nu}  \p_\nu \hmu + u_\lam  F^{\lam \mu}) \ .
\end{gather}

At order $O(a^2)$ we have at zeroth order in derivatives,
\be
\begin{split}
 \tilde I^{(2)}_0=& {2 i }  \int d^dx \, T (x)  \left[ \eta ( \partial_{<\mu}X_{a\nu>})( \partial_{<\rho} X_{a\sigma>}  ) \eta^{\mu\rho}\eta^{\nu\sigma} \right. \\
 & + { \lambda_1 \ov 2} \Delta^{\mu\rho}  (2 u^\nu \partial_{(\mu}X_{a\nu)}) ( 2u^\sigma \partial_{(\rho}X_{a\sigma)})  +{\lambda_{2} \ov 2} \Delta^{\mu\nu}\p_\mu \vp_a \p_\nu \vp_a  -  \lam_{12} \Delta^{\mu\rho} (2 u^\nu\partial_{(\mu}X_{a\nu)}) \p_\rho \vp_a  \\
& + {f_{11} \ov 2 }(u^\mu \partial X_{a\mu})^2 + {f_{22} \ov 2}  (\Delta^{\mu\nu}\partial_\mu X_{a\nu})^2 + {f_{33} \ov 2} (\p \vp_a)^2   \\
& -  f_{12}  \Delta^{\mu\nu}\partial_\mu X_{a\nu} u^\rho\partial X_{a\rho} -  f_{23} ( \p \vp_a) \Delta^{\mu\nu}\partial_\mu X_{a\nu}  -  f_{13}  u^\mu\partial X_{a\mu} ( \p \vp_a)
 \bigg] \
\end{split}
\label{3qua}
\ee
where  we have used the non-equilibrium fluctuation-dissipation relations~\eqref{1tv}--\eqref{2sc}.

%Note that various coefficients in~\eqref{4kee} and~\eqref{3qua} are related by the local KMS conditions,
%which to leading order around equilibrium are given in Sec.~\ref{sec:fullfdt}.

Notice that in~\eqref{3qua}, other than $X_a^\mu, \vp_a$, the dynamical variables
appear through standard hydrodynamical variables $u^\mu, \mu $ and $\tau$. Also recall that $u^\mu, \mu $
are derived variables constructed from $\sig^a (x), \vp (x)$ as discussed in Sec.~\ref{sec:forph}.
Below we will  refer to $u^\mu, \mu $ and $\tau$ as hydro variables, and  $X_a^\mu, \vp_a$ as noises.

\subsection{Stochastic hydrodynamics}

Approximating all the hydro variables by their background values, we obtain
a Gaussian action for the noises $X_{a}^\mu$ and $\vp_a$.
 As in Sec.~\ref{sec:kpz}, introducing the Legendre conjugates $\rt^{\mu \nu}$ and ${\rm j}^\mu$ for $\p_{(\mu}X_{a \nu)}$ and $\p_\mu \vp_a$ respectively, the equations of motion for $X_a^\mu$ and $\vp_a$ become
\be
\p_\mu \le(T^{\mu \nu}_{\rm hydro} + {\rm t}^{\mu \nu} \ri) =0 , \qquad \p_\mu \le(J^\mu + {\rm j}^\mu \ri) = 0,
\ee
where $\rt^{\mu \nu}$ and $ {\rm j}^\mu$ can be interpreted as the noise contribution to the full stress tensor and current respectively, and  satisfy Gaussian distributions. More explicitly, around equilibrium values, i.e. $\tau =\vp =0$ and $u^\mu = (1, \vec 0)$, we find the path integrals for $\rt^{\mu \nu}$ and ${\rm j}^\mu$ have the form
\be \label{gao0}
 \int D \rt^{\mu \nu} D  {\rm j}^\mu \, \exp \le[- { \beta_0 \ov 4 } \le({1 \ov 2 \eta} \rt_{<ij>}^2 +  \sum_{a,b=1}^2 \Lam^{-1}_{ab} v_{ai} v_{bi} + \sum_{a,b=1}^3 F^{-1}_{ab} \phi_a \phi_b \ri) \ri]  \
\ee
where
\bega
\Lam = \bma \lam_1  & \lam_{12} \cr
\lam_{12} & \lam_2 \ema, \qquad v_{1i} =  \rt_{0i}, \qquad v_{2i} = {\rm j}_i , \\
F = \bma f_{11} & - f_{12} &  f_{13} \cr - f_{12} & f_{22} &  f_{23}  \cr
f_{13} & f_{23} & f_{33} \ema, \quad
\phi_1 = \rt_{00} , \quad \phi_2 = {1 \ov d-1} \sum_i \rt_{ii} , \quad \phi_3 = {\rm j}_0 \ .
\end{gather}
All coefficients in~\eqref{gao0} should be understood as equilibrium values.
%which can be obtained by the Legendre transform of~\eqref{3qua}. \HL{work out $\xi$ }

Beyond the quadratic approximation, as in the vector case again, there appears to be no benefit to introducing the Legendre conjugate for $X_a^\mu$ and $\vp_a$. The action~\eqref{xfull} provides an interacting effective field theory among hydro variables and noises.

\subsection{Entropy current}

Now consider the $O(a)$ action~\eqref{4kee} in the ideal fluid limit, i.e.
\be \label{ieen}
\tilde I_0^{(1)} = \int d^d x \,  \le[ T^{\mu \nu}_0  \p_\mu X_{a \nu}
+ J^\mu_0 \p_\mu \vp_a  \ri]   \equiv \int d^dx \, \tilde \sL_0^{(1)},
\ee
with
\be
T^{\mu \nu}_0 = \ep_0 u^\mu u^\nu + p_0 \De^{\mu \nu}, \qquad
J^\mu_0 = n_0 u^\mu,
\ee
which are respectively $\hat T^{\mu \nu}_{\rm hydro}$ and $\hat J^\mu_{\rm hydro}$ at zeroth order in the derivative expansion.

The ideal fluid action~\eqref{ieen} has an ``accidental'' symmetry: it is invariant under
\be \label{utrn}
\de X_{a\mu} = \lam e^\tau u_\mu , \qquad \de \vp_a = \lam \hmu
\ee
for some constant infinitesimal parameter $\lam$, as %Under~\eqref{turn},
\be \label{1totd}
\de \tilde \sL_0^{(1)} =  \lam T^{\mu \nu}_0 \p_\mu(e^\tau u_\nu) + \lam J^\mu_0 \p_\mu \hmu
=\lam \p_\mu (p_0 e^\tau u^\mu)
\ee
is a total derivative. To see this, note that
\be
(\ep_0 u^\mu u^\nu + p_0 \De^{\mu \nu}) \p_\mu \le(e^\tau u_\nu \ri) + J_0^\mu \p_\mu \hmu
= - \ep_0 u^\mu \p_\mu e^\tau + p_0 e^\tau \p_\mu u^\mu + J_0^\mu \p_\mu \hmu
\ee
and~\eqref{1totd} follows, since from~\eqref{thermo} we have
\be
dp_0 = - (\ep_0+ p_0) d \tau + n_0 e^{-\tau} d \hmu \quad \to \quad
d (p_0 e^\tau) u^\mu = - \ep_0 u^\mu d e^\tau + J_0^\mu d \hmu  \ .
\ee
The conserved Noether current $S^\mu$ corresponding to~\eqref{utrn} can be written as
\be
S^\mu %= p \beta^\mu  -  T^{\mu \nu}_0 \de X_{a\nu} -  J^\mu_0 \de \phi_a
=  p_0 e^\tau u^\mu  -  T^{\mu \nu}_0 e^\tau u_\mu -  J^\mu_0 \hmu,
\ee
which is precisely the standard covariant form of the entropy current~\cite{Israel:1979wp}.
The entropy current has previously appeared as a Noether current in~\cite{Haehl:2015pja,deBoer:2015ija}. In fact this connection was central to developing the framework
proposed in~\cite{Haehl:2015pja}.

It can now be readily checked that~\eqref{utrn} is no longer a symmetry either beyond the leading order in the derivative expansion in $\tilde I^{(1)}$ or of $\tilde I^{(2)}_0$.
We have also not been able to find a generalization of~\eqref{utrn} under which
the action is invariant beyond $\tilde I^{(1)}_0$.
That~\eqref{utrn} is present only for $\tilde I^{(1)}_0$ is consistent with the physical expectation that a conserved entropy current is an accident
at the ideal fluid level. With noises or dissipations, we do not expect a conserved entropy current.

It is natural to ask what happens to the entropy current beyond the ideal fluid level at $O(a)$.
The local KMS condition will ensure that it has a non-negative divergence from the following
reasoning. As discussed in Sec.~\ref{sec:consh}, the partition function prescription of~\cite{Banerjee:2012iz,Jensen:2012jh} arises as a subset of the local KMS condition at $O(a)$. It has been shown
by~\cite{Bhattacharyya:2013lha,Bhattacharyya:2014bha}  that constraints from the partition function prescription
are equivalent to equality-type requirements  from the non-negative divergence of the entropy current to all orders in derivatives.  As seen in Sec.~\ref{sec:trans} the inequality constraints from non-negative divergence of the entropy current follow in our context from the well-definedness of the integration measure. We have examined this to first derivative order. In~\cite{Bhattacharyya:2013lha,Bhattacharyya:2014bha} it has been argued these first order inequalities are the only inequality constraints coming from the entropy current to all derivative orders. Thus at $O(a)$, the entropy current (suitably corrected at each derivative order) will have a non-negative divergence to all orders in derivatives.
At $O(a^2)$ level, where noises are included, we do not expect the divergence of the entropy current should be non-negative as noises are random fluctuations.

\subsection{Two-point functions}

Now let us consider~\eqref{kee} and~\eqref{aa0} in the small amplitude expansion
in the sources and dynamical fields. More explicitly, we write
\be
X^\mu_s (\sig) = \de_a^\mu \sig^a + \pi^\mu (\sig) + \cdots, \qquad
g_{s \mu \nu} (x) = \eta_{\mu \nu} + \ga_{s \mu \nu} (x),
\ee
and expand~\eqref{kee} and~\eqref{aa0} to quadratic order in $\ga_{\mu \nu}, A_\mu$ and $\pi^\mu, \vp, \tau$ with dynamical and source fields considered to be of the same order.
It is then straightforward, but a bit tedious, to integrate out the dynamical fields to obtain
the generating functional for all retarded and symmetric two-point functions among components of the stress tensor and current in the hydrodynamical regime.

 One can readily verify that with thermodynamical relations~\eqref{thermo}, the Onsager relations~\eqref{11ons}, and the local FDT relations~\eqref{1tv}--\eqref{2sc}, the full quadratic Green functions satisfy the FDT relations~\eqref{fdt4} and~\eqref{1onsa}.

The explicit quadratic action and the final expressions are a bit long. Here we will first outline the general structure and then present the final expression of the generating functional for a neutral fluid.

We will take the spatial momentum $\vk$ of external fields to be along the $z$ direction, i.e. $k_z =q$ and $k_\al =0$ with $\al$ denoting all the transverse spatial directions. Then the background metric and gauge fields can be separated into three sectors
\bega
 {\rm tensor}: \qquad  \hat \ga_{\al \beta} = \ga_{\al \beta}  - {1 \ov d-2} \ga \de_{\al \beta}, \\
{\rm vector:} \qquad  a_\al = \ga_{0 \al}, \; \; b_\al = \ga_{z \al} , \; \; A_\al, \\
 {\rm scalar:} \qquad \ga_{00}, \; \ga_{0z}, \; \ga_{zz}, \; \ga = \sum_\al \ga_{\al \al} , \; A_0, \; A_z,  \
\end{gather}
where we have again suppressed $1,2$ subscripts. Again, below, $r,a$ will be used to denote the symmetric and antisymmetric combinations of these variables.

After integrating out the dynamical modes, the final generating functional should be diffeomorphism and gauge invariant, i.e. invariant under
\be \label{1infg}
\delta \gamma_{\mu \nu}=-2\partial_{(\mu}\xi_{\nu)}-\xi^\lam \partial_\lam \gamma_{\mu \nu}-2\gamma_{\lam(\mu}\partial_{\nu)}\xi^\lam + \cdots,\qquad \delta  A_\mu=-\partial_\mu\sigma -\partial_\mu \xi^\lam  A_\lam-\xi^\lam\partial_\lam A_\mu + \cdots,
\ee
for arbitrary infinitesimal fields $\xi^\mu$ and $\sig$. Then to quadratic order in external fields,
the final generating functional can be written as
\be
W = W_1 + \tilde W_2 + W_2,
\ee
where $W_1$ is linear in the external fields, i.e. giving one-point functions
\be
W_1 = \frac{i}{2} \ep_0 \gamma_{a00}+\frac{i}{2} p_0 (\gamma_{azz} + \ga_a)+ i n_0 A_{a0},  \
\ee
with $\ep_0, p_0, n_0$ all constants. Clearly $W_1$ is invariant under the linear part of~\eqref{1infg}.
Its variations under the quadratic part of~\eqref{1infg} are canceled by the variations of
the quadratic piece $\tilde W_2$ under the linear part of~\eqref{1infg}.
The other quadratic piece, $W_2$, is invariant under the linear part of~\eqref{1infg} by itself, and thus must
be expressed in terms of the following (linear) gauge invariant combinations:
\be
\hat \ga_{\al \beta} , \quad Z_\al = q a_\al + \om b_\al , \quad A_\al ,  \quad
Z = q^2 \ga_{00} + 2 \om q \ga_{0z} + \om^2 \ga_{zz} , \quad \ga ,
\quad E_z = \om A_z + q A_0, \
\ee
where we have again suppressed $r,a$ indices.

Let us now give the explicit expressions for $\tilde W_2$ and $W_2$ for a neutral fluid.
For the tensor sector, we have $\tilde W_2^{\rm tensor} =0$ and
\be
W_2^{\rm tensor} = -{i \ov 2} p_0 \hat \ga_{a\al \beta }\hat \ga_{r \al \beta} -  {\eta T_0 \ov 2 } \hat \ga_{a \al \beta}^2 - {i \ov 2} \eta \hat \ga_{a \al \beta}\partial_0 \hat \ga_{r \al \beta} \ ,
\ee
where we have used the first equation of~\eqref{1tv}.

For the vector sector, we have
\be
\tilde W_2^{\rm vector} = - i \ep_0 a_{a\alpha} a_{r \alpha}  - i  p_0 b_{a\alpha} b_{r\alpha},
\ee
and
\be
W_2^{\rm vector} = i \frac{\eta }{- i \om + q^2 D }  Z_{a\alpha} Z_{r\alpha} -
\frac{\eta T_0}{\om^2 + q^4 D^2}
 Z_{a\alpha}^2,
\ee
where we have kept only the leading term in the numerators in the small $\om$ and $q$ expansion, and the momentum diffusion constant $D$ takes its expected value:
\be
D =  {\eta \ov \ep_0 + p_0}  \ .
\ee

For the scalar sector, we have
\bega
\tilde W_2^{\rm scalar}=  {i \ep_0 \ov 4} \le[\gamma_{a00}\gamma_{r 00} - {1 \ov \om^2}
(q \ga_{a00} + 2 \om \ga_{a0z}) (q \ga_{r00} + 2 \om \ga_{r0z}) \ri] \cr
 - {i p_0 \ov 4} \le[\gamma_{azz}\gamma_{rzz}  - (\gamma_{azz}-\gamma_{a00})\gamma_r -\gamma_a(\gamma_{rzz}-\gamma_{r00}) - {1 \ov q^2} (\om \ga_{azz} + 2 q \ga_{a0z}) (\om \ga_{rzz} + 2 q \ga_{r0z}) \ri],
 \label{90}
\end{gather}
and
\be
\begin{split}
W_2^{\rm scalar}=& i K_1 \gamma_a \gamma_r + i K_2 Z_a Z_r+ i K_3 (\gamma_a Z_r +Z_a\gamma_r)
- \ha  G_1 \gamma_a^2   -
\ha G_2 Z_a^2  -  G_3 Z_a  \gamma_a,
\end{split}
\label{91}
\ee
where
\bega
K_1=\frac{- (d-2)(\varepsilon_0+p_0) c_s^2 \omega^2 + (d-4)p_0 (\om^2 - c_s^2 q^2)}{4 (d-2)R}, \cr
G_1=\frac{\ze  %+ (d-1) f_{14} c_s^2 )
\omega^4+  {2  \eta \ov (d-1) (d-2)} (\om^2 - (d-1) c_s^2 q^2)^2}{2\beta_0 R^* R}, \cr
K_2=- \frac{p_0 \omega^2+ \ep_0 c_s^2 q^2}{ 4 q^2 \omega^2 R},\qquad
K_3=- \frac{(\ep_0 + p_0) c_s^2 }{4 R}, \cr
G_2=\frac{  \ze + {2(d-2) \ov d-1} \eta }{2\beta_0 R^* R},\qquad G_3=\frac{\ze \omega^2
- {2 \eta \ov d-1}(\om^2 - (d-1) c_s^2 q^2)}{2 \beta_0 R^* R},
\end{gather}
and
\be R= \omega^2 - c_s^2 q^2 % (1 -  i r_1  \omega)
+ i {1 \ov \ep_0 + p_0} \le({2 (d-2) \ov d-1} \eta + \ze \ri)  \omega q^2+O(\omega^4,\omega^2 q^2,q^4) \ .
\ee
In the above expressions,
\be
c_s^2 = {\p_\tau p_0 \ov \p_\tau \ep_0}
\ee
is the sound velocity. Clearly the expressions exhibit the expected sound pole and attenuation constant.
One can also check that the apparent singularity at $\om =0$ and $q=0$ in~\eqref{90} and~\eqref{91} cancel.

\section{Discussion} \label{sec:discussion}

We conclude this paper by mentioning some future directions.

Firstly, it would be interesting to explore the physical implications of the new constraints for hydrodynamical equations of motion from the generalized Onsager relations proposed in this paper. We already saw that these relations lead to nontrivial new constraints
for the vector theory starting at the second derivative order for cubic terms. For a full charged fluid,
these relations will also lead to new constraints at second derivative order. It would be of clear interest to work them out explicitly and to understand their physical implications. We also hinted in Sec.~\ref{sec:consh} that local KMS condition may give rise to new inequality constraints at higher derivative orders. It would also be interesting to explore it further.

Secondly, the discussion of the bosonic action can be generalized in many different respects,
to more than one conserved currents or non-Abelian global symmetries,
parity and time reversal violations, inclusion of a magnetic field, anomalies, non-relativistic systems, superfluids, as well as
anisotropic and inhomogeneous systems. Also important is to generalize it to situations with additional gapless modes, such as systems near a phase transition or with a Fermi surface.

Thirdly, it is clearly of importance to use our formalism to
study effects of hydrodynamical fluctuations in various physical contexts\footnote{See e.g.~\cite{Kovtun:2011np,Kovtun:2012rj,Gripaios:2014yha,Endlich:2010hf} for some recent discussions of the effects of fluctuations.}, in particular to non-equilibrium situations. Furthermore, it would be very interesting to understand physical implications of ``ghost'' fields.

Finally, the relation between supersymmetry and the KMS conditions should be understood better.
Even for the theory of a single vector current, our understanding of the role of supersymmetry at both the classical statistical and quantum level can be much improved.  At the classical statistical level, do the local KMS conditions combined with supersymmetry ensure all the KMS conditions at all loop levels?
 While it is tempting to conjecture the answer is the affirmative we do not yet have a full proof. At the quantum level, how should the $\hbar$ deformed ``supersymmetric'' algebra %~\eqref{11alg}
\be
[\de, \bar \de] = \bar \ep \ep \, 2 \tanh {i \beta_0  \p_t \ov 2}
\ee
be generalized to nonlinear level? Another important problem is to write down the fermonic part
of the full charged fluid action. This is straightforward to do in a small amplitude expansion at quadratic, cubic, or higher orders, as in the theory of a single vector current, but the number of terms greatly proliferate and the analysis gets tedious. It is certainly more desirable to write down a full nonlinear fermonic action. This appears to require a supergravity
theory at the classical statistical level due to the time diffeomorphism in the fluid spacetime, and a ``quantum deformed'' supergravity theory at the quantum level.

\vspace{0.2in}   \centerline{\bf{Acknowledgements}} \vspace{0.2in}
We thank E.~Altman, T. Banks,  S.~Dubovsky, G.~Festuccia, P.~Gao, F.~Haehl, U.~Heinz, L.~Hui, N.~Iqbal, K.~Jensen, M.~Kardar, R.~Loganayagam, J.~Maldacena, S.~Minwalla, A.~Nicolis, M.~Rangamani, K.~Rajagopal, N. Seiberg, A.~Starinets, D.~T.~Son,  Y.~Wang, A.~Yarom, and B.~Zwiebach for correspondence and discussion. Work supported in part by funds provided by the U.S. Department of Energy
(D.O.E.) under cooperative research agreement DE-FG0205ER41360.

\appendix

\section{Explicit forms of various response and fluctuation functions} \label{app:functions}

At two point level, we have
\bega
G_{ra} (t_1, t_2) = G_R (t_1, t_2) \equiv i \th (t_{12}) \vev{[\sO (t_1), \sO (t_2) ]}, \\
G_{ar} (t_1, t_2) = G_{ra} (t_2, t_1) = G_A (t_1, t_2) \equiv -i \th (t_{21}) \vev{[\sO (t_1), \sO (t_2) ]}, \\
G_{rr} (t_1, t_2) = G_S (t_1, t_2) \equiv \ha \vev{\{\sO (t_1), \sO (t_2)\}},
\end{gather}
where $t_{12} = t_1 - t_2$, and $[\cdots]$ and $\{\cdots\}$ denote commutators and anticommutators respectively.
From~\eqref{dfun3}, at three point level,
\begin{align}
G_{raa}(1,2,3)&= - \theta(t_{12})\theta(t_{23})\langle\left[\left[\mathcal{O}(1),\mathcal{O}(2)\right],\mathcal{O}(3)\right]\rangle\\
&- \theta(t_{13})\theta(t_{32})\langle\left[\left[\mathcal{O}(1),\mathcal{O}(3)\right],\mathcal{O}(2)\right]\rangle,\\
G_{rra}(1,2,3)&= {i \ov 2} \theta(t_{13})\theta(t_{23})\langle\left[\left\{\mathcal{O}(2),\mathcal{O}(1)\right\},\mathcal{O}(3)\right]\rangle\\
&+ {i \ov 2} \theta(t_{13})\theta(t_{32})\langle\left\{\left[\mathcal{O}(1),\mathcal{O}(3)\right],\mathcal{O}(2)\right\}\rangle\\
&+ {i \ov 2} \theta(t_{31})\theta(t_{23})\langle\left\{\left[\mathcal{O}(2),\mathcal{O}(3)\right],\mathcal{O}(1)\right\}\rangle,\\
G_{rrr}(1,2,3)&={1 \ov 4} \theta(t_{21})\theta(t_{31})\langle\left\{\mathcal{O}(1),\left\{\mathcal{O}(2),\mathcal{O}(3)\right\}\right\}\rangle\\
&+{1 \ov 4} \theta(t_{12})\theta(t_{32})\langle\left\{\mathcal{O}(2),\left\{\mathcal{O}(3),\mathcal{O}(1)\right\}\right\}\rangle\\
&+{1 \ov 4} \theta(t_{13})\theta(t_{23})\langle\left\{\mathcal{O}(3),\left\{\mathcal{O}(1),\mathcal{O}(2)\right\}\right\}\rangle \ .
\end{align}
Other orderings can be obtained by switching the arguments of $\sO$'s, e.g.
\be
G_{rar}(1,2,3) = G_{rra} (1,3,2) \ .
\ee

\section{Fluctuation-dissipation theorem at general orders} \label{app:fdt}

In this appendix, we first review and slightly extend the formulation of KMS conditions at general
orders developed in~\cite{Wang:1998wg}, and then use the formalism to prove the relation~\eqref{allc2}.

\subsection{Properties of various Green functions}

We can expand $W$ and $W_T$ defined respectively in~\eqref{pager1} and~\eqref{defwt}  as
\begin{gather}
\label{w12}
W = \sum_{n=1}^\infty {(-1)^{n_2} i^n \ov n !}
G_{a_1 i_1 \, a_2 i_2 \, \cdots a_n i_n} \phi_{a_1 i_1} \cdots \phi_{a_n i_n} , \\
W_T = \sum_{n=1}^\infty {(-1)^{n_1} i^n \ov n !}
\tilde G_{a_1 i_1 \, a_2 i_2 \, \cdots a_n i_n} \phi_{a_1 i_1} \cdots \phi_{a_n i_n},
\label{tw12}
\end{gather}
where $i_k$ label different operators, $a_i =1, 2$, and $n_{1,2}$ are the number of $1$ and $2$ indices respectively.
In the above equations, integrations over the positions of $\phi$'s  should be understood.
Below, we will use a simplified notation to denote $G_{a_1 i_1 \, a_2 i_2 \, \cdots a_n i_n} $ as $G_{\al I}$,
with $G_{\bar \al I}$ denoting the corresponding Greens function obtained from $G_{\al I}$ by switching $1 \lra 2$.
By definition, in coordinate space
\be \label{idef}
G^*_{\al I} (x) = G_{\bar \al I} (x), \qquad \tilde G_{\al I}^* (x) = \tilde G_{\bar \al I} (x)
\ee
and in momentum space
\be \label{idef11}
G^*_{\al I} (k) = G_{\bar \al I} (-k), \qquad \tilde G_{\al I}^* (k) = \tilde G_{\bar \al I} (-k)
\ee
where we use $x$ and $k$ to collectively denote $x_1, x_2 \cdots$ and $k_1, k_2 , \cdots$ respectively.

 It is also convenient to introduce
\be \label{1eod}
G_{\al I}^{(e)} = \ha (G_{\al I} + G_{\bar \al I}), \qquad G_{\al I}^{(o)} = {1 \ov 2i} (G_{\al I} - G_{\bar \al I}),
\ee
and similarly for $\tilde G$. From~\eqref{idef}, $G^{(e)}_{\al I}$ and $G^{(o)}_{\al I}$ are real in coordinate space, and in
momentum space satisfy
 \be
 G_{\al I}^{(e) *} (k )=  G_{\al I}^{(e) } (-k), \qquad
 G_{\al I}^{(o) *} (k )=  G_{\al I}^{(o) } (-k )  \ .
 \ee
%\be
% G_{\al I}^{(e) *} (k_1 , \cdots , k_n )=  G_{\al I}^{(e) } (-k_1 , \cdots , -k_n ), \qquad
% G_{\al I}^{(o) *} (k_1 , \cdots , k_n )=  G_{\al I}^{(o) } (-k_1 , \cdots , -k_n )  \ .
% \ee
 Note that $G_{\al I}^{(e)}$ ($G_{\al I}^{(o)}$) is symmetric (antisymmetric) under $1\lra 2$
and thus contains an even (odd) number of $a$-operators, i.e.
\be
 G_{\al I}^{(o)} = \sum_{n_a \, {\rm odd}} G_{\al_1 \cdots \al_n} , \qquad
 G_{\al I}^{(e)} = \sum_{n_a \, {\rm even}} G_{\al_1 \cdots \al_n},
 \ee
 where $\al_i = a,r$ and $n_a$ is the number of $a$ indices.
Since
\be\label{aazo}
0 = G_{a \cdots a}  = \sum_\al (-1)^{n_2} G_{ \al I} = \sum_\al (-1)^{n_2}   \bca G_{\al I}^{(e)} & n \;{\rm even}, \cr
   i G_{\al I}^{(o)} & n \;{\rm odd},
   \eca
\ee
we conclude from~\eqref{3top} that
\be \label{ope0}
0 = \sum_\al (-1)^{n_2}   \bca G_{\al I}^{(e)} & n \;{\rm even}, \cr
   G_{\al I}^{(o)} & n \;{\rm odd}.
   \eca \
   \ee
There is a parallel relation for $\tilde G$.

Note that the response functions can be expressed as
\be \label{ope1}
G_{r a \cdots a} = \bca {(-1)^{n-1 \ov 2} \ov 2} \sum_{a_i} (-1)^{n_2} \le(G^{(e)}_{1a_1 \cdots a_{n-1}} +  G^{(e)}_{2a_1 \cdots a_{n-1}}  \ri)  = (-1)^{n-1 \ov 2}  \sum_{a_i} (-1)^{n_2} G^{(e)}_{1a_1 \cdots a_{n-1}} & n \; {\rm odd}, \cr \cr
 {(-1)^{n \ov 2} \ov 2} \sum_{a_i} (-1)^{n_2} \le(G^{(o)}_{1a_1 \cdots a_{n-1}} +  G^{(o)}_{2a_1 \cdots a_{n-1}}  \ri)
 =   (-1)^{n \ov 2}  \sum_{a_i} (-1)^{n_2} G^{(o)}_{1a_1 \cdots a_{n-1}} &  n \; {\rm even},
 \eca
\ee
where $a_i = 1,2$ and $n_2$ counts the number of $2$-index among $a_1, \cdots a_{n-1}$.

\subsection{KMS conditions in terms of correlation functions}

From the expansion~\eqref{w12}--\eqref{tw12}, the KMS conditions~\eqref{newfdt}
can be written in momentum space as\footnote{Here we  use the momenta of $\phi$'s to denote $G$. For example $\int dx_1 dx_1 G (x_1, x_2) \phi (x_1) \phi (x_2) =  \int dk_1 dk_1 G (k_1, k_2) \phi (k_1) \phi (k_2)$. Thus, $G(k_1, k_2)$ is the Fourier transform of $G (x_1, x_2)$ using an opposite convention.
 }
\be \label{kms1}
G_{\al I} (k)=  e^{-\beta \Om_2}  \tilde G_{\bar \al I} (k) , \qquad G_{\bar \al I} = e^{\beta \Om_2}  \tilde G_{\al I},
\ee
where $\Om_{2}$ denote the sum of all frequencies of $2$-operators as indicated by index $\al$.~\eqref{kms1} can further be written in terms of~\eqref{1eod}
as
\be \label{fdt5}
G_{\al I}^{(e)}+ \tilde G_{\al I}^{(e)}=  - i \coth {\beta \Om_2 \ov 2} \le(G_{\al I}^{(o)}  + \tilde G_{\al I}^{(o)}  \ri), \qquad
G_{\al I}^{(e)} - \tilde G_{\al I}^{(e)}= - i \tanh {\beta \Om_2 \ov 2} \le(G_{\al I}^{(o)}  - \tilde G_{\al I}^{(o)}  \ri) \ .
\ee
Note that the above equations relate correlation functions containing an even number of $a$-operators to those
containing an odd number of $a$-operators, and thus can be considered generalized fluctuation-dissipation theorems.

Now consider the case that the system is $\sP \sT$ invariant. From~\eqref{cpti1}, we then have
\be
%\tilde G_{\al I} (x_1, \cdots,  x_n) = \eta_I G^*_{\al I} (- x_1, \cdots, -x_n) = \eta_I G_{\bar \al I} (-x_1 \cdots -x_n)
\tilde G_{\al I} (x) = \eta_I G^*_{\al I} (- x) = \eta_I G_{\bar \al I} (-x)
, \qquad
\eta_I = \prod_{k} \eta_{i_k}^{PT}, \
\ee
where we have used~\eqref{idef}.
In momentum space, we then have
\be
%\tilde G_{\al I} (k_1, \cdots,  k_n)
%= \eta_I G_{\al I}^* (k_1, \cdots, k_n) =  \eta_I G_{\bar \al I} (-k_1, \cdots, -k_n) \ .
\tilde G_{\al I} (k)
= \eta_I G_{\al I}^* (k) =  \eta_I G_{\bar \al I} (-k) \ .
\ee
With $\eta_i^{PT} =1$, then equation~\eqref{kms1} becomes
\be  \label{fdt7}
G_{\al I} (k) = e^{-\beta \Om_2} G_{\al I} (-k)
\ee
%we then have
%\be
%\tilde G_{\al I}^{(e)} (k) =  G_{\al I}^{(e)*}  (k) = G_{\al I}^{(e)} (- k) , \qquad
%\tilde G_{\al I}^{(o)} (k) =  - G_{\al I}^{(o)*}  (k) = - G_{\al I}^{(o)} (- k)  \ .
%\ee
and~\eqref{fdt5} becomes
\be \label{fdt6}
{\rm Re} \, G_{\al I}^{(e)} =   \coth {\beta \Om_2 \ov 2} {\rm Im} \,  G_{\al I}^{(o)}  , \qquad
{\rm Im} \, G_{\al I}^{(e)} = - \tanh {\beta \Om_2 \ov 2} {\rm Re} \,  G_{\al I}^{(o)}  \ .
\ee

Now let us discuss some immediate implications of~\eqref{fdt7}--\eqref{fdt6}.
%From~\eqref{fdt7} we immediately conclude that

%From~\eqref{fdt6} we conclude that

\ben

\item All correlation functions of $\sO_{Ai} (x) \equiv \sO_{1i} (t, \vx) -  \sO_{2i} (t - i \beta, \vx)$
among themselves are zero, i.e.
\be  \label{ggg1}
G_{A \cdots A} (x) \equiv \vev{\sO_{A i_1} (x_1) \cdots \sO_{A i_n} (x_n)} = 0 \ .
\ee
  To see this note that  $G_{A \cdots A}$ can be written in momentum space as
\be
G_{A \cdots A} (k) = \sum_{\al} (-1)^{n_2} e^{\beta \Om_2} G_{\al I} (k) = \sum_{\al} (-1)^{n_2} G_{\al I} (-k) = G_{a\cdots a} (-k) = 0
\ee
where in the second equality we have used~\eqref{fdt7} and in the third equality used~\eqref{aazo}.
Note that in momentum space
\be
\sO_{A} (\om) = \le(1 - e^{-\beta \om} \ri) \sO_r +  \ha \le(1 + e^{-\beta \om} \ri) \sO_a
=  \ha \le(1 + e^{-\beta \om} \ri) \tilde \sO_A (\om)
\ee
with
\be
\tilde \sO_A (x) = \sO_a + 2 \tanh {i \beta_0 \p_t \ov 2} \sO_r  \ .
\ee
Thus correlation functions of $\tilde \sO_A $ with themselves are also all zero.
In the $\hbar \to 0$ limit discussed in Sec.~\ref{sec:susy} and Sec.~\ref{sec:hbar},
\be \label{gggf}
\tilde \sO_{A} (x) =  \sO_a  (x)+ i \beta_0 \p_t  \sO_r (x) \ .
\ee
Note that for two-point functions~\eqref{ggg1} is the full condition, but this is not the case for $n \geq 3$.

\item  $\Om_2 =0$ automatically for $\al = 2,\cdots 2$. In order for~\eqref{fdt6} to be nonsingular, we need
 \be \label{con1}
{\rm Im} \, G_{2 \cdots 2 I}^{(o)}   = 0, \qquad  {\rm Im} \, G_{2 \cdots 2 I}^{(e)} %- \tilde G_{2 \cdots 2 I}^{(e)}
=0 \ .
\ee

\item Taking $\Om_2 \to 0$, we conclude that
\be \label{con2}
{\rm Im} \, G_{\al I}^{(o)}  (\Om_2 =0)  = 0, \qquad  {\rm Im} \, G_{\al I}^{(e)} (\Om_2 =0) %- \tilde G_{2 \cdots 2 I}^{(e)}
=0 \ .
\ee

\item Consider the $\om_i \to 0$ limit for all $i$. For all $\al$, then,
 \be \label{con3}
{\rm Im} \, G_{\al I}^{(o)} (\om_i \to 0)  = 0, \qquad  {\rm Im} \, G_{\al I}^{(e)} (\om_i \to 0)  %- \tilde G_{2 \cdots 2 I}^{(e)}
=0 \ .
\ee
\een

\subsection{Implications for response functions} \label{app:fdtproof}

%Let us now look at the implications of these conditions for response functions.
%From~\eqref{con3}  for all $n$
%\be \label{gew}
 %{\rm Im} \, G_{r a \cdots a}  (\om_i \to 0) = 0 \ ,
%\ee
%i.e. the static limit of response functions must be real, which is certainly expected physically.
%Equations~\eqref{con1}--\eqref{con2} lead to more subtle conditions.
Denoting
\be \label{1defkk}
K_1 = G_{ra \cdots a}, \qquad K_2 = G_{ara \cdots a} , \qquad \cdots  \qquad
K_n = G_{a \cdots a r},  \
\ee
we now show that when taking any $n-2$ frequencies to zero, e.g.
\be \label{1allc2}
K_1 = K_2^* , \qquad \om_3, \om_4, \cdots , \om_n \to 0  \ .
\ee
For definiteness, let us take $n$ even. From~\eqref{ope1}, we then find that
\bega
K_1 = (-1)^{n \ov 2}  \sum_{a_i} (-1)^{n_2} \le(G^{(o)}_{11 a_1 \cdots a_{n-2}}  - G^{(o)}_{12 a_1 \cdots a_{n-2}} \ri), \\
K_2 = (-1)^{n \ov 2}  \sum_{a_i} (-1)^{n_2} \le(G^{(o)}_{11 a_1 \cdots a_{n-2}}  + G^{(o)}_{12 a_1 \cdots a_{n-2}} \ri),
\end{gather}
and
\bega \label{o1}
K_1 + K_2 =  2 (-1)^{n \ov 2}  \sum_{a_i} (-1)^{n_2} G^{(o)}_{11 a_1 \cdots a_{n-2}} , \\
K_1 - K_2 = -2 (-1)^{n \ov 2}  \sum_{a_i} (-1)^{n_2} G^{(o)}_{12 a_1 \cdots a_{n-2}}  \ .
\label{o2}
\end{gather}
For $\om_3, \cdots, \om_n =0$, using~\eqref{con1}--\eqref{con2}, we have
\be \label{o5}
{\rm Im} G^{(o)}_{11 a_1 \cdots a_{n-2}}  = 0 , \qquad {\rm Im} G^{(e)}_{11 a_1 \cdots a_{n-2}}  = 0,
\ee
which when applied to~\eqref{o1} leads to
\be \label{o3}
{\rm Im} (K_1 + K_2 ) = 0  \ .
\ee
Taking the real part of~\eqref{o2}, and using~\eqref{fdt6}, we then find that
\be \label{o4}
 {\rm Re} (K_1 - K_2) = 2 \coth {\beta \om_2 \ov 2} (-1)^{n \ov 2}  \sum_{a_i} (-1)^{n_2} {\rm Im} G^{(e)}_{12 a_1 \cdots a_{n-2}} \ .
 \ee
Now, from~\eqref{ope0}, we find that
\be
  \sum_{a_i} (-1)^{n_2} \le[G^{(e)}_{11 a_1 \cdots a_{n-2}}  -  G^{(e)}_{12 a_1 \cdots a_{n-2}}  \ri] = 0,
 \ee
 which when used in~\eqref{o4} (recall~\eqref{o5}) leads to
 \be \label{o6}
  {\rm Re} (K_1 - K_2) = 0 \ .
 \ee
 From~\eqref{o3} and~\eqref{o6}, we then find~\eqref{1allc2}.

From~\eqref{1allc2}, and permutations of it, it then follows that
\be \label{1allc}
 K_1 = K_2 = \cdots = K_n \equiv K, \quad {\rm Im} \, K = 0, \qquad {\rm all} \; \om_i \to 0   \ .
\ee
%Equation~\eqref{allc} says in the static case there is no retardation effect, and thus there should be no difference with regard to the location of $r$.

\section{KMS conditions for tree-level generating functional} \label{app:fdtar}

In this appendix, we show that in the vector theory~\eqref{1newc} local KMS conditions lead to KMS conditions for the full generating functional at tree-level. Recall that
\be
\label{1class}
W_{\rm tree}  [\phi_r, \phi_a]  \equiv i  I_{\rm on-shell} [\phi_r , \phi_a]
= i I [\chi_a^{\rm cl}, \chi_r^{\rm cl}; \phi_r, \phi_a],
\ee
where $\chi^{\rm cl} [\phi_r, \phi_a]$ is the solution to the equations of motion.
Below we will use $\chi$ and $\phi$ to collectively denote the dynamical and background fields.

For this purpose, we first note a general result regarding an on-shell action:
suppose an action has a symmetry
\be \label{1sym}
I [\chi; \phi] = I [\tilde \chi; \tilde \phi],
\ee
where variables with a tilde are related to the original variables by some transformation. Then
\be \label{1onsh}
I_{\rm on-shell}  [\phi] = I_{\rm on-shell}  [\tilde \phi] \ .
\ee
To see this, note that equation~\eqref{1sym} implies
\be
\tilde \chi^{\rm cl} [\phi] = \chi^{\rm cl} [\tilde \phi] \ ,
\ee
and thus
\be
I_{\rm on-shell}  [\phi]  = I [\chi^{\rm cl} [\phi]; \phi] = I [\tilde \chi^{\rm cl} [\phi]; \tilde \phi] =
I [\chi^{\rm cl} [\tilde  \phi]; \tilde \phi]  = I_{\rm on-shell}  [\tilde \phi] \ .
\ee

Now, for the theory~\eqref{1newc} of a single vector current, the local KMS conditions are
\be \label{1sou}
I_s [A_{1} , A_{2}] = - I_s [\tilde A_{1} , \tilde A_{2} ] , \qquad
\tilde A_{1 \mu} = A_{1 \mu} (-x), \; \tilde A_{2\mu} = A_{2 \mu} (-t - i \beta_0, -\vx) \ .
\ee
%where $\tilde A_{1 \mu} = A_{1 \mu} (-x), \,  \tilde A_{2\mu} = A_{2 \mu} (-t - i \beta_0, -\vx)$.
Given that $B_{\mu} = A_\mu + \p_\mu \vp$, the above equation implies that
\be
I [B_1, B_2 ] = I [\tilde B_1, \tilde B_2],
\ee
and thus
\be
I [\vp_1,\vp_2; A_1, A_2 ] = I [\tilde \vp_1,\tilde \vp_2; \tilde A_1, \tilde A_2 ],
\ee
where tildes again act as in~\eqref{1sou} and now $I$ is the full bosonic action.
From~\eqref{1onsh}, we then conclude that the local KMS conditions lead to KMS conditions for the tree-level generating functional.

\section{Derivative expansion for vector theory at cubic order} \label{app:example}

As an illustration of imposing the local KMS conditions at linear level, let us consider~\eqref{13r}
up to second order in derivatives in $K$, first order in derivatives in $H$ and zeroth order in derivatives in $G$.
The most general Lagrangian, then, which is rotationally invariant and satisfies~\eqref{1daug} can be written as
\begin{gather}
\sL_{aaa} ={a \ov 3!} B_{a0}^3 + {b  \ov 2} B_{a0} B_{ai}^2,    \\
\sL_{aar} = i \le[{\ba \ov 2} B_{a0}^2 B_{r0}  %+  {\ba_1 \ov 2} B_{a0}^2 \p_0 B_{r0}
 +  {\bar d \ov 2} B_{ai}^2   B_{r0} + B_{a0} (\bar c_1 \p_i B_{ai} B_{r0} + \bar c_2 B_{ai} \p_i B_{r0})
 + \bar f B_{a0} B_{ai} \p_0 B_{ri}  \ri], \\
\sL_{arr} = {\ta \ov 2}  B_{a0} B_{r0}^2  +  {\tb \ov 2} \p_i  B_{ai} B_{r0}^2 +
\tilde c_i  B_{a0} B_{r0} \p_0 B_{ri} +
 \te  B_{ai}  B_{r0} \p_0 B_{ri} + \tf_i B_{aj}  B_{r0}  F_{rij}
\cr
+ {\tg \ov 2}  B_{a0} (\p_0 B_{ri})^2  + {\tilde h \ov 2} B_{a0} F_{rij} F_{rij} +
 \tk  B_{ai} \p_0 B_{rj} F_{rij},
\end{gather}
where $a, b, c_1, c_2, \bar f, \tg, \tilde h, \tk$ are constants and
\bega
\bar a = \bar a_0 - i \om_3 \ba_1 , \quad \bd = \bd_0 - i \om_3 \bd_1 , \quad
\tb = \tb_0 - i \om_1 \tb_1, \quad \tc_i = i (\tilde c_2 k_{2i} + \tilde c_3 k_{3i}), \cr
\te = \te_0 - i \te_2 \om_2 - i \te_3 \om_3,  \qquad
\tf_i = i (\tf_2 k_{2i} + \tf_3 k_{3i}) , \cr
 \tilde a = \tilde a_0 - i \tilde a_1 \om_1 + \tilde a_2 (k_2^2 + k_3^2) + \ta_3 k_2 \cdot k_3 + \tilde a_4 (\om_2^2 + \om_3^2) + \tilde a_5 \om_2 \om_3 \ .
\end{gather}
Let us first look at the static conditions~\eqref{allc} which imply that
\be
\ta_2 = \ta_3 , \qquad \tf_2 = \tf_3  = - 2 \tilde h, \qquad \tb_0 =0 \ .
\ee
With time-dependent sources, equation~\eqref{allc2} further requires that
\be
\tc_3 = \tb_1, \qquad \tc_2 =0 \ .
\ee
Imposing the full FDT we find in addition that (in the $\hbar \to 0$ limit)
\begin{gather}
\bar a_0 = 2\frac{ \tilde a_1}{\beta}, \qquad \bar a_1 = -3\frac{\tilde a_4 - \tilde a_5}{\beta},  \qquad \bar d_0 = -2\frac{\tilde e_0}{\beta}, \qquad \bar d_1 = -\frac{2\tilde e_2 - \tilde e_3+\tilde g}{\beta},  \cr
 \bar f = -\frac{2 \tilde e_2 - \tilde e_3 + \tilde g}{\beta}, \quad  \bar c_1 = \bar c_2 = 0, \quad a=-6\frac{\tilde a_4 - \tilde a_5}{\beta^2}, \quad  b=-2\frac{ 2\tilde e_2 -  \tilde e_3 + \tilde g}{\beta^2}.
\end{gather}

%\be
%\bar a_0 = - {2 \ov \beta} \tilde a_1 , \quad a = {6 \ov \beta^2} (\tilde a_4 - \tilde a_3) , \quad
%\bar c_1 = \bar c_2 = 0, \quad \bar d = - {2 \ov \beta} \tilde e, \quad
%\bar f = - {1 \ov \beta} \tilde g = b \beta
%\ee

\section{Useful formulas}

\subsection{Integrability conditions} \label{app:inter}

From~\eqref{tbe0}, we have the integrability conditions
\begin{gather} \label{int1}
(- b v_i u^\nu + \lam_i{^\nu}) \p_\nu (b u^\mu) = b u^\nu \p_\nu (- b v_i u^\mu + \lam_i{^\mu} ), \\
(- b v_i u^\nu + \lam_i{^\nu}) \p_\nu (- b v_j u^\mu + \lam_j{^\mu} )  = (- b v_j u^\nu+ \lam_j{^\nu}) \p_\nu (- b v_i u^\mu + \lam_i{^\mu} ).
\label{int2}
\end{gather}
From~\eqref{int1} we get
\begin{gather} \label{int11}
\p v_i = - {1 \ov b^2} \lam_i{^\mu} \p_\mu b + {1 \ov b} \lam_i{^\mu} \p u_\mu,  \\
\p \lam_i{^\mu} =   \lam_i{^\nu} \nab_\nu u^\mu % - c_i \p u^\mu
+ u^\mu \lam_i{^\nu} \p u_\nu,
\label{lamg}
\end{gather}
where we have defined
\be
\p \equiv u^\mu \nab_\mu \ .
\ee
From~\eqref{tbe1}, we get
\begin{gather}
\label{int3}
\p_\nu \lam^i{_\mu} - \p_\mu \lam^i{_\nu} = 0 \\
\p_\mu \le({u_\nu \ov b} - v_i \lam^i{_\nu} \ri) = \p_\nu \le({u_\mu \ov b} - v_i \lam^i{_\mu} \ri)  \ .
%- {1 \ov b} \p_\nu b (u_\mu + w_\mu) + \p_\nu (u_\mu + w_\mu) + {1 \ov b} \p_\mu b (u_\nu + w_\nu) -
 %\p_\mu (u_\nu + w_\nu) = 0
\label{int4}
\end{gather}
%where equation~\eqref{int4} can be further written as
%\be \label{int5}
%- {1 \ov b} \p_\nu b (u_\mu + w_\mu) + {1 \ov b} \p_\mu b (u_\nu + w_\nu) + \p_\nu u_\mu   -
% \p_\mu u_\nu  + \lam^i{_\mu} \p_\nu c_i -  \lam^i{_\nu} \p_\mu c_i = 0 \ .
 %\ee

\subsection{Variations with respect to background metric and gauge field} \label{app:vary}

Here we list the variation of various quantities with respect to the external metric and gauge field.
For a single segment under variation of $g_{1\mu \nu}$, we have (with the subscript $1$ and $\de g_{1\mu \nu}$ suppressed)
\be
\de b = -{b \ov 2} u^\mu u^\nu %\de g_{\mu \nu}
, \quad
\de u^\rho = - {\de b \ov b} u^\rho = \ha u^\mu u^\nu  u^\rho   , \quad
% \de c_i = -{c_i \ov 2} u^\mu u^\nu - u^{(\mu} \lam_i{^{\nu)}} ,
 \de v_i = {1 \ov b} u^{(\mu} \lam_i{^{\nu)}}  , \quad
 \de \lam_i{^\rho} = {u^\rho } u^{(\mu} \lam_i{^{\nu)}} \ .
\ee
Including both segments under variations of $g_{1 \mu \nu} (X)$ we have
\begin{gather}
\de E_r = -{b \ov 4}  u^\mu u^\nu , \quad
\de \sqrt{a_r} = {1 \ov 4} \sqrt{a_r}   a_r^{ij} \lam_i{^\mu}  \lam_j{^\nu}, \quad
 \de a_{r ij} = \ha  \lam_i{^\mu}  \lam_j{^\nu} \quad \de E_a = -\ha u^\mu u^\nu,  \cr
\de  v_{ri} = \ha \de v_{ai} =   {1 \ov 2 b} u^{(\mu} \lam_i{^{\nu)}} , \qquad \de \chi_a %= \ha \de \tr \Xi
= \ha a^{ij} \lam_i{^\mu} \lam_j{^\nu} = \ha \De^{\mu \nu}, \cr
\de \mu_r = \ha \de \mu_a %= {1 \ov 4} e^\tau u^\rho \hat A_\rho u^\mu u^\nu
 = {1 \ov 4} \mu u^\mu u^\nu  , \quad
\de \fb_{ri} = \ha \de \fb_{ai} %= \ha u^{(\mu} \lam_i{^{\nu)}} u^\rho \hat A_\rho
  = \ha \mu u^{(\mu} \lam_i{^{\nu)}},
\end{gather}
where we have again suppressed $\de g_{1\mu \nu}$ and the index $1$ (all variables without an explicit subscript $r$ or $a$ should
be understood as having index $1$). The variation of $\Xi$ will be treated separately below.
Also note that under variation of $\de A_{1\mu}$, we find (again suppressing the subscript $1$)
\be
\de \mu_r = \ha \de \mu_a = \ha  u^\mu , \qquad \de \fb_{ri} = \ha \de \fb_{ai} = \ha \lam_i{^\mu} \ .
\ee

Now let us consider the variation of $\Xi$ under $\de g_{1\mu \nu}$, which is tricky due to the logarithm. As discussed in the main text, both the action and the stress tensor are organized as expansions of $a$-variables, it is thus enough for us to work out the variation as an expansion of $\Xi$. For this purpose,
let us first introduce
\be
\de_1 \equiv \hat a^{-1}_1 \de \hat a_1 = a^{ik}_1 \lam_{1k}{^\mu} \lam_{1j}^\nu - {\De^{\mu \nu}_1 \ov d-1} \de_i^j  \ .
\ee
Then expanding both sides of
\be
\hat a_2^{-1} \de \hat a_1 = e^\Xi \de_1 = \de e^{\Xi}
\ee
in $\Xi$, we find that
\be
\de \Xi = \de_1 + \ha [\Xi, \de_1] + O(\Xi^2)  \ .
\ee
Similarly, under a variation of $g_2$ we find that
\be
\de \Xi = - \de_2 + \ha [\Xi, \de_2] + O(\Xi^2) \ .
\ee

%\subsection{Relations to standard hydrodynamical variables}

%Using the integrability conditions (F1),(F2) in \cite{38}, we find
%\be \ft_{ij} = 2 e^{-\tau}\lambda_{[i}^\mu\lambda_{j]}^\nu\nabla_\mu u_\nu=2  e^{-\tau}\lambda_{i}^\mu\lambda_{j}^\nu\omega_{\nu\mu}.
%\ee

\section{Structure of stress tensor and current at order $O(a^0)$} \label{app:proof}

In this appendix, we prove that at leading order in $a$ expansion, the stress tensor and current
can be expressed in terms of velocity-type variables $u^\mu, \mu, \tau$ to all
derivative orders.

The stress tensor at $O(a^0)$ can be obtained by varying the action with respect to  $g_{1\mu\nu}$ and setting the $a$-type fields to zero. At this order, there is only one set of background fields and dynamical variables (see~\eqref{yyy}).  The $r$-subscripts can thus be dropped. From~\eqref{stress1}, we then find
\bega
  \hat T^{\mu \nu} (x) =  \le(\mu {\de \sL \ov \de \mu_a} - {\de \sL \ov \de E_a} \ri) u^\mu u^\nu +  {\de \sL \ov \de \chi_a} \De^{\mu \nu}  \cr
+ 2 {\de \sL \ov \de \Xi^i{_j}} \le(\lam^{i (\mu} \lam_j{^{\nu)}} - {\De^{\mu \nu} \ov d-1} \de_i^j  \ri) +
2   \le( \mu {\de \sL \ov \de \fb_{ai}} + {1 \ov E} {\de \sL \ov \de v_{ai}}  \ri) u^{(\mu} \lam_i{^{\nu)}}, \
\label{stress2}
\end{gather}
where  we have used~\eqref{onlam}. Similarly, the current can be written as
\be \label{cur2}
  \hat J^\mu =  {\de \sL \ov \de \mu_a}  u^\mu +
  {\de \sL \ov \de \fb_{ai}}  \lam_i{^\mu}   \ .
 \ee

We will now show that for the most general $\sL$ invariant under~\eqref{sdiff}--\eqref{tdiff}
and~\eqref{daug}, only velocity-type variables $u^\mu, \tau, \mu$ and their derivatives will occur in~\eqref{stress2}--\eqref{cur2}. %; \HL{(ii) all possible
%combinations of  $u^\mu, \tau, \hat \mu$ and their derivatives will occur. }

For this purpose, let us consider a general tensor under spatial diffeomorphisms~\eqref{sdiff}, invariant under~\eqref{tdiff} and~\eqref{daug}, which are constructed out of $r$-variables. Below we will refer to such a quantity as a spatial tensor. From our discussion of covariant derivatives
in Sec.~\eqref{sec:cord}, a spatial tensor of any rank can be constructed by acting with
$D_0,  D_i$ on the following basic objects:
\be \label{bloc}
  \tau,\quad  \mu,\quad D_i E,\quad D_0 \fb_{i}   ,\quad a_{ij},\quad \sB_{ij} , \quad
  \tilde R_{ijk}{^l} , \quad \ft_{ij}  \ .
\ee
%with indices will be raised by $a^{ij}$.
Recall that, acting on a vector $\vp_j$,
\be
D_i \vp_j = d_i \vp_j - \tilde \Ga^k_{ij} \vp_k,
\ee
with $d_i \equiv \p_i + v_{ri} \p_0$ and
\be \label{newch}
\tilde\Gamma^i_{jk} \equiv \ha a^{il} \left(d_j a_{kl}+ d_k a_{jl}
-d_l a_{jk}\right)  =  - \lam_k{^\mu} \lam_j{^\nu} \nab_\mu \lam^i{_\nu},
\ee
where we have used the integrability condition~\eqref{int3} in obtaining the last expression.
Similarly, with the help of various integrability conditions~\eqref{int11}--\eqref{int3}, we find
\bega\label{d11}
D_i E =  \lam_i{^\mu} \p u_\mu % - \p_\mu \tau )
, \\
\label{d12}
D_0 \fb_i =  \lam_i{^\mu} \le(\nab_\mu  \mu +  \mu \p u_\mu - u^\nu F_{\mu \nu} \ri) \\
\sB_{ij} = \lam_i{^\mu} \lam_j{^\nu} \le(F_{\mu \nu}  +  \mu (\nab_\mu u_\nu - \nab_\nu u_\mu) \ri)  \\
\label{d14}
\ft_{ij} = 2 \lambda_{i}^\mu\lambda_{j}^\nu\omega_{\nu\mu} , \quad
\omega^{\mu\nu}=-\Delta^{\mu\alpha}\Delta^{\nu\beta}\nabla_{[\alpha}u_{\beta]}  \\
\tilde R_{ijk}^{\ \ \ l}=\lambda_i{^\mu}\lambda_j{^\nu}\lambda_k{^\rho}\lambda^l{_\beta}\left[R_{\mu\nu\rho}^{\ \ \ \beta}+2\nabla_{[\mu}u^\beta\nabla_{\nu]} u_\rho-2\nabla_{[\mu}u_{\nu]}\nabla_\rho u^\beta\right] \ .
\label{d13}
\end{gather}
From~\eqref{d11}--\eqref{d13}, all quantities in~\eqref{bloc} are either scalars such as $\tau, \mu$, or tensors of the following form:
\be \label{bloc1}
\vp_{i} = \lam_i{^\mu} \vp_\mu, \qquad  \vp_{ij} = \lam_i{^\mu} \lam_j{^\nu}  \vp_{\mu \nu},
\ee
with $\vp_\mu, \vp_{\mu \nu}$ expressed in terms of velocity-type variables only (for $a_{ij}$ the corresponding $\vp_{\mu \nu}$ is $\De_{\mu \nu}$). Now one can show that acting with $D_0$ and $D_i$ on tensors of the form~\eqref{bloc1}, one again obtains a tensor of the form
\be \label{tens1}
\lambda_{i_1}^{\mu_1}\cdots\lambda_{i_n}^{\mu_n}\vp_{\mu_1\cdots\mu_n},
\ee
with $\vp_{\mu_1\cdots\mu_n}$ expressed in terms of velocity-type and background variables only.
Since $D_0$ and $D_i$ satisfy the Leibniz rule, it is enough to demonstrate their actions on a scalar $\vp$ and a vector $\vp_i$. It can be readily found then that
\be \label{cod0}
D_0 \vp =   \p \vp, \qquad D_i \vp = \lam_i{^\mu} \nab_\mu \vp \qquad D_0 \vp_i =
 \lam_i{^\mu} (\p \vp_\mu + \vp_\nu \nab_\mu u^\nu + \vp_\nu u^\nu \p u_\mu),
\ee
and
\be \label{codi}
D_i \vp_j =  \lam_i{^\mu} \lam_j{^\nu} \nab_\mu (\De_\nu{^\rho}  \vp_\rho)  \ .
  \ee
  To derive~\eqref{codi}, it is convenient to use the identity
  \be \label{cobm}
D_i \lam_j{^\mu} \equiv \lam_i{^\nu} \nab_\nu \lam_j^\mu -\tilde \Ga^k_{ij} \lam_k{^\mu} = \lam_i{^\al} \lam_j{^\beta} \nab_\al  \De_\beta{^\mu},
\ee
which follows from~\eqref{newch}.  With all tensors of the form~\eqref{tens1}, any scalar constructed out of them will then be in terms of velocity-type modes only, and any vector or two-tensors will also be of the form~\eqref{bloc1}.  Plugging these forms into~\eqref{stress2}--\eqref{cur2}, we then find that the stress tensor and current will have the form
\be
\hat T^{\mu \nu} = \ep u^\mu u^\nu + p \De^{\mu \nu} + %\De^{\mu \nu, \rho \sig}
t^{\mu \nu}+ u^{(\mu} q^{\nu)}
%\De^{\nu) \rho} q_\rho ,
\qquad \hat J^\mu = n u^\mu + \De^{\mu \nu} j_\nu,
\ee
where
\bega
\ep = \mu {\de \sL \ov \de \mu_a} - {\de \sL \ov \de E_a} , \qquad p =    {\de \sL \ov \de \chi_a} , \qquad  t^{\mu \nu} = 2  \lam^{i (\mu} \lam_j{^{\nu)}} {\de \sL \ov \de \Xi^i{_j}}  , \\
q^\mu = 2   \lam_i{^\mu} \le( \mu {\de \sL \ov \de \fb_{ai}} + {1 \ov E} {\de \sL \ov \de v_{ai}}  \ri), \qquad
n =  {\de \sL \ov \de \mu_a} , \qquad j^\mu =  \lam_i{^\mu}   {\de \sL \ov \de \fb_{ai}}
\end{gather}
are all expressed in terms of velocity-type variables.

We believe the converse statement is likely also true, i.e. any combinations of velocity-type variables can be obtained from variation of $I$ at order $O(a^0)$. This amounts to showing that any
tensors defined in $X^\mu$-space built out of $u^\mu, \tau, \mu$ and their covariant derivatives
can be expressed in terms of $D_0, D_i$ acting on quantities in~\eqref{bloc}. We will leave this for the future.
%For this purpose, let us note that from~\eqref{cod0}  longitudinal and transverse derivatives of $\vp = \tau, \hmu$
%can all be expressed as
%\be
%\p \vp =  e^{-\tau} D_0 \vp  , \qquad  \De^{\mu \nu} \nab_\nu \vp = \lam^{i \mu} D_i \vp \ .
%\ee
%Let us consider covariant derivative of $u_\mu$ which can be decomposed as
%\be
%\nab_\mu u_\nu =
%\ee

\end{document}